\documentclass[journal]{IEEEtranTIE}
\usepackage{amsmath,amsfonts}
\usepackage{array}
\usepackage[caption=false,font=normalsize,labelfont=sf,textfont=sf]{subfig}
\usepackage{textcomp}
\usepackage{stfloats}
\usepackage{url}
\usepackage{verbatim}
\usepackage{graphicx}
\usepackage{caption}

\UseRawInputEncoding
\usepackage{cite}
\graphicspath{{./Figures}}

\begin{document}
	
	\title{Scalable Cyber-Physical Testbed for  Cybersecurity Evaluation of Synchrophasors in Power Systems}
	
	\author{{Shuvangkar Chandra Das and Tuyen Vu\\
	
	\textit{Clarkson University}}
		% <-this % stops a space
		%\thanks{This paper was produced by the IEEE Publication Technology Group. They are in Piscataway, NJ.}% <-this % stops a space
		\thanks{Contact: tvu@clarkson.edu}}
	
	% The paper headers
	% \markboth{Journal of \LaTeX\ Class Files,~Vol.~14, No.~8, August~2021}%
	% {Shell \MakeLowercase{\textit{et al.}}: A Sample Article Using IEEEtran.cls for IEEE Journals}
	
	% \IEEEpubid{0000--0000/00\$00.00~\copyright~2021 IEEE}
	% Remember, if you use this you must call \IEEEpubidadjcol in the second
	% column for its text to clear the IEEEpubid mark.
	
	\maketitle
	\vspace{-20pt}
	\begin{abstract}
		This paper presents a real-time cyber-physical (CPS) testbed for power systems with different real attack scenarios on the synchrophasors-phasor measurement units (PMU). The testbed focuses on real-time cyber-security emulation with components including a digital real-time simulator, virtual machines (VM), a communication network emulator, and a package manipulation tool. The script-based VM deployment and the software-defined network emulation facilitate a highly-scalable cyber-physical testbed, which enables emulations of a real power system under different attack scenarios such as Address Resolution Protocol (ARP) poisoning attack, Man In The Middle (MITM)  attack, False Data Injection Attack (FDIA), and Eavesdropping Attack. The common synchrophasor, IEEE C37.118.2 named pySynphasor has been implemented and analyzed for its security vulnerabilities. The paper also presented an interactive framework of injecting false data into a realistic system utilizing the pySynphasor module. The framework can dissect and reconstruct the C37.118.2 packets, which expands the potential of testing and developing PMU-based systems and their security in detail and benefits the power industry and academia. A case for the demonstration of the FDIA attack on the linear state estimation together with the bad-data detection procedure are presented as an example of the testbed capability.
	\end{abstract}
	
	
	\section{Introduction}
	\IEEEPARstart{T}{he} power system is the most extensive interconnected machine made by humans. Modern civilization entirely depends upon electricity. Therefore, any compromise or disturbance of the power system significantly impacts the economy. Hence, real-time monitoring plays a vital role in modern cyber-physical power system\cite{prabhuStateoftheartReviewSynchrophasor2017}. Synchrophasor technology opens a new horizon in power systems by collecting real-time GPS time-stamped current and voltage phasors. A typical synchrophasor technology consists of Phasor Measurement Units(PMU), Phasor Data Concentrator(PDC), communication network, and control center\cite{khanThreatAnalysisBlackEnergy2016}.
	
	State estimation is a mathematical tool to compute the current states of the network based on redundant noisy measurements introduced by Fred Schweppe in 1968\cite{wuPowerSystemState1990}. Three types of state estimation are commonly used in power systems: AC state estimation, DC state estimation, and linear state estimation. AC state estimation involves finding the buses' complex voltage by iteratively solving the Weighted Least Square(WSL) optimization problem. So, it is computationally very intensive because of the iterative solution mechanism.
	
	Linear state estimation reduces the computation complexity of AC state estimation by utilizing synchrophasor technology. PMU is an integral part of synchrophasor technology which directly measures current and voltage phasors employing GPS as reference time. PDC receives data from multiple PMUs and then aggregates data based on GSP timestamp\cite{khanThreatAnalysisBlackEnergy2016}. IEEE standard C37.244-2013\cite{ieeepowerandenergysocietyIEEEStdC372013} suggested two communication protocols such as IEEE C37.118.2-2011 and IEC 61850-90-5 for PMU and PDC communication. However, IEEE C37.118.2 became the de-facto communication protocol for synchrophasor communication because of its compact packet size and low bandwidth requirement\cite{khanAnalysisIEEEC372016}.
	
	The transmission system is primarily balanced, so PMUs built for the transmission system are single phases. PMUs that are used in transmission is not suitable for distribution system due to the cost and technical constraints. Also, distribution systems have many different features, including radial topology, high resistance to reactance ratio, and three-phase unbalanced system\cite{wangRevisedBranchCurrentbased2004}. Therefore, $\mu$PMU was developed, focusing distribution system. It is a high precision synchrophasor device with comparatively low cost that works well in distribution system\cite{chenStateEstimationDistribution2015}.
	
	Modern power system primarily uses Open System Interconnection (OSI) protocols to communicate with all the control and measurement devices. Making the protocol works flawlessly rather than focusing on security was the only concern at the time of designing these protocols. As a result, many vulnerabilities of these standard protocols, such as man in the middle attack using the vulnerability of ARP cache poisoning and through ICMP packet manipulation  \cite{nayak2010} , have been enlisted and documented up to date. Moreover, the technology lifetime for cyber-physical power system is around 15-20 years, whereas conventional IT infrastructure changes within 3-5 years\cite{yangManinthemiddleAttackTestbed2012}. As a result, the power system communication infrastructure is more vulnerable to attack.

	Almost any system is vulnerable to attack. OpenSSH project was designed from the ground up with security as the primary concern. Every line of code was written with security as the top priority, and countless developers audited the changes daily. After all these efforts, a project like OpenSSH has a remotely exploitable vulnerability. Now, it is easy to imagine the condition of other systems which were not designed, focusing the security as the primary concern rather than making the system fully functional was the main focus. 
	
	Many types of cyber-attack in power systems are listed in the paper\cite{khanAnalysisIEEEC372016}. A few attacks already observed are Command manipulation, Code manipulation, malware injection, GPS spoofing, false data injection, denial of service, fuzzing, rouge node, and channel jamming. Still, there might be zero-day attacks. False Data Injection Attack (FDIA) is one of the major attacks on modern power systems where adversary craft fully injects malicious data into the system to compromise a part of the system or whole network. 2015 Ukraine blackout is the real implication of FDIA attack on large scale power system\cite{liang2015UkraineBlackout2017}. The attacker used BlackEnergy Version 3 malware to access the system and injected false command, which trips seven 110kV and 233kV substation breakers, resulting in blackout for more than 225k people for 6+ hours\cite{khanThreatAnalysisBlackEnergy2016}.

	\begin{figure*}[ht]
		\centering
		\includegraphics[width=14cm]{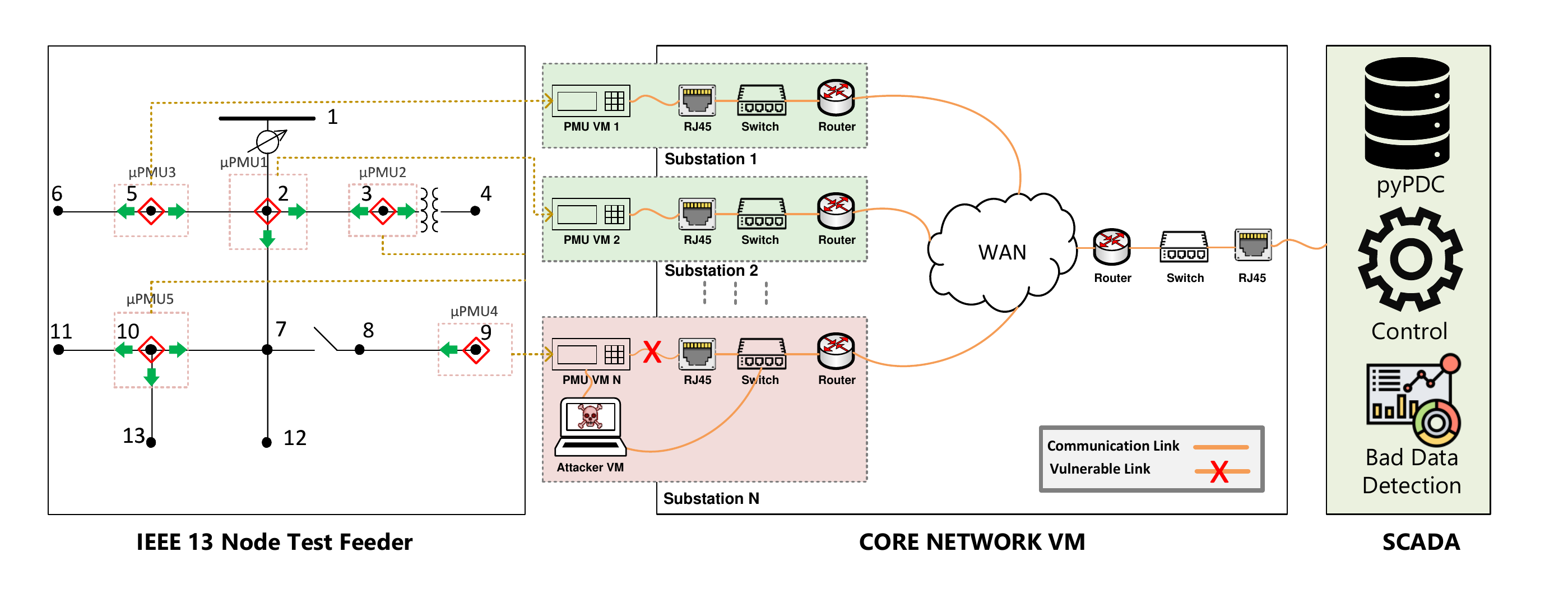}
		\caption{Cyber-Physical Testbed  Architecture}
		\vspace{-15pt}
		\label{fig:testbed_architecture}
	\end{figure*}
	
	Many widely used major communication protocols have vulnerabilities. Few vulnerabilities of the IEEE C37.118 protocol has been enlisted in the paper\cite{khanIEEEC3711822016}; Authentication attack, establishing PMU and PDC communication without any authentication;  MITM attack, involves hijacking session, altering, dropping and injecting C37.118.2 packets; Reply attack, involves recording packet and reply it multiple times to hide the real scenario; DoS attack, involves overwhelming target by high speed bulk packets which result in communication loss of PMU, PDC and control center. GPS spoofing attack on the PMU device has been presented in the paper\cite{fanSynchrophasorDataCorrection2018}. Unlike military GSP, the civilian GSP signal can be predicted by a low-cost GPS receiver, so it is easy to forge the matched version of the corresponding GPS signal by the attacker. Finally data tampering attack  on C37.118.2 protocol has been presented in \cite{paudelDataIntegrityAttacks2016}, \cite{singhStealthyCyberAttacks2016}.
	
	The testing of new applications must be done in environments that can characterize both the physical system and the cyber network due to the interconnected nature of the physical and cyber components. Testbeds are used to run strict and reproducible tests to verify new controls and applications and, most importantly, to find out security vulnerabilities. It is vital to understand how an attacker crafts different tools and vulnerabilities to perform a successful attack. Honeypot is the one way to study the attacks. It is used to analyze and learn new malware and generate anti-virus signatures. A low-interaction honeypot for detecting unauthorized traffic for the distribution system has been presented in the paper\cite{koltysSHaPeHoneypotElectric2015}. The problem with the low-interaction honeypot is that it emulates a small amount of internet protocols and services.
	For this reason, there is a tension between scalability and fidelity in a honeypot-based intrusion detection system. So, the testbed plays a vital role in this case as it exactly mimics the real system. Hence, fidelity and scalability can both be controlled. 
	
	The author in paper\cite{johnsonAssessingNetworkCybersecurity2020} designed a testbed in a co-simulation environment for DER focusing on power system performance, security trade-offs, network segmentation, and encryption. The author in the paper\cite{wlazloManinTheMiddleAttacksDefense2021} investigated on DNP3 for the case of MITM attack in emulation environments. They utilized Scapy extension of DNP3 developed in the paper\cite{rodofileRealTimeInteractiveAttacks2015} to design the attack. The testbed paper\cite{khanDemonstratingCyberphysicalAttacks2018} demonstrated how a hacker could develop a custom-made tool for performing a stealthy MITM attack against a synchrophasor device. Researchers in\cite{cuiCyberPhysicalSystem2020} developed a modular architecture utilizing software-defined network, virtual machine, and pyPMU\cite{sandiPypmuOpenSource2016} to monitor data acquisition and closed-loop control in a wide area network. A hardware-based testbed was presented in\cite{adhikariCyberphysicalPowerSystem2014} to develop an intrusion detection system. The testbed was modeled using a real-time digital simulator (RTDS), relays, PMU, PDC, and PC running a Snort intrusion detection system. In the paper\cite{khanDemonstratingCyberphysicalAttacks2018}, the authors developed a testbed scenario for attacking synchrophasor communication. They used Scapy to implement the MITM attack and a custom python script to pack and unpack the IEEE C37.118.2 packets. They have also developed the testbed based on a microgrid scenario with a single PMU device and implemented a management server in Raspberry Pi. The survey paper\cite{cintugluSurveySmartGrid2017} on smart grid listed a comprehensive review of different aspects of the smart grid, including physical power infrastructure, communication network, security and privacy, smart grid protocol, and cloud computing. A Fuss testing platform was developed in Queen's University cybersecurity testbed\cite{yangCybersecurityTestbedIEC2015} using RTDS, actual IEDs, and merging units for testing IEC 61850 protocol.
	
	It is obvious to understand the significance of testbed in studying power systems holistically. Yet, the majority of testbeds mentioned before were designed to evaluate and verify specific tasks. These testbeds are not practically scalable. Also, these testbeds focus on conventional SCADA-based systems. Moreover, only a few testbed\cite{khanDemonstratingCyberphysicalAttacks2018,cuiCyberPhysicalSystem2020,adhikariCyberphysicalPowerSystem2014} experimented on the IEEE C37.118.2 protocol and synchrophasor technology. However, these testbeds are unsuitable for large system deployment and vulnerability testing of synchrophasor technology. In addition, there are no open-source tools for analyzing the IEEE C37.118 protocol in python, according to our knowledge from literature reviews. Therefore, an emerging need for an IEEE C37.118 open-source tool that can perform different types of vulnerability testing, injection testing, and eavesdropping, false data injection attack on the phasor measurements.

	This research will demonstrate scalable cyber-physical system design in a simulation environment using different open-source tools like CORE network emulation, VirtualBox, Scapy, Vagrant, and Scapy. We will also demonstrate how to emulate cyber faults on a synchrophasor-based cyber-physical system. So the main contribution of this research goes to 
	\begin{enumerate}
		\item Designing a scalable cyber-physical distribution system that incorporates designing the physical layer, cyber layer, an attacker layer employing different tools like Opal-RT, VirtualBox, Vagrant and CORE  network emulator, Scapy, NetfilterQueue, and pyPMU.
		\item Implementation of the open-source python module named pySynphasor for  IEEE C37.118 protocol on top of Scapy Framework. pySynphasor can dissect and build IEEE C37.118 packet just by a few lines of commands.
		\item Demonstration of different cyber attacks such as MITM, FDIA, FCIA, and Eavesdropping in the developed testbed by leveraging pySynphasor. 
		\item Development of python-based simple PDC module named pyPDC capable of collecting and aggregating data from multiple PMUs. 
	\end{enumerate}
	
	We will also present linear state estimation-based attack detection techniques for FDIA, enabling the industry stakeholders and researchers to test similar MITM attacks on real systems and find the best detection technique for real scenarios.

	\begin{figure}[t]
		\centering
		\includegraphics[width=7cm]{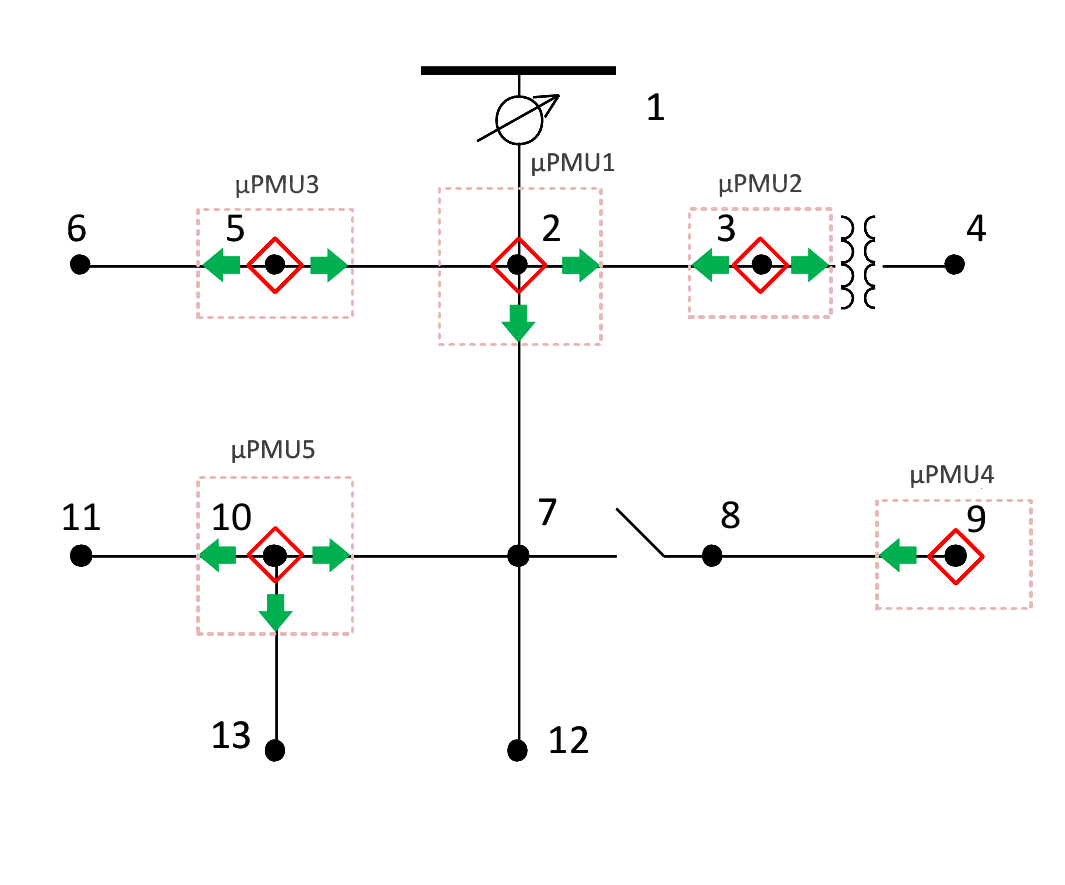}
		\caption{PMU Placement for IEEE 13 Node Test Feeder}
		\vspace{-12pt}
		\label{fig:pmu_placement}
	\end{figure}
	
	\begin{figure}
		\centering
		\includegraphics[width=\linewidth]{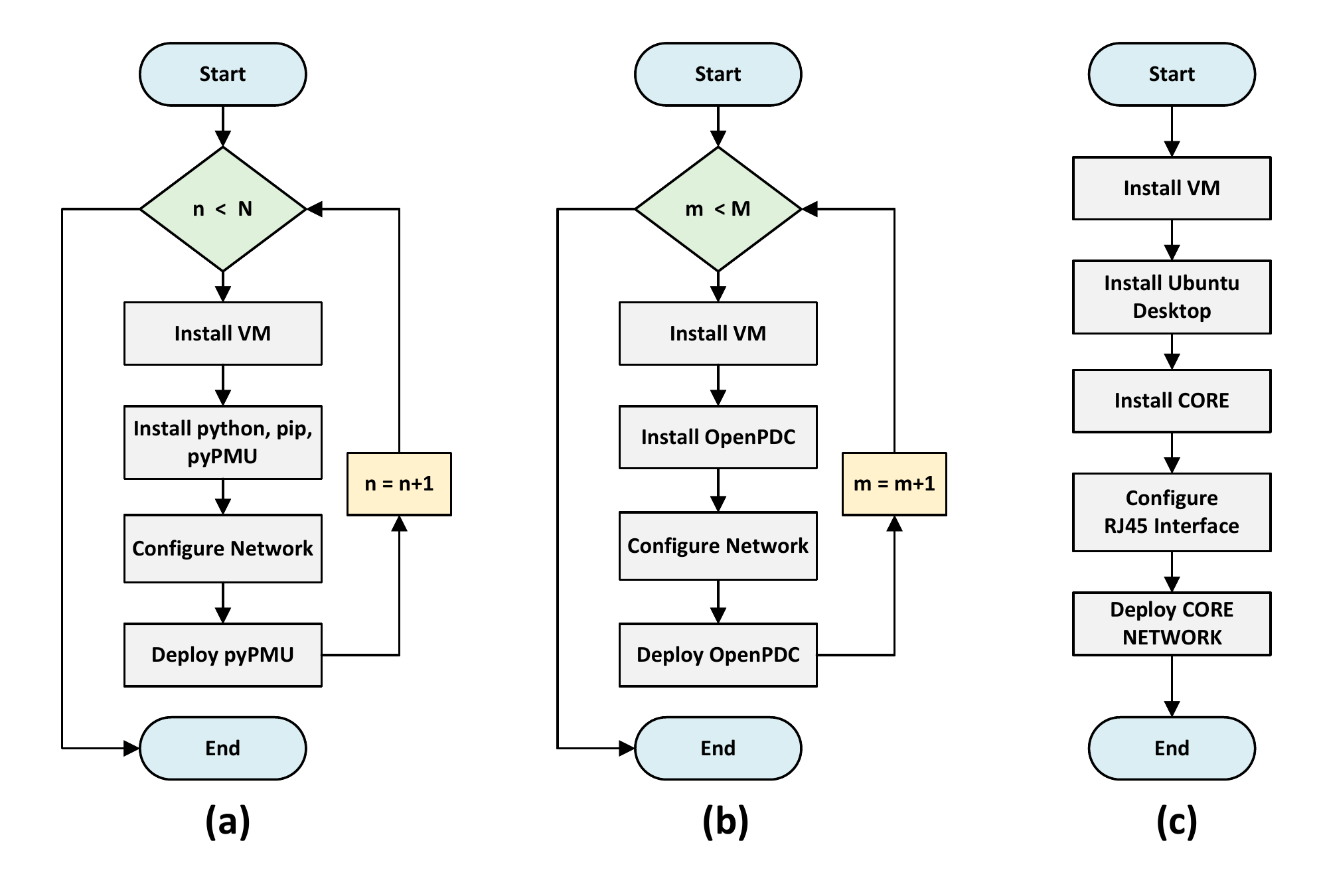}
		\caption{Vagrant VM Deployment}
		\vspace{-12pt}
		\label{fig:Vagrant_VM_Deployment}
	\end{figure}
	
	\section{Method}
	
	The electric power system is intrinsically a cyber-physical system (CPS), with power flowing in the physical system and information flowing in the cyber network. Therefore, A cyber-physical power system testbed equipped with cyber and physical layers is extremely important for studying the cyber and physical faults in a simulated environment. Due to the enormous nature of the power system, a highly scalable testbed can ideally mimic the real power system with different scenarios. The victim and attacker's points of view must be considered while designing a smart-grid testbed. The victim's point of view consists of designing the physical, cyber, and detection layers. The attacker's point of view consists of designing different attack scenarios on the communication protocols and understanding the network topology to implement attacks effectively. Figure \ref{fig:testbed_architecture} depicts the designed testbed representing both the victim and attacker's points of view. Therefore, the testbed consists of three layers. (I) Physical Layer (II) Cyber Layer and (III) Attack Layer. The Physical layer consists of IEEE 13-Node Test Feeder, PMU, control center with PDC, and linear state estimation with bad data detection technique. The cyber layer connects all the equipment utilizing a software-defined emulated network. Finally, the attack layer demonstrates deploying different cyber faults in a real system. The attacker layer also demonstrates how to implement FDIA attacks utilizing different vulnerabilities in the smart grid.

	\subsection{Physical Layer Design}
	The state of a power system can be expressed by the complex voltages of all buses\cite{chenStateEstimationDistribution2015}. State-estimation involves solving the network Quasi-Static model to find the states $x$ utilizing the digital and analog measurements $z_i$ from the system\cite{monticelliElectricPowerSystem2000}. Power system measurement data  $z_i$  can be expressed with relation to the system states $x$ and measurement error $e_i$.
	
	\begin{equation}
		\label{eq1}
		z_i  = h_i(x)+e_i
	\end{equation}
	
	Here, subscript $i$ represents the $i^{th}$ meter and $h(.)$  is the measurement function which expresses the relationship between measurements and state, $x$. For AC state estimation, this function is non-linear, requiring an iterative method like Newton-Raphson. On the other hand, this function is linear in the linear state estimation problem. As a result, the solution can be obtained just by matrix multiplication.
	
	A Weighted Least Square(WLS) optimization problem can be formulated to solve this problem. By minimizing $J(x)$, we effectively choose $x$ that best "fits" the measurements. 
	
	\begin{equation}
		\label{eq2}
		min \text{  } J(x) = \sum_i^{N_m}\frac{(z_i - h_i(x))^2}{\sigma_i^2}
	\end{equation}

	Here, $\sigma_i^2$ represents the variance of the meter measurement. For linear state estimation, both of the equation can be written in matrix form:
	\begin{equation}
		\label{eq3}
		\vec{z} = H\vec{x} + \vec{e}
	\end{equation}
	
	\begin{equation}
		\label{eq4}
		min \text{  } \vec{j} = (\vec{z}-H\vec{x})W(\vec{z}-H\vec{x})^T
	\end{equation}

	Here, $W = R^{-1}$ and $R = diag\{\sigma_1^2,\sigma_2^2,\sigma_3^2,\ldots\}$ 
	The optimization problem can be solved by first order optimal condition. So, the estimated value of $\vec{x}_{est}$ can be obtained by  $\frac{\partial \vec{j}}{\partial \vec{x}} = 0$ \cite{khanThreatAnalysisBlackEnergy2016,chenStateEstimationDistribution2015}
	\begin{equation}
		\label{eq5}
		\vec{x}_{est} = (H^TWH)^{-1}H^TW\vec{z}
	\end{equation}
	
	Linear state estimate is just a matrix multiplication with contrast to AC state estimation but we need to build the $H$ matrix from system network\cite{monticelliElectricPowerSystem2000}.
	$$ H = \begin{bmatrix}II \\M\end{bmatrix}$$
	$$M = yA+y_s $$
	Here $II$ represents voltage measurement bus incident matrix, $A$ current measurement bus incident matrix, $y$ represents series admittance matrix, $y_s$ shunt admittance matrix. These matrices were formed following the instruction from thesis\cite{taraliBadDataDetection2012}.
	
	%To measure the performance of state estimation two measurement metrices  are used such as  Mean Absolute Percentage Error(MAPE) and Mean Absolute Error(MAE). Mean Absolute Percentage Error(MAPE) is used to measure accuracy of Voltage Magnitude.
	%$$MAPE = \frac{1}{N}\left(\sum_{i=1}^N \frac{|V_i^{ex} - V_i^{est}|}{V_i^{ex}} \times 100\%\right)$$
	%Mean Absolute Error(MAE) is used to measure the accuracy of phase angle
	%$$MAE = \frac{1}{N}\left(\sum_{i=1}^N |\theta_i^{ex} - \theta_i^{est}|\right)$$
	%Here superscript $(.)^{ex}$ represents the exact value and $(.)^{est}$ represent the estimated value and $N$ represents total number of buses.

	\begin{figure}[t]
		\centering
		\includegraphics[width=7cm]{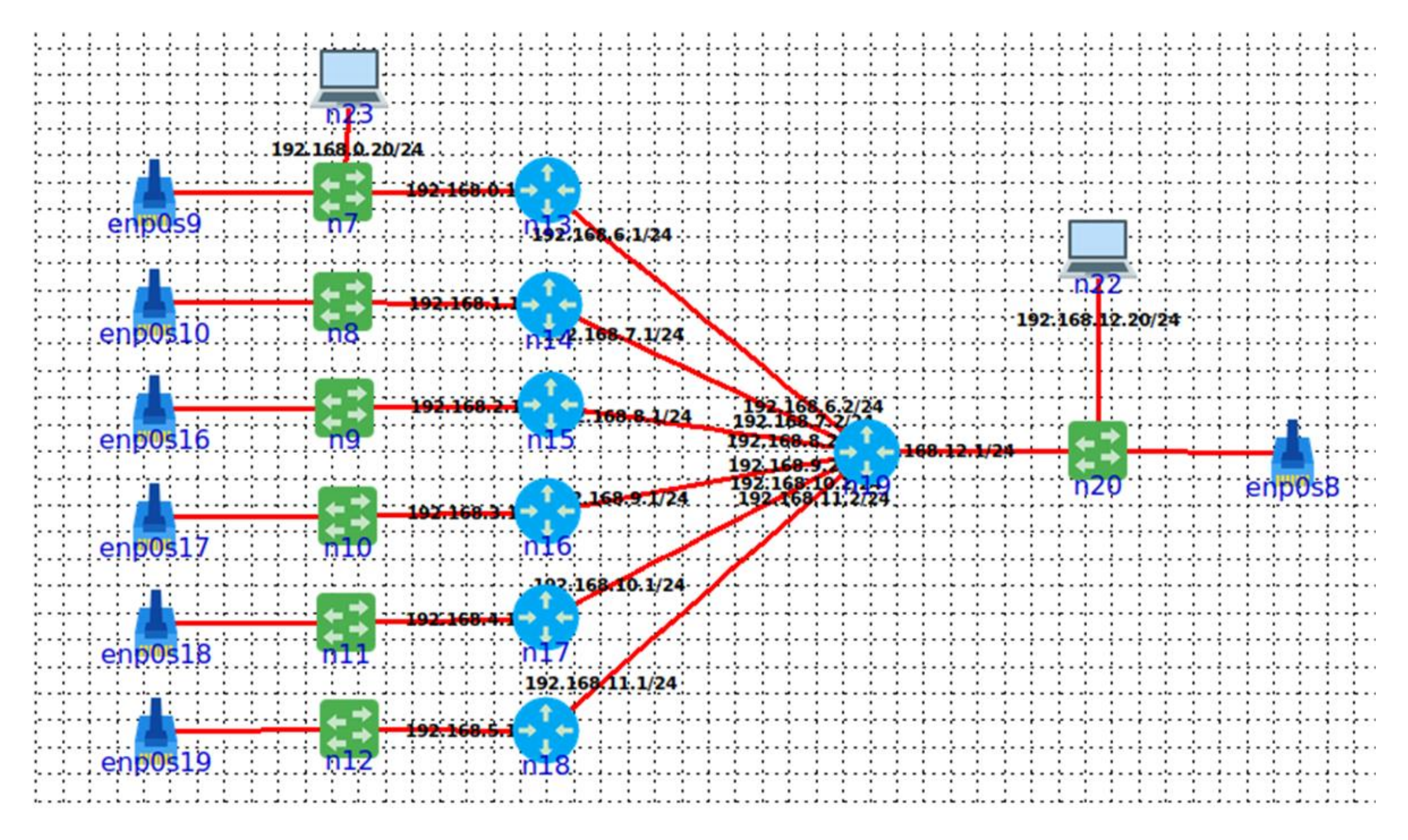}
		\caption{CORE Network Design}
		\vspace{-12pt}
		\label{fig:CORE_Network_Design}
	\end{figure}

	One of the main goals of state estimation is to find out the bad meter measurements and eliminate that measurements. As the state estimation problem is intrinsically over-determined, there are more measurements than the state. So, bad measurements can be eliminated. For linear state estimation, the presence of bad data detection problem can be formulated as follows\cite{zhangDesignTestingImplementation2017a}:
	\begin{equation}
		\label{eq:bad_data_detedtion}
		j = \sum_j^{N_m} \frac{e_j^2}{\sigma_i^2} = (\vec{z}-H\vec{x_{est}})W(\vec{z}-H\vec{x_{est}})^T
	\end{equation}
	$j$ follows the Chi-Square distribution, $\chi^2(K)$. Here, $K = N_m-Ns$, is known as degree of freedom, where,  $N_m$ Number of total measurements  and $N_s$, Number of total states \cite{bandakPOWERSYSTEMSSTATE2013}. The presence of bad data can be estimated by checking the condition if $j<T_j$ , No bad data in the system. $j\ge T_j$,  else, Bad data exists. Threshold value, $T_j$ can be obtained from the Chi-Square table utilizing the degree of freedom(d.f) $K$ and meter accuracy, $\sigma$. 
	
	After detecting the presence of bad data in the measurement, it is also possible to  identify the bad or poisoned meter by hypothesis testing.
	\begin{equation}
		\label{eq:meter_error}
		\sigma_Y = \left|\frac{z_i - h_i(\hat{x})}{\sigma_i} \right|
	\end{equation}
	if $\sigma_Y >3$, it identifies that  $i^{th}$ meter measurement is bad. The same method that is used for bad meter detection can also be utilized to detect false data injection attack on meter measurements. 
	
	\begin{figure}
		\centering
		\includegraphics[width=\linewidth]{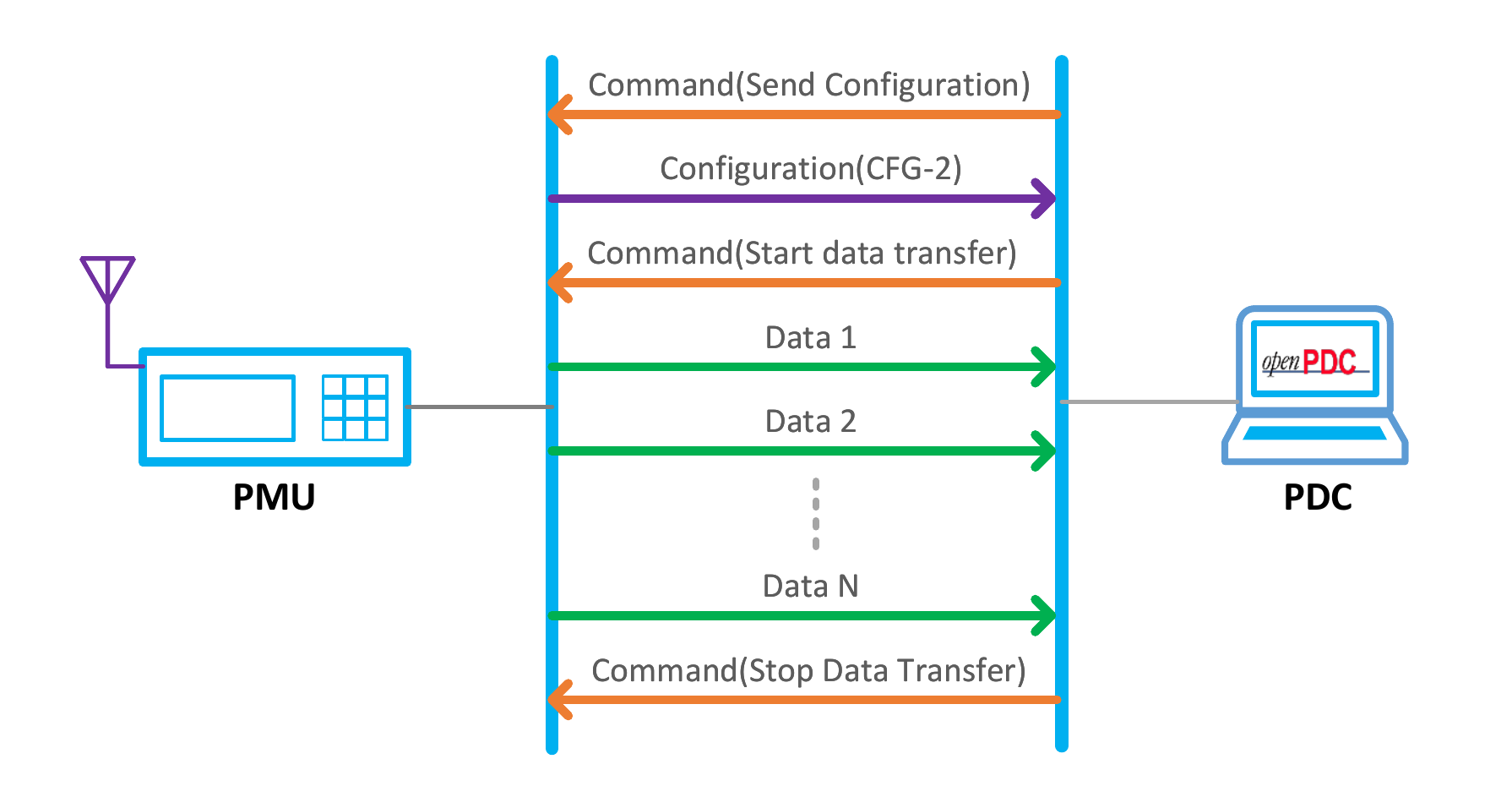}
		\caption{IEEE C37.118.2 Packet Transfer Mechanism}
		\vspace{-12pt}
		\label{fig:Synphasor_protocol}
	\end{figure}

	\vspace{-10pt}
	\subsection{Design the IEEE 13-Node Test Feeder}

	%\begin{figure}[!t]
	%	\centering
	%	\begin{subfigure}
		%		{
			%		\includegraphics[width=\linewidth]{pmu_placement.png}
			%		\caption{PMU Placement for IEEE 13 Node Test Feeder}
			%		\label{fig:pmu_placement}
			%		}
		%	\end{subfigure}
	%	
	%	\begin{subfigure}
		%		{
			%		\centering
			%		\includegraphics[width=\linewidth]{Vagrant_VM_Deployment.png}
			%		\caption{Vagrant VM Deployment}
			%		\label{fig:PMU and PDC Automatic Deployment Sequence}
			%		}
		%	\end{subfigure}
	%	
	%\end{figure}

	%\begin{figure}
	%	\includegraphics[width=\linewidth]{pmu_placement.png}
	%	\caption{PMU Placement for IEEE 13 Node Test Feeder}
	%	\label{fig:pmu_placement}
	%\end{figure}

	We developed the testbed focusing on the IEEE 13 Node test feeder, a three-phase unbalanced distribution system operating in 4.16kV\cite{kerstingRadialDistributionTest2001}. The distribution system is designed in MATLAB Simulink and deployed in Opal-RT real-time simulator. Placing the $\mu$PMU is an optimization problem to minimize the number of PMUs. The paper \cite{chenStateEstimationDistribution2015} identified the minimum number of PMUs for IEEE 13 node test feeder. Following that,  we placed 5 $\mu$PMUs in nodes 2, 3, 5, 9, and 10. These 5 $\mu$PMUs measure voltage at five buses and current at nine lines in a total of 13 voltages and 20  current measurements for three phases. One phase is considered in the case of a transmission system due to its balanced nature. However, due to the unbalance in phases, the distribution system poses another challenge. The authors in paper\cite{chenStateEstimationDistribution2015} demonstrated that the three-phase unbalanced system could be decoupled into three-state estimation problems and possibly calculated by parallel computing. Figure \ref{fig:pmu_placement} demonstrates the $\mu$PMU placement with current and voltage measurements. The red diamond indicates the complex voltage measurements, and the green arrow indicates the complex current measurements. The yellow dashed border indicates the $\mu$PMU—a single $\mu$PMU measures multiple complex voltages and multiple line currents. We also added standard normal distributed noise with the simulated PMU measurements to mimic the real behavior. 
	\begin{equation}
		\label{eq:noise_modeling}
		\begin{aligned}
			\mathbf{N} &= \sigma R(z) +\mu \\	
			S_N &= S+\mathbf{N}
		\end{aligned}
	\end{equation} 
	Here, $\mathbf{N}$ represents the generated noise  utilizing the standard normal distribution function $R(z)$, variance $\sigma$ and mean $\mu$. $S_N$ is the noisy measurement that mimics the real meter measurements. 
	
	\begin{figure}[t]
		\centering
		\includegraphics[width=\linewidth]{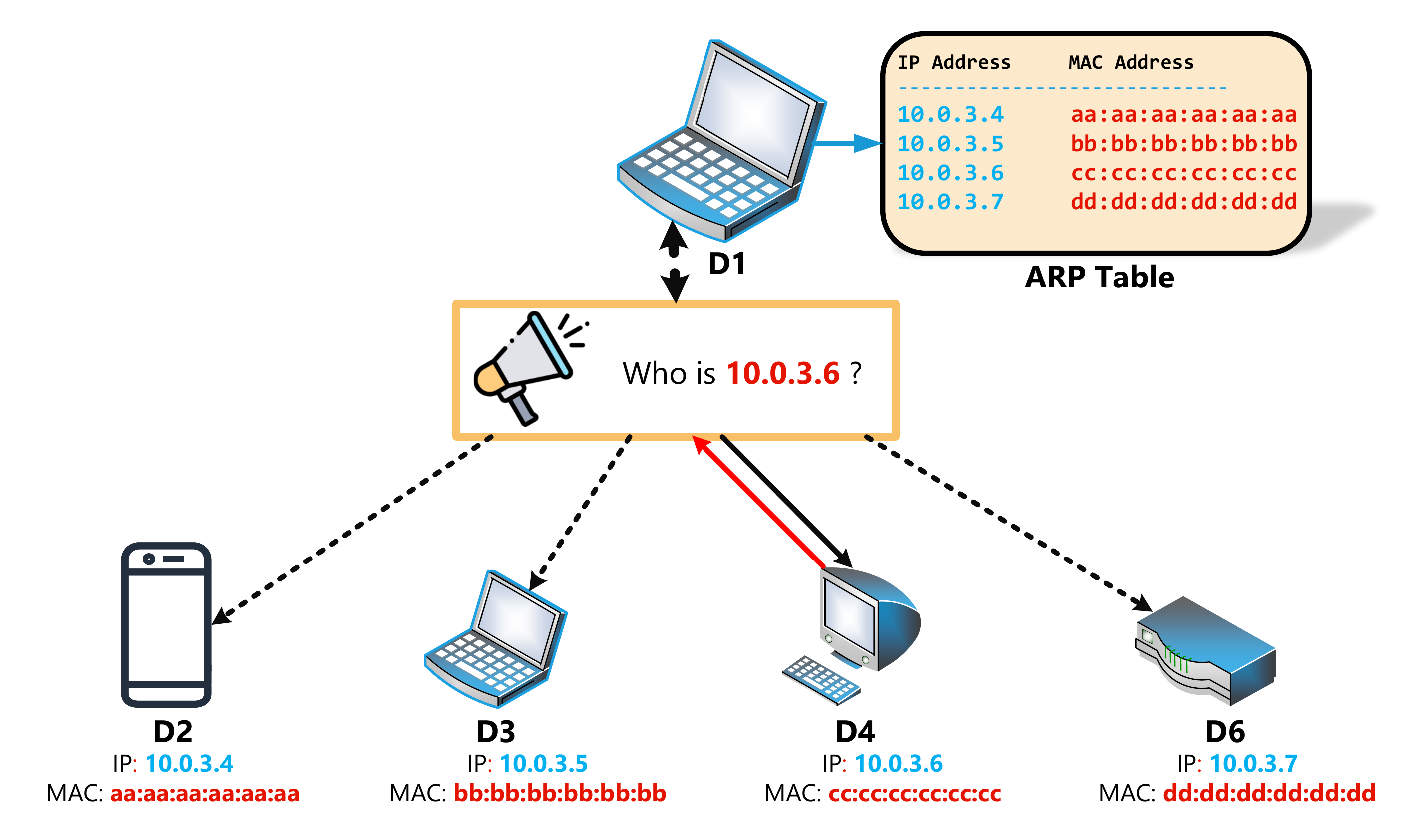}
		\caption{Mechanism of ARP Protocol}
		\vspace{-12pt}
		\label{fig:ARP Protocol}
	\end{figure}
	
	\begin{figure}
		\centering
		\includegraphics[width=\linewidth]{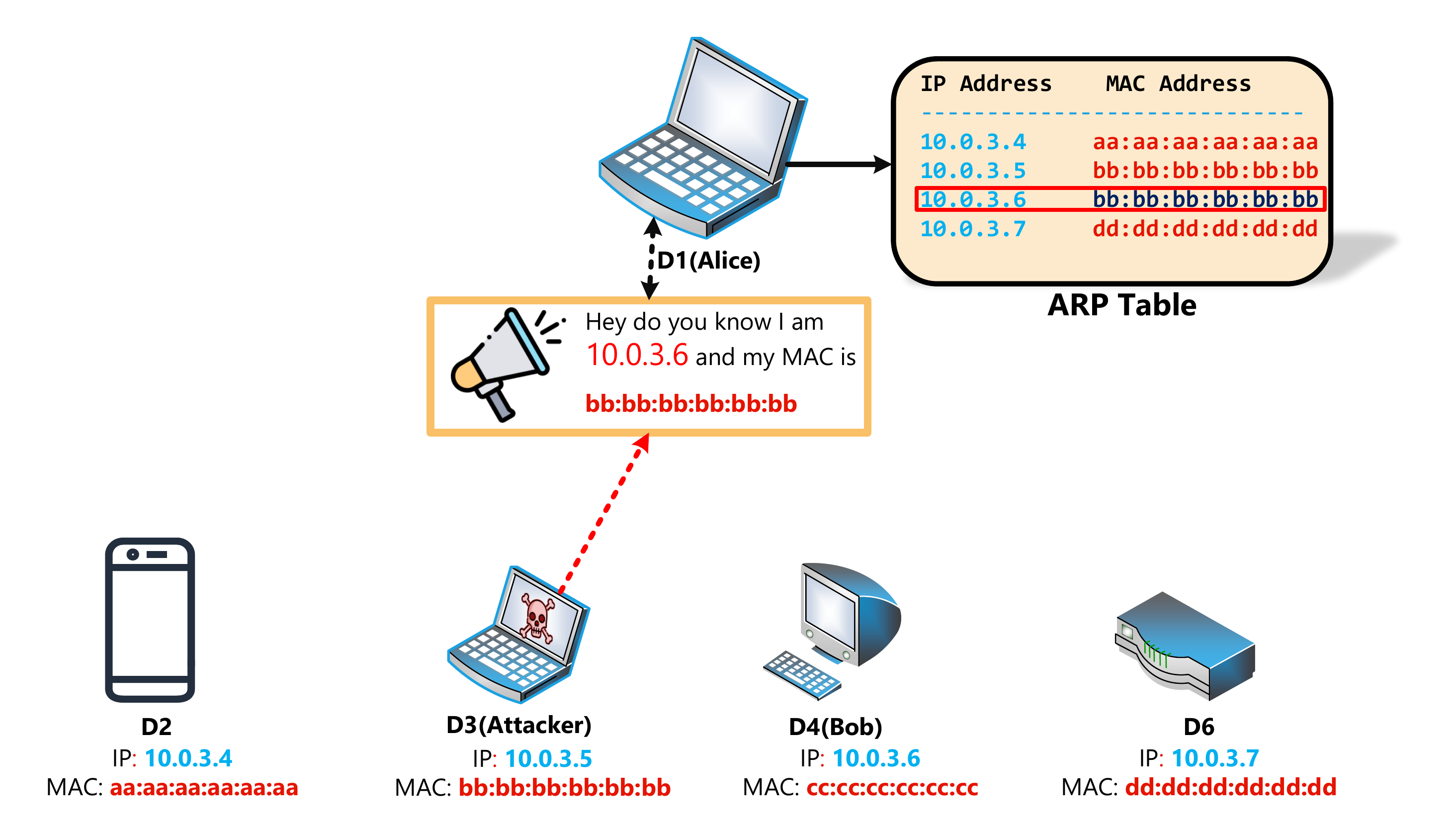}
		\caption{ARP Poisoning Mechanism}
		\vspace{-12pt}
		\label{fig:ARP Poisoning}
	\end{figure}
	
	\subsection{Implement Synchrophasor Devices}
	
	Complex voltage and current measurements must be aggregated to the control center to perform linear state estimation. This task is accomplished through synchrophasor technology. it consists of Phasor Measurement Unit(PMU), Phasor Data Concentrator(PDC), communication network and control center\cite{khanThreatAnalysisBlackEnergy2016}. Real-time GPS time-stamped electrical quantities are measured and transmitted to the control center through a suitable communication protocol. The latest synchrophasor standard has been split into two standards, IEEE Std 37.118.1-2011, which covers measurement provision, and IEEE Std 37.118.2™-2011, which covers data communication\cite{ieeepowerandenergysocietyIEEEStdC372011}. IEEE C37.118.2 protocol is an application layer protocol that is accepted by industries for phasor data transferring between PMU and PDC. Although IEEE standard C37.244-2013\cite{ieeepowerandenergysocietyIEEEStdC372013} deals with PDC, it did not impose strictly to use of IEEE C37.118.2 protocol for phasor data transfer. This standard recommended IEEE C37.118.2 or IEC 61850-90-5 for PDC data transfer. Both of the protocols have their unique features and limitations\cite{khanThreatAnalysisBlackEnergy2016}.

	Four message types are defined in the IEEE C37.118.2 standard\cite{ieeepowerandenergysocietyIEEEStdC372011} such as data, configuration, header, and command. A sample communication scenario of IEEE C37.118.2 protocol has been presented in the figure \ref{fig:Synphasor_protocol}. In the synchrophasor protocol, PMU functions as a server, and PDC functions as a client. In the server-client paradigm, a client initiates communication session. Therefore, PDC starts a session by sending a command that requests the PMU to send configuration message. Without a configuration packet, PDC cannot interpret the data packet. Then After receiving the configuration packet, PDC sends another command to start data transmission. PDC transmits synchrophasor data continuously at a fixed rate until further PDC commands stop data transmission.

	We deployed all the $\mu$PMU, PDC, and network layer on the top of the virtual machine(VM), which gives us the advantage of scaling. So, We need to deploy four types of VMs to develop the whole system. Moreover, different types of VM has separate dependencies and setup mechanism, which is one of the challenges for scaling the system. To scale up and automate the VM deployment process, we leverage the power of Vagrant. It is developed by HashiCorp to create and manage a portable virtual machine environment. Vagrant utilizes Ruby programming language to write instructions for deploying all virtual machines in a single file named Vagrantfile\cite{hashimotoVagrantRunningCreate2013}.

	We leverage an open-source Python library named  pyPMU\cite{sandiPypmuOpenSource2016} to develop the $\mu$PMU. All the dependencies have been installed inside the virtual machine to mimic a real PMU. The $\mu$PMU  is performing two significant tasks. First, it collects the phasor measurements from the distribution system that is running inside the Opal-RT; second, pyPMU encodes phasor measurements into the IEEE C37.118.2 packet. Finally, it is waiting for the command from PDC to initiate communication, just like in figure \ref{fig:Synphasor_protocol}. Manual installing many PMUs is a cumbersome process, So we utilized the Vagrant automation tool to automate the virtual machine deployment process. 
	
	All the steps of $\mu$PMU installation is depicted in figure \ref{fig:Vagrant_VM_Deployment}(a), and the script of the implementation can be found in the GitHub link: "https://github.com/shuvangkar". From VM deployment to dependencies installation, all the steps are handled by the vagrant script. As the Vagrant script is based on the Ruby environment, it supports basic programming syntax like for loop. The support of the loop in the Vagrant script elevates the scaling issue significantly. Hence, we implemented a base PMU device and utilized a "for loop" to deploy the n number of PMUs. The command sequence of PMU deployment is essential because all the dependencies should be installed before changing the network card. Each PMU requires two network cards to work correctly. One network card works in bridge mode that connects with Opal-RT, and another network card works in internal network mode that connects the PMU with the smart-grid network. Then dependencies for the PMU, such as pyPMU, are installed in the script. Finally, static IP addresses are correctly set into both network cards to connect the physical layer with the cyber layer.

	% Need to have a paragraph on pySynphasor
	The pySynphasor module can be found in the link: "https://github.com/shuvangkar" 
	
	As the pySynphasor can dissect and build IEEE C37.118.2 packets, we developed a simple PDC application utilizing that. We named the PDC application pyPDC. The pyPDc can receive data from multiple PMUs. The pyPDC is available at the following link: "https://github.com/shuvangkar" The pyPDC being a python application gives us the advantage of deploying everything in a python environment utilizing vagrant script. The steps for PDC deployment is shown in the figure \ref{fig:Vagrant_VM_Deployment}(b).
	
	The vagrant script's VM setting with dependencies is written down sequentially. Afterward, a simple vagrant command "vagrant up" deploys the whole testbed except the CORE network. The script installs all the PMUs and PDC with dependencies and necessary setups within a few minutes. This automatic installation opens the real potential of the testing cyber-physical system in a hardware-in-loop and co-simulation environment as it does not require purchasing costly PMU and PDC hardware to test more extensive distribution or transmission systems. Both the PMU and PDC have critical network configurations in the testbed, which will be discussed in the subsection \ref{Cyber Layer Modeling}.  
	
	\begin{figure*}[t]
		\centering
		\includegraphics[width=14cm]{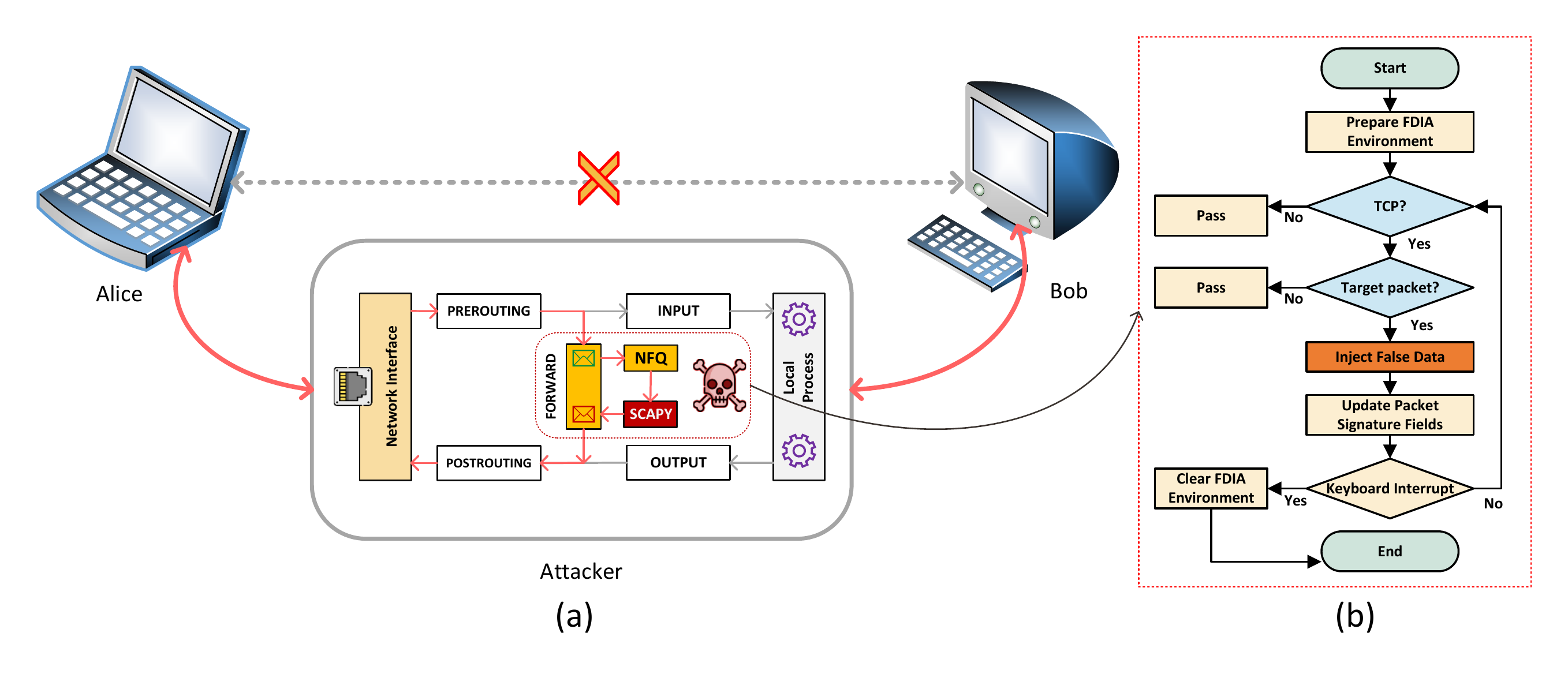}
		\caption{FDIA Attack Mechanism}
		\vspace{-12pt}
		\label{fig:FDIA Attack Mechanism}
	\end{figure*}

	\subsection{Cyber Layer Modeling}\label{Cyber Layer Modeling}
	
	Designing a highly scalable cyber-physical system in the virtual environment is one of the main goals of the research. So, a software-defined network that runs in real-time is required to connect all the virtual machines. CORE(Common Open Research Emulator) is an open-source network emulator tool that runs on top of Linux. It runs in real-time and connects many nodes, taking advantage of the Linux network namespace. CORE has python API and a graphical user interface for building the emulated network.

	Figure \ref{fig:CORE_Network_Design} depicts the emulated network designed in the CORE emulator. The network is composed of routers, switches, and RJ45 connectors. Let us consider the router n13. The network under the router n13 is considered the substation network. Under the substation router, a switch connects all the devices in the substation, such as PMU, local PDC, relay, and RJ45 connector. The RJ45 connector is the interface between the virtual network and PMU. Therefore, the RJ45 connector is the main bridge between the independent host and emulated network. The independent host can be any physical or virtual machine like PMU VM. 
	
	The VirtualBox network needs to be configured in a particular way to connect any host with the CORE virtual network through the RJ45 connector. For that reason, configuring the network in VirtualBox is one of the crucial factors for modeling the network. VirtualBox has seven network modes such as (1) Not Attached, (2) NAT, (3) NAT Network, (4) Bridge Network, (5) Internal Network, (6) Host-only network, and  (7) Generic Network\cite{virtualboxChapterVirtualNetworking}. Different networking modes has different application and different advantages and disadvantages. Internal network mode is suitable for this application as it allows connecting a particular VM with the CORE RJ45 ethernet port. While building the CORE VM, Promiscuous mode must be allowed while configuring the network interface cards. 
	
	Promiscuous mode allows the incoming traffic to pass the physical network adapter and reach the CORE virtual network adapter. So, without enabling this mode, CORE will not receive the packet from the rest of the PMU and PDC VMs. The IP address of the PMU VMs has to be set according to the network address of the substation router. Otherwise, the VM cannot connect with the CORE router. For example, the network IP of the link connecting the enpos9 ethernet port is 192.168.0.1. The allowable IP range is 192.168.0.2-192.168.0.255 for the PMU VM connected with the enpos9 RJ45 connector. The same rules apply to all the VM connected with the CORE network through the RJ45 connectors. Also, 192.168.0.1 will be the gateway address of the PMU. The network configuration mentioned above is automated in the Vagrant script. We only have to manually deploy the CORE network and configure and connect the VM with the RJ45 ethernet interface, just like the figure \ref{fig:Vagrant_VM_Deployment}(c). It is also possible to automate the CORE VM deployment using CORE python API.

	\begin{figure*}[ht]
		\centering
		\includegraphics[width=\linewidth]{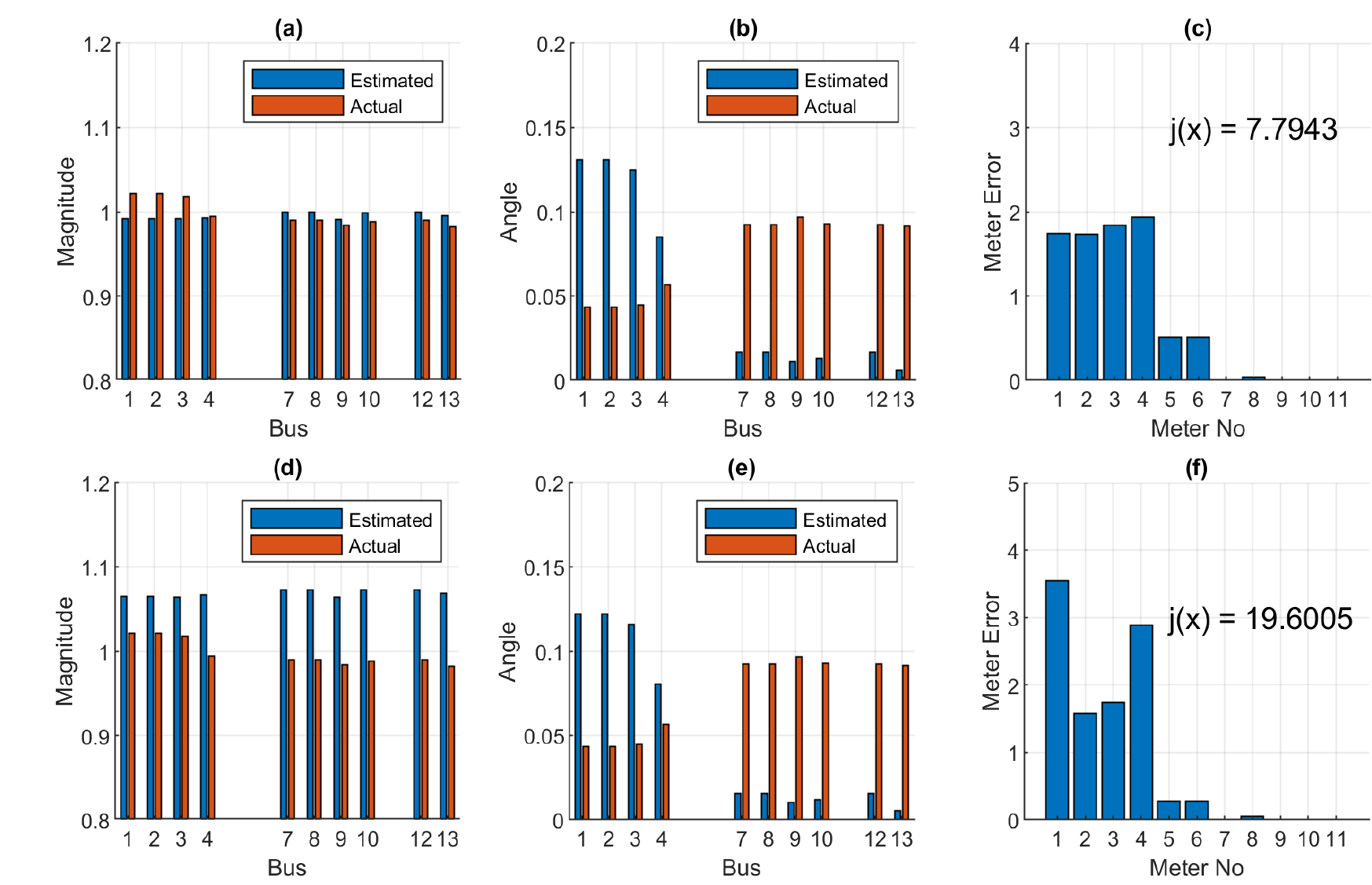}
		\caption{State Estimation}
		\vspace{-12pt}
		\label{fig:state estimation}
	\end{figure*}

	\subsection{Attacker Modeling}
	MAC address is the unique identifier assigned to the network interface card (NIC). The IP address is a 32-bit address that identifies a device in a local area network in the IPv4 system. The problem with IP addresses is that they might change depending on the network setting. So, to identify a device in the network, the MAC address plays a vital role because of its uniqueness. ARP protocol resolves MAC address from IP address in the local area network. Therefore, the ARP protocol is the bridge between layer 2(MAC address) and layer 3(IP address).
	
	Consider a scenario in figure \ref{fig:ARP Protocol} where device D1 needs to send data to device D4. D1 only knows the IP address of the device D4. Therefore, D1 needs to know the MAC address of D4 to send data. Then, D1 sends a broadcast message to the local area network asking MAC address of D4. Afterward, all the devices in the local network will get the broadcast request. Only D4 will respond by attaching its MAC address. After receiving the MAC address, D1 caches the MAC address of D4 along with the IP address in its ARP table.

	\begin{figure*}[ht]
		\centering
		\includegraphics[width=\linewidth]{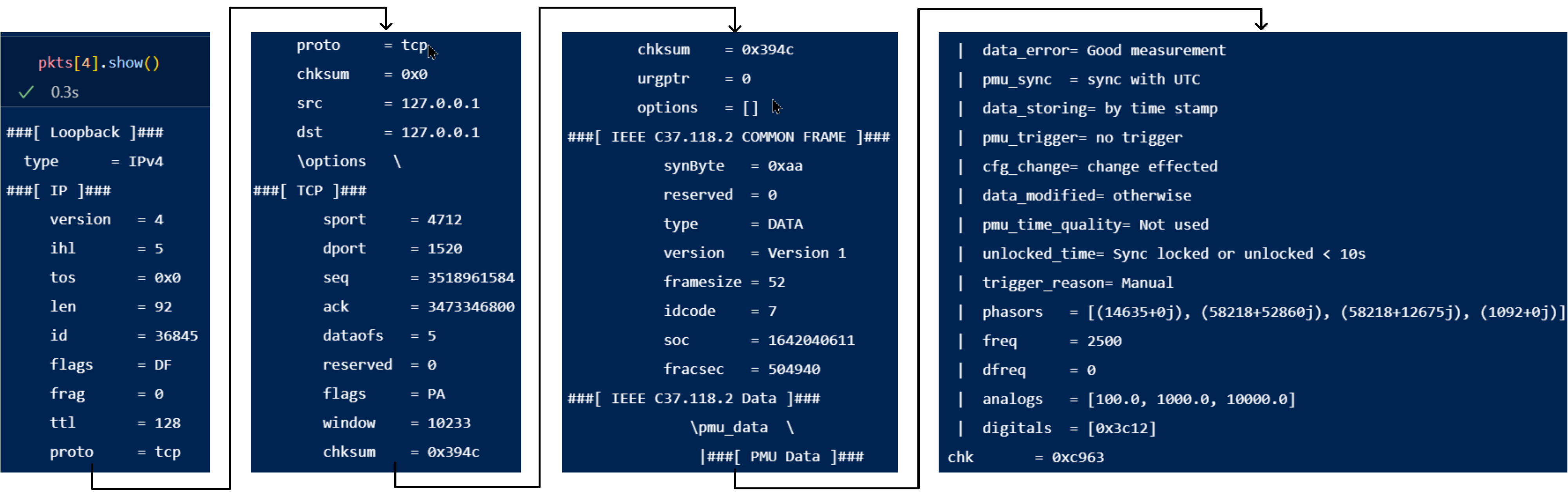}
		\caption{Dissection of IEEE C37.118.2 data packet using pySynphasor module}
		\vspace{-15pt}
		\label{fig:pySynphasor_data_dissection}
	\end{figure*}
	
	ARP protocol does not verify the responder address before caching into the ARP table. So, any device can send an ARP reply message without a query. That means it is not mandatory for D1 to ask for the D4 MAC address to get an ARP reply. Any host in the local area network can generate an ARP reply without inquiry. The adversary uses this vulnerability to pretend to be someone in the network. In the figure \ref{fig:ARP Poisoning}, attacker D3 sends an ARP reply to device D1, attaching the IP address of device D4 and MAC address of its own without any initial inquiry from device D1. So, the attacker, D3, is pretending to be D4. As a result, the ARP table of D1 will be poisonous. This technique is known as ARP poisoning. Because of ARP poisoning, whenever D1 sends any packet to D4, it will be forwarded to the attacker machine(D3).
	
	Now, all the packets from D1 to D4 pass to the attacker machine(D3) after deploying the  ARP poisoning script. We have written an ARP poisoning script in Scapy. The aprspoof is another ARP poisoning tool with the built-in command to perform ARP poisoning. Both tools work according to the mentioned method in figure \ref{fig:ARP Poisoning}. After deploying the ARP poisoning script, all the packets from D1(Alice) to D4(Bob) will be passed to the D3(Attacker). However, Bob(D4) will not receive the packet unless the Attacker enables the packet forwarding. Therefore, Alice and Bob can establish communication after enabling packet forwarding in the attacker machine. In the case of the Man in The Middle(MITM) attack, the attacker convinces two victims that they are directly transferring data with each other\cite{khanIEEEC3711822016}. In this way, the MITM attack has been implemented in our testbed. 
	
	FDIA attack is the most dangerous attack that can disrupt the regular operation and lead to blackout, just like the 2015 Ukraine attack. In this attack, the adversary craft fully changes the meter measurement. We utilized Linux iptables, NetfilterQueue, and pySynphasor to implement FDIA in our testbed. The iptables is the basic firewall program of the Linux operating system. We developed a python script that combines all the tools and automates the FDIA attack process. The packet filtering mechanism provided by iptables is organized into three different kinds of structures; (1)tables, (2)chains, and (3)targets. The table allows processing packets in specific ways.
	Filter table decides whether a packet should be allowed to reach the destination, Mangle table allows to alter packet header, NAT table allows to route packet to different hosts on NAT network, Raw table is a stateful firewall. Tables have a chain attached to them, and the chain inspects traffic at a different point. Prerouting chain applies when a packet arrives. The input chain applies just before the local process. Forward chain applies to the packet that are routed through the current host. The Postrouting chain applies when the packet leaves the network interfaces. Finally, the target decides the fate of a packet, such as allowing, rejecting packets, or passing the packet to a queue. We created iptables rules such as way so that any packet passing through the FORWARD chain;  will pass to the NetfilterQueue buffer(NFQ) just like depicted in figure \ref{fig:FDIA Attack Mechanism}(a). 
	
	After reading the packet from the NFQ buffer, the attacker inject false data in the measurements. As the pySynphasor module can intuitively build and dissect the IEEE C37.118.2 packet, It plays a vital role in injecting false data. It reads the network packet and builds an internal object structure in the Scapy framework. That is known as the internal representation of the packet. The internal representation is helpful in the Scapy framework as it allows the user to modify and redesign the packet. After modifying the packet, Scapy again builds the packet for the network. This network representation is the machine representation of a packet in Scapy terminology. Another critical point while rebuilding the packet after the injection is updating the packet fields that hold the packet's signature in some form, such as IP length, TCP length, TCP checksum, and IEEE C37.118.2 CRC value. 
	
	Figure \ref{fig:FDIA Attack Mechanism}(a) depicts the direction and flow of the packet in Linux kernel space and user space. The red line indicates the path of the synchrophasor packets while passing through the attacker's machine. Figure \ref{fig:FDIA Attack Mechanism}(b) presented the flowchart of the script that we have developed for implementing the FDIA attack in the synchrophasor system. The script begins with preparing the FDIA environment that includes (1) enabling packet forwarding, (2) resetting iptables (3) adding new iptables rules. Then, the script filters the TCP packet that is aimed at the target machine. Afterward, the script injected false measurements. The next part is a little tricky. As the packet changes after the injection, a few fields such as IP length, TCP length, IP checksum, TCP checksum, and IEEE C37.118.2 CRC need to recalculate again. Otherwise, the receiver machine will discard the packet. 
	For a successful FDIA, the steps are  (1) Capture packet from Linux Forward chain and pass it to NetfilerQueue buffer, (2) When packets are available on the buffer, accept them and inject false data using pySynphasor. (3) Recalculate checksum as packets are changing after injecting false measurements. (4) Release the packet to the Linux Forward chain.

	% \begin{figure}[!ht]
		% 	\centering
		% 	\includegraphics[width=\linewidth]{FDIA_Algorithms.png}
		% 	\caption{CORE Network Design}
		% 	\label{fig:ARP Protocol}
		% \end{figure}

	\section{Results}
	\subsection{State Estimation in Normal Conditions} \label{sub:state_estimation_in_normal_conditions}
	
	The distribution system is intrinsically unbalanced. This unbalanced system can be realized as the three separate equations\cite{chenStateEstimationDistribution2015} for designing state estimation estimation problem. These three separate equations can be solved by parallel processing, reducing computation time. So, we formulated the three-state estimation problems and calculated the voltages and phase angles of the individual phases. Figure \ref{fig:state estimation}(a) and \ref{fig:state estimation}(b) presents the estimated and actual  bus voltage and phase angle  of phase A for IEEE 13 node test feeder. The results were generated based on the following scenarios. For phase A, buses 5,6 and 11 are not present. Therefore, 4 PMU measurements were used while calculating states. Seven current and four voltage measurements are comprised in total of $N_m=11$ measurements. Bus 7 and 8 are connected with a breaker, and bus 12 has no load. So, we need to estimate the voltage for bus 2,3,4,7,9,10,13. Therefore, the total number of states to estimate is $N_n = 7$. The noise was added with the meter measurement utilizing the equation \ref{eq:noise_modeling}. $\sigma = 0.05$ and mean $\mu = 0$ were considered while adding noise for this result. 
	
	For detecting presence of bad data, we utilized the equation \ref{eq:bad_data_detedtion} and Chi-square distribution table. For degree of freedom $K = N_m-N_n = 11-7 = 4$ and $\sigma = 0.05$, the threshold value $T_j = 9.49$ obtained from the Chi-square distribution table. Figure \ref{fig:state estimation}(c) represents the meter error  for all 11 measurement calculated using the hypothesis testing equation \ref{eq:meter_error}. The calculated value of $j=7.7943<T_j$, that indicates there is no bad data in meter measurement. Also hypothesis testing results for detecting bad meter was presented in figure \ref{fig:state estimation}(c). For all 11 meters, the value of $\sigma_Y<3$ that indicates all meter measurements are good. It also concludes that there is no attack on meter measurement in this case.
	
	\begin{figure}[t]
		\centering
		\includegraphics[width=7cm]{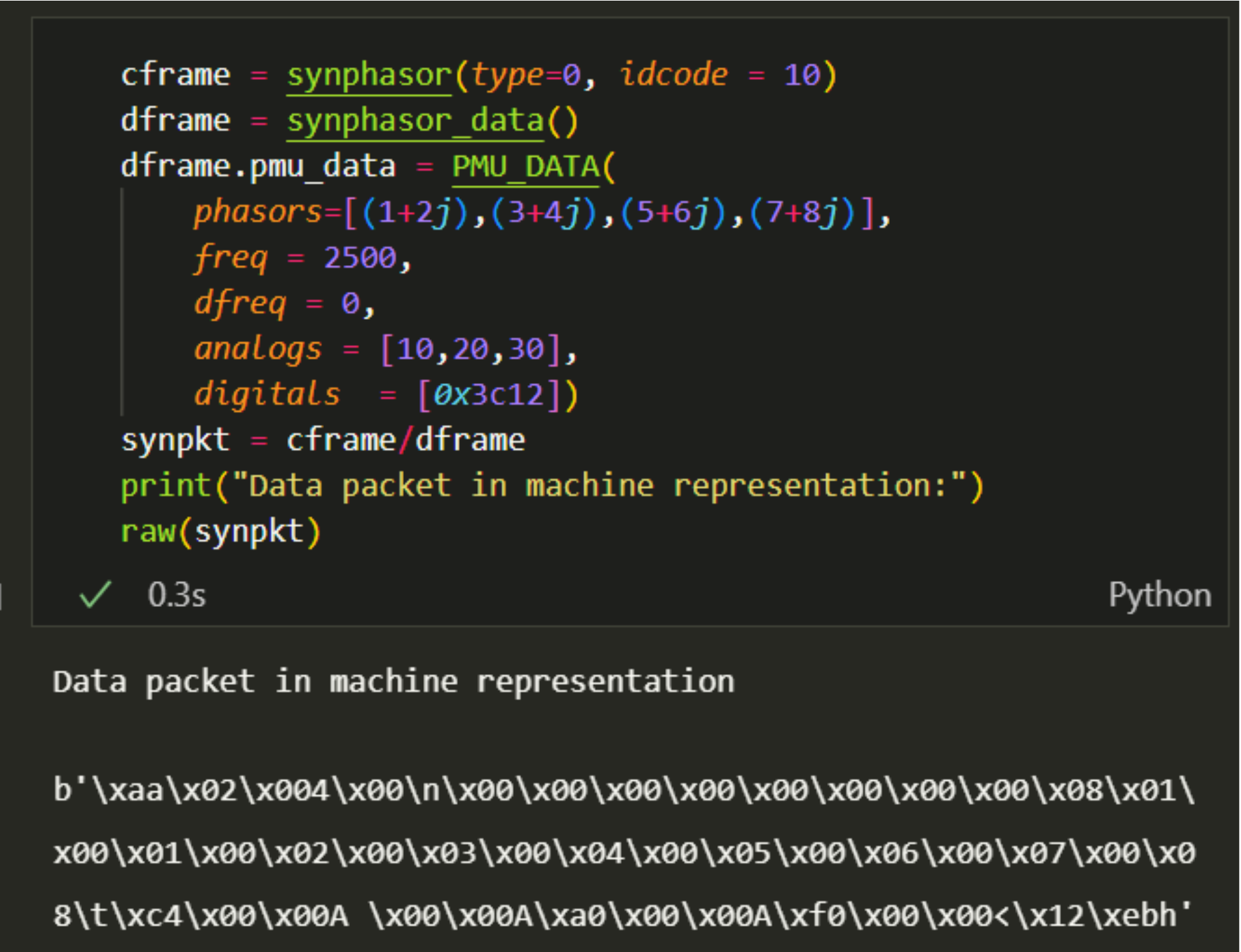}
		\caption{Building IEEE C37.118.2 data packet using pySynphasor module}
		\vspace{-12pt}
		\label{fig:pySynphasor_data_building.png}
	\end{figure}
	
	\subsection{pySynphasor Packet Building and Dissection} \label{sub:pysynpharos_result}
	pySynphasor module is built on top of Scapy. Scapy is an interactive packet manipulation program based on python. It has three types of packet representation: internal, machine, and human. Machine representation is the actual raw packet that will be sent through the network. Using internal representation Scapy manipulates the packet. Human representation is the human-readable representation in plain text. So, this representation is easy to deal with for injecting packets. Figure \ref{fig:pySynphasor_data_dissection} represents the human-readable representation of IEEE C37.118.2 data packets. These packets were collected using the Wireshark tool. Wireshark stores these packets from PMU and PDC communication in a pcap file format. Then, pySynphasor reads the pcap file and dissects the packet just by one line of code. The figure \ref{fig:pySynphasor_data_dissection} demonstrates that the packet dissection is done just by applying show() method on packet. It presents the packet in a human-readable format. The data packet has five sub-segments: an Ethernet Header, IP header, TCP header, IEEE C37.118.2 common frame, and IEEE C37.118.2 data frame. Data frame represent the phasors measurements in complex number format. Similarly, pySynphasor is capable of dissecting command, configuration, and header packet of IEEE C37.118.2 protocol.

	% Opal-RT sends the phasors measurement data to the $\mu$PMU, then it builds IEEE C37.118.2 packets from the measurements and sends the packet to the control center PDC. The pySynphasor has three types of representations for any packet: internal, machine and human-readable format. pySynphasor can dissect the network packet in a human-readable format by a simple python command. Figure xx(a) depicts the human-readable representation of the synchrophasor packet by the pySynphasor module. The module smoothens the analysis of the synchrophasor network packet in a real system. The figure xx(b) demonstrates the same packet in Wireshark capture for side by side comparison with pySynphasor dissection.   It is evident that the pySynphasor accurately dissect the IEEE C37.118.2 packets.

	The module is also capable of building the IEEE C37.118.2 packet from raw phasor measurements. The figure \ref{fig:pySynphasor_data_building.png} demonstrated an example of intuitively building the  IEEE C37.118.2 data packet just using a few lines of the python script. 
	
	\subsection{CORE Network Connection Result}\label{sub:core_network_result}
	\begin{figure}[t]
		\centering
		\includegraphics[width=7cm]{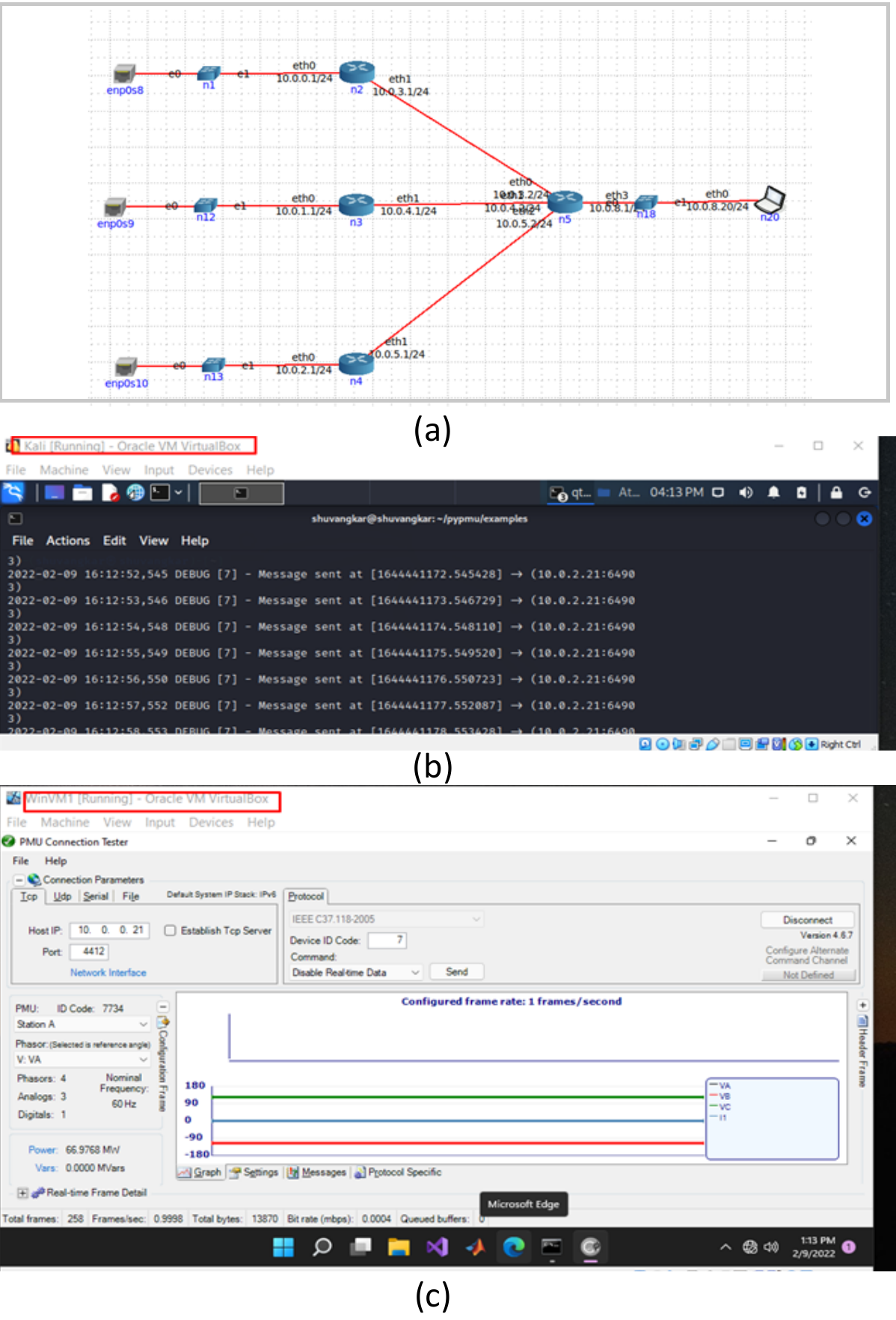}
		\caption{CORE Connection Results}
		\label{fig:CORE_Connection_Result}
		\vspace{-15pt}
	\end{figure}
	
	The figure \ref{fig:CORE_Connection_Result} presents an example of how CORE connects the PMU and PDC that were deployed in the virtual machine. Figure \ref{fig:CORE_Connection_Result}(a) depicts a sample CORE network deployed in VM for connecting PMU and PDC. Figure \ref{fig:CORE_Connection_Result}(b) presents PMU deployed in a VM  that transfers phasor measurement after a specific interval. The PMU device deployed here is built using the pyPMU python module. This VM is connected to the RJ45 ethernet interface enp0s8 in the CORE network. Figure \ref{fig:CORE_Connection_Result}(c) is another VM connected with the enp0s10 RJ45 network interface. PMU Connection Tester application is deployed in this VM for verifying the connection with PMU through the CORE network. PMU Connection Tester plotted four phasors measurements when it received data from the PMU device. All the phasors measurements are flat, because PMU sends a constant value. So these results verify the successful connection of PMU and PDC through the CORE virtual network in our proposed testbed. 
	
	%The round trip delay presented in the paper for the real hardware
	
	\begin{figure*}[ht]
	\centering
	\includegraphics[width=\linewidth]{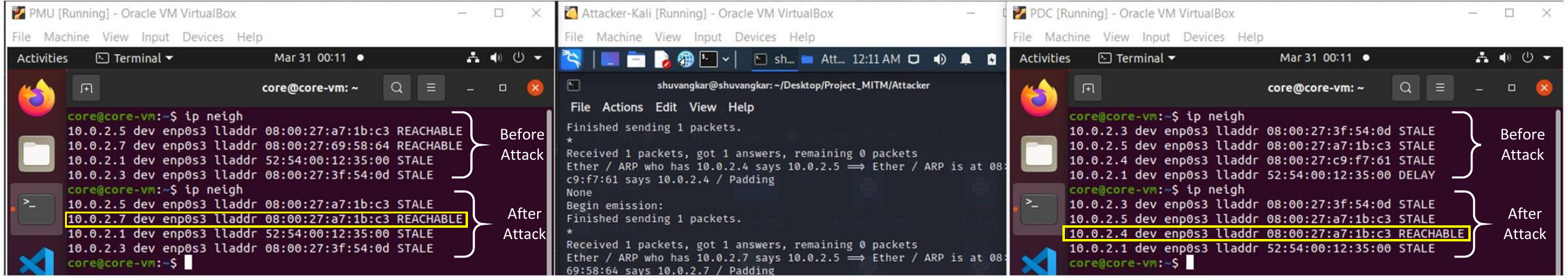}
	\caption{ARP Poisoning Result}
	\vspace{-20pt}
	\label{fig:arp_poisoning_result}
	\end{figure*}
	
	\vspace{-6pt}
	\subsection{ARP Poisoning Result} \label{sub:arp_poisoning_result}
	ARP poisoning mechanism is used to implement MITM attack in our smart-grid testbed. The primary technique of this type of implementation is that the attacker has to poison two victims' ARP table. The figure \ref{fig:arp_poisoning_result} presents the results of ARP poisoning where PMU and PDC are deployed in a local network in two separate VMs. The figure \ref{fig:arp_poisoning_result} presents the ARP table of both victims before and after the poisoning. If we look into the ARP table of the PMU device,  The IP address of PDC is 10.0.2.7, and the MAC address is 08:00:27:69:58:64 before the attack. The MAC address changed to 08:00:27:a7:1b:c3 in the PMU ARP table after the poisoning; that is the MAC address of the attacker machine. That means the PMU ARP table is poisoned, and all packets from PMU to PDC will go through the attacker machine. The same thing happened to the PDC ARP table.
	
	%Make a table for ARP poisoning
	
	%In the figure xx, there is a Wireshark capture between Alice and Bob. It shows that, whenever Alice is sending any packet to the Bob, it is going to the Attacker machine. 
	%In our testbed Alice is be PMU or PDC and Bob is the substation router. 

	%\subsection{MITM and Eavesdropping Results}
	%
	%
	%In the method section, we described how the attacker deploying the MITM attack to get access to the PMU and PDC packets. Figure xx, demonstrates the eavesdropping of the synchrophasor packets by the man in the middle attack where the PMU is sending data to the PDC. PMU and PDC are communicating normally. The attacker poison the PMU and PDC ARP table and deployed the attack-script depicted in figure xx and figure xx. After all the packet from PMU are passing through the attack script. In the figure, the attacker script is running inside the kali VM. The attacker script can eavesdrop the synchrophasor data in real-time. The figure demonstrates that the attacker can dissect the synchrophasor packets from the network binary data utilizing pySynphasor module. 
	
	%\begin{figure*}[h]
	%	\centering
	%	\includegraphics[width=\linewidth]{Result_MITM.png}
	%	\caption{MITM and Eavesdropping}
	%	\label{fig:Result_ARP_Poisoning}
	%\end{figure*}
	
	\subsection{C37.118.2 FDIA Results} \label{sub:fdia_results}
\begin{figure*}[h]
		\centering
		\includegraphics[width=\linewidth]{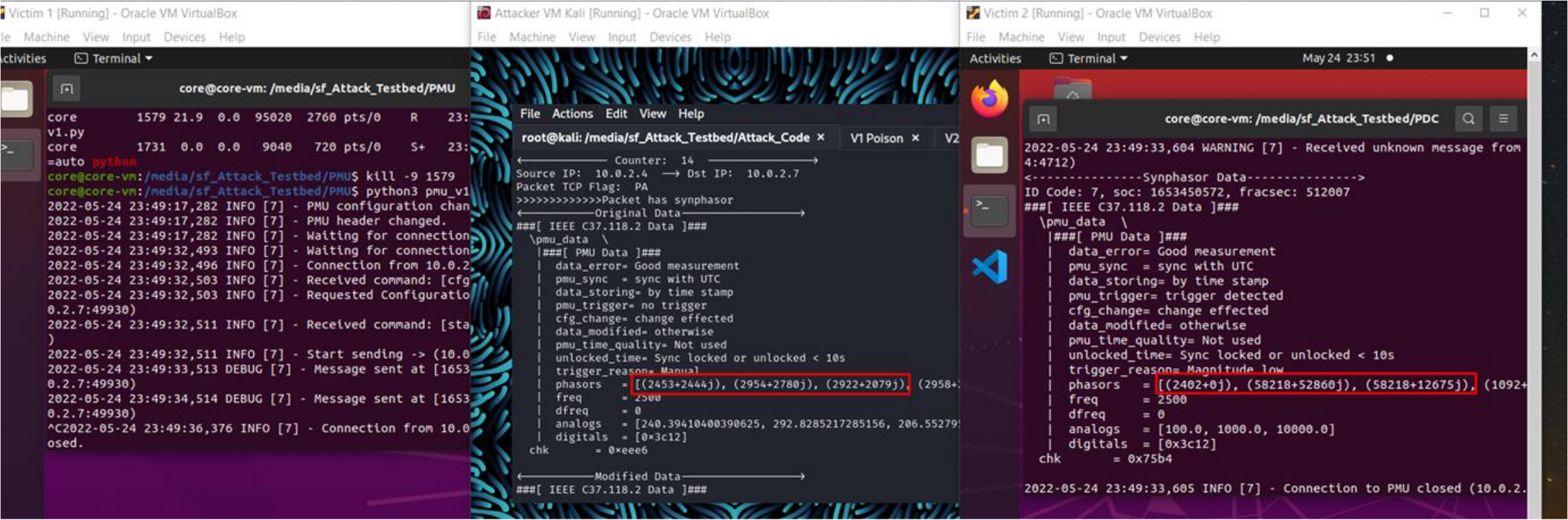}
		\caption{Real implications of False Data Injection Attack}
		\vspace{-15pt}
		\label{fig:result_fdia}
	\end{figure*}

	After ARP poisoning, the attacker machine has access to the victim packet. It created a temporary queue for manipulating and forwarding the packet afterward. In the figure \ref{fig:FDIA Attack Mechanism},  we explained how the attacker gets access to the synchrophasor data from Linux kernel space to the user space through NetfilterQueue and iptables package. Afterward, utilizing the pySynphasor module, the attacker dissects the synchrophasor packet. The figure \ref{fig:result_fdia} presents scenarios of  MITM and FDIA attack. Here, PMU and PDC are transferring packets normally. The left VM represents the PMU, and the right side represents the PDC and Center VM, which is the attacker machine. PMU and PDC are transferring data normally. By poisoning the PMU and PDC ARP table, the attacker now steals the session of the PMU and PDC. After deploying the attack script, it has access to the synchrophasor data. In this way, the attacker is eavesdropping on the synchrophasor packet.The center VM  in figure \ref{fig:result_fdia} presents  dissection of one of the synchrophasor data packet. pySynphasor module paved the way for real-time dissection of synchrophasor packets. The figure \ref{fig:result_fdia} also presents a scenario of false data injection attack. In the attacker VM, we can observe that the phasor measurements were [(2453+2444j), (2954+2780j), (2922+2079j)] before the attack, and after the attack, it was injected to  [(2402+0j), (58218+ 2860j),(58218+12675j)]. The packet injection mechanism is just a few lines of pySynphasor commands, just like the figure in \ref{fig:pySynphasor_data_building.png}. As the packet contents are changing, it is mandatory to update the fields that keep track of the packet signature, such as TCP and IP packet length, checksum, and IEEE C37.118.2 CRC value. Updating of these signature fields is performed in the script. After the injection, the script forwards the packet to the PDC. This is how the testbed smoothens the different attack scenarios for testing large system deployment.

	%We developed an attack script for manipulating the synchrophasor data. After manipulating the packet, the packet TCP, IP and synchrophasor packets fields such as IP and TCP packet length, checksum changed. So we are updating all the fields that hold the signature of packet integrity after manipulating the packet. By this way the receiver PDC consider it as a valid packet. The figure xx demonstrates that the original data that the PMU send to PDC as xx and attacker inject a false data in its field and changed it to xx. So the testbed can experiment such cyber-physical scenario in real-time. 
	
	\subsection{Attack Detection} \label{sub:attack_detection}
	The testbed paves the way for testing different types of cyber attack detection in smart grid systems. The detection mechanism can be classified into packet-based and physics-based. Linear state estimation-based bad data detection technique is physics-based attack detection. Figure \ref{fig:state estimation}(d), \ref{fig:state estimation}(e) and \ref{fig:state estimation}(f) represents one of the scenarios of linear state estimation based bad data detection technique. Where the meter 1 measurement was poisoned by the attacker and bad data detection algorithms were performed in the control center. We mentioned earlier that for the degree of freedom $K=7$ and $\sigma=0.05$ The threshold value, $T_j=9.49$. But in the figure \ref{fig:state estimation}(f), the value of $j = 19.6005>T_j$. That indicates that there is bad data present in the measurement. The \ref{fig:state estimation}(f) also demonstrates the result of hypothesis testing on meter measurements. The error value $\sigma_Y>3 $ for meter 1. It proves that the attacker poisoned the meter 1 measurement.

	\section{Discussion}
	
	%Summarise key findings
	Our main goal was to present a scalable cyber-physical testbed for smart grid system so that the testbed can be utilized to experiment different detection mechanisms for bigger system that will prevent future attack just like 2015  blackout in Ukraine due to the False Command Injection Attack(FCIA) \cite{liang2015UkraineBlackout2017}. In the method section, we described how to develop such high fidelity testbed that can mimic a real system. We not only focused on designing the testbed but also demonstrated how to design real-time attack on the testbed so that the system can be studied from both the attacker's point of view and the victim's point of view. Another major part of the testbed was the demonstration of the pySynphasor module. It can be utilized in many ways because of its complete capabilities of building and dissecting the IEEE C37.118.2 protocol. We developed a simple PDC named as pyPDC utilizing the module; we also designed an FDIA attack utilizing the module. It is also possible to design different types of attacks in IEEE C37.118.2 protocols. One of the potential testings might be implementing fuzz testing on synchrophasor devices. Therefore, pySynphasor has many usage cases to explore in future research. 
	
	Then, in the result section, we presented how the whole testbed works. In the \ref{sub:state_estimation_in_normal_conditions} subsection, state estimation and bad data detection results for IEEE 13 node test feeder has been presented. In the \ref{sub:core_network_result} subsection, we presented how software emulated the network connecting the whole testbed. The whole smart grid is ready and equipped with a cyber layer and physical layer.In the \ref{sub:pysynpharos_result} subsection, we presented how pySynphasor can build and dissect IEEE C37.118.2 protocol intuitively. Performing cyber attacks was one of the main goals of the testbed. In the \ref{sub:arp_poisoning_result} subsection, we presented how to access the PMU data packet by applying the ARP poisoning mechanism. In the \ref{sub:fdia_results} subsection, we presented how to inject false data on the phasor measurements utilizing the pySynphasor module. Finally, we focused on the detection mechanism. So, subsection \ref{sub:attack_detection} presented how linear estate estimation based bad data detection can identify false data injection attack on synchrophasor data.

	%Any unexpected results
	
	%Limitations and weakness in the results
	The synchrophasor standard suggested  IEEE C37.118.2-2011 and IEC 61850-90-5 protocols for PMU and PDC communication. But, we only explored the IEEE C37.118.2 protocol and did not address  IEC 61850-90-5 protocol in the testbed. 
	
	%Potential follow up research

	%Conclude with restatement of most important findings

	%Our main goal was to demonstrate a scalable cyber-physical testbed for smart grid that incorporates synchrophasor and IEEE C37.118.2 communication protocol. We also focused on how such testbed can be utilized to figure out the vulnerabilities in bigger system by experimenting with different types of detection mechanisms. So we developed a high fidelity testbed that can mimic a real system.
	%
	%In the xx subsection we build a IEEE 13 node test feeder that we deployed in Opal-RT and demonstrate linear state estimation results that estimate the node complex voltages. Then we demonstrated how to build the PMU, PDC and cyber layer utilizing virtual machines and CORE network emulator. At this point the whole smart grid is ready equipped with cyber layer and physical layer. One of the fundamental parts of the testbed was to perform dissection on IEEE C37.118.2 protocol and demonstrate the effectiveness of the pySynphasor module. section xx describes that how a single python command can dissect the synchrophasor network packet on the fly and presents results in human-readable format. Performing cyber attack was one of the main goals of the testbed. Section xx we demonstrated the results of different types of attacks and how to perform those attacks using different tools. Finally we focused on detection mechanism . The figure xx demonstrate how such types of attack can be detected utilizing bad data detection technique.  
	
	\section{Conclusion}
	The development of a scalable cyber-physical testbed for smart grid has been presented. The effectiveness of such testbed has been demonstrated to identify the best detection mechanism for a more extensive system. We also presented the mechanism of building and dissection synchrophasor network packets that is also useful in detecting packet-based detection. Finally, we presented a physics-based detection mechanism such as state estimation and bad data detection. Although the system is highly scalable, it can be more robust than the virtual machine by the docker container. So building the whole testbed on top of the docker container will be follow-up research. Designing the testbed around IEC61850 will be another improvement of current research.

	% \balance % uncomment this can remove two blanks last pages.
	\bibliographystyle{IEEEtran}
	%\bibliography{References}
	\bibliography{References}

% Generated by IEEEtran.bst, version: 1.14 (2015/08/26)
\begin{thebibliography}{10}
\providecommand{\url}[1]{#1}
\csname url@samestyle\endcsname
\providecommand{\newblock}{\relax}
\providecommand{\bibinfo}[2]{#2}
\providecommand{\BIBentrySTDinterwordspacing}{\spaceskip=0pt\relax}
\providecommand{\BIBentryALTinterwordstretchfactor}{4}
\providecommand{\BIBentryALTinterwordspacing}{\spaceskip=\fontdimen2\font plus
\BIBentryALTinterwordstretchfactor\fontdimen3\font minus
  \fontdimen4\font\relax}
\providecommand{\BIBforeignlanguage}[2]{{%
\expandafter\ifx\csname l@#1\endcsname\relax
\typeout{** WARNING: IEEEtran.bst: No hyphenation pattern has been}%
\typeout{** loaded for the language `#1'. Using the pattern for}%
\typeout{** the default language instead.}%
\else
\language=\csname l@#1\endcsname
\fi
#2}}
\providecommand{\BIBdecl}{\relax}
\BIBdecl

\bibitem{prabhuStateoftheartReviewSynchrophasor2017}
M.~S. Prabhu and P.~K. Nayak, ``A state-of-the-art review on synchrophasor
  applications to power network protection,'' vol. 436, pp. 531--541.

\bibitem{khanThreatAnalysisBlackEnergy2016}
R.~Khan, P.~Maynard, K.~McLaughlin, D.~Laverty, and S.~Sezer, ``Threat analysis
  of {{BlackEnergy}} malware for synchrophasor based real-time control and
  monitoring in smart grid,'' pp. 1--11.

\bibitem{wuPowerSystemState1990}
\BIBentryALTinterwordspacing
F.~F. Wu, ``Power system state estimation: A survey,'' vol.~12, no.~2, pp.
  80--87. [Online]. Available:
  \url{https://www.sciencedirect.com/science/article/pii/014206159090003T}
\BIBentrySTDinterwordspacing

\bibitem{ieeepowerandenergysocietyIEEEStdC372013}
{IEEE Power and Energy Society}, ``{{IEEE Std C37}}.244™-2013,'' pp. 1--65.

\bibitem{khanAnalysisIEEEC372016}
\BIBentryALTinterwordspacing
R.~Khan, K.~Mclaughlin, D.~Laverty, .~Sezer, R.~Khan, K.~Mclaughlin,
  D.~Laverty, and S.~Sezer, ``Analysis of {{IEEE C37}}. 118 and {{IEC}}
  61850-90-5 synchrophasor communication frameworks,'' p. 2016. [Online].
  Available: \url{https://ieeexplore.ieee.org/abstract/document/7741343/}
\BIBentrySTDinterwordspacing

\bibitem{wangRevisedBranchCurrentbased2004}
H.~Wang and N.~Schulz, ``A revised branch current-based distribution system
  state estimation algorithm and meter placement impact,'' vol.~19, no.~1, pp.
  207--213.

\bibitem{chenStateEstimationDistribution2015}
X.~Chen, K.~J. Tseng, and G.~Amaratunga, ``State estimation for distribution
  systems using micro-synchrophasors,'' in \emph{2015 {{IEEE PES Asia-Pacific
  Power}} and {{Energy Engineering Conference}} ({{APPEEC}})}, pp. 1--5.

\bibitem{nayak2010}
G.~N. Nayak and S.~G. Samaddar, ``Different flavours of {{Man-In-The-Middle}}
  attack, consequences and feasible solutions,'' vol.~5, pp. 491--495.

\bibitem{yangManinthemiddleAttackTestbed2012}
Y.~Yang, K.~McLaughlin, T.~Littler, S.~Sezer, E.~G. Im, Z.~Q. Yao,
  B.~Pranggono, and H.~F. Wang, ``Man-in-the-middle attack test-bed
  investigating cyber-security vulnerabilities in smart grid {{SCADA}}
  systems,'' vol. 2012.

\bibitem{liang2015UkraineBlackout2017}
G.~Liang, S.~R. Weller, J.~Zhao, F.~Luo, and Z.~Y. Dong, ``The 2015 {{Ukraine
  Blackout}}: {{Implications}} for {{False Data Injection Attacks}},'' vol.~32,
  no.~4, pp. 3317--3318.

\bibitem{khanIEEEC3711822016}
R.~Khan, K.~McLaughlin, D.~L. â. C. o.~I. …, and u.~2016, ``{{IEEE
  C37}}.118-2 {{Synchrophasor Communication Framework Overview}}, {{Cyber
  Vulnerabilities Analysis}} and {{Performance Evaluation}}.''

\bibitem{fanSynchrophasorDataCorrection2018}
\BIBentryALTinterwordspacing
X.~Fan, L.~Du, D.~D. I. T. o.~S. Grid, and u.~2017, ``Synchrophasor data
  correction under {{GPS}} spoofing attack: {{A}} state estimation-based
  approach.'' [Online]. Available:
  \url{https://ieeexplore.ieee.org/abstract/document/7839276/}
\BIBentrySTDinterwordspacing

\bibitem{paudelDataIntegrityAttacks2016}
\BIBentryALTinterwordspacing
S.~Paudel, P.~Smith, T.~Z. f. I. . S. C.~S. …, and u.~2016, ``Data integrity
  attacks in smart grid wide area monitoring.'' [Online]. Available:
  \url{https://www.scienceopen.com/hosted-document?doi=10.14236/ewic/ICS2016.9}
\BIBentrySTDinterwordspacing

\bibitem{singhStealthyCyberAttacks2016}
\BIBentryALTinterwordspacing
V.~Singh, A.~O. . N.~A. …, and u.~2016, ``Stealthy cyber attacks and impact
  analysis on wide-area protection of smart grid.'' [Online]. Available:
  \url{https://ieeexplore.ieee.org/abstract/document/7747927/}
\BIBentrySTDinterwordspacing

\bibitem{koltysSHaPeHoneypotElectric2015}
\BIBentryALTinterwordspacing
K.~Kołtyś and R.~Gajewski, ``{{SHaPe}}: {{A Honeypot}} for {{Electric Power
  Substation}},'' pp. 37--43. [Online]. Available:
  \url{https://www.infona.pl//resource/bwmeta1.element.baztech-dd7fe369-6682-4a1e-bd10-e89d3723cbb6}
\BIBentrySTDinterwordspacing

\bibitem{johnsonAssessingNetworkCybersecurity2020}
J.~Johnson, I.~Onunkwo, P.~C. I. C.-P. …, and u.~2020, ``Assessing {{DER}}
  network cybersecurity defences in a power-communication co-simulation
  environment.''

\bibitem{wlazloManinTheMiddleAttacksDefense2021}
\BIBentryALTinterwordspacing
P.~Wlazlo, A.~Sahu, Z.~Mao, H.~Huang, A.~Goulart, K.~Davis, and S.~Zonouz,
  ``Man-in-{{The-Middle Attacks}} and {{Defense}} in a {{Power System
  Cyber-Physical Testbed}}.'' [Online]. Available:
  \url{http://arxiv.org/abs/2102.11455}
\BIBentrySTDinterwordspacing

\bibitem{rodofileRealTimeInteractiveAttacks2015}
N.~R. Rodofile, K.~Radke, and E.~Foo, ``Real-{{Time}} and {{Interactive
  Attacks}} on {{DNP3 Critical Infrastructure Using Scapy}}.''

\bibitem{khanDemonstratingCyberphysicalAttacks2018}
R.~Khan, K.~McLaughlin, J.~H.~D. Laverty, H.~David, and S.~Sezer,
  ``Demonstrating {{Cyber-Physical Attacks}} and {{Defense}} for
  {{Synchrophasor Technology}} in {{Smart Grid}}.''

\bibitem{cuiCyberPhysicalSystem2020}
H.~Cui, F.~Li, K.~T. I. E.~S. Integration, and u.~2020, ``Cyber‐physical
  system testbed for power system monitoring and wide‐area control
  verification.''

\bibitem{sandiPypmuOpenSource2016}
\BIBentryALTinterwordspacing
S.~Šandi, B.~Krstajić, T.~P. .~t. Telecommunications, and u.~2016,
  ``Pypmu—open source python package for synchrophasor data transfer.''
  [Online]. Available:
  \url{https://ieeexplore.ieee.org/abstract/document/7818916/}
\BIBentrySTDinterwordspacing

\bibitem{adhikariCyberphysicalPowerSystem2014}
\BIBentryALTinterwordspacing
U.~Adhikari, T.~Morris, S.~P. . I. P.~G. Meeting, and u.~2014, ``A
  cyber-physical power system test bed for intrusion detection systems.''
  [Online]. Available:
  \url{https://ieeexplore.ieee.org/abstract/document/6939262/}
\BIBentrySTDinterwordspacing

\bibitem{cintugluSurveySmartGrid2017}
\BIBentryALTinterwordspacing
M.~Cintuglu, O.~Mohammed, K.~A. â. S.~. Tutorials, and u.~2016, ``A survey on
  smart grid cyber-physical system testbeds.'' [Online]. Available:
  \url{https://ieeexplore.ieee.org/abstract/document/7740849/}
\BIBentrySTDinterwordspacing

\bibitem{yangCybersecurityTestbedIEC2015}
\BIBentryALTinterwordspacing
Y.~Yang, H.~Jiang, K.~McLaughlin, L.~G. . I. P.~. …, and u.~2015,
  ``Cybersecurity test-bed for {{IEC}} 61850 based smart substations,'' pp.
  1--5. [Online]. Available:
  \url{https://ieeexplore.ieee.org/abstract/document/7286357/}
\BIBentrySTDinterwordspacing

\bibitem{monticelliElectricPowerSystem2000}
A.~Monticelli, ``Electric power system state estimation,'' vol.~88, no.~2, pp.
  262--282.

\bibitem{taraliBadDataDetection2012}
\BIBentryALTinterwordspacing
A.~Tarali, ``Bad data detection in two stage estimation using phasor
  measurements.'' [Online]. Available:
  \url{http://hdl.handle.net/2047/d20002926}
\BIBentrySTDinterwordspacing

\bibitem{zhangDesignTestingImplementation2017a}
\BIBentryALTinterwordspacing
L.~Zhang, A.~Bose, A.~Jampala, V.~Madani, and J.~Giri, ``Design, {{Testing}},
  and {{Implementation}} of a {{Linear State Estimator}} in a {{Real Power
  System}},'' vol.~8, no.~4, pp. 1782--1789. [Online]. Available:
  \url{http://ieeexplore.ieee.org/document/7373664/}
\BIBentrySTDinterwordspacing

\bibitem{bandakPOWERSYSTEMSSTATE2013}
C.~E. Bandak, ``{{POWER SYSTEMS STATE ESTIMATION}}.''

\bibitem{kerstingRadialDistributionTest2001}
W.~Kersting, ``Radial distribution test feeders,'' in \emph{2001 {{IEEE Power
  Engineering Society Winter Meeting}}. {{Conference Proceedings}} ({{Cat}}.
  {{No}}.{{01CH37194}})}, vol.~2, pp. 908--912 vol.2.

\bibitem{ieeepowerandenergysocietyIEEEStdC372011}
I.~P. a.~E. Society, ``{{IEEE Std C37}}.118.1™-2011,'' pp. 1--61.

\bibitem{hashimotoVagrantRunningCreate2013}
\BIBentryALTinterwordspacing
M.~Hashimoto, \emph{Vagrant: Up and Running: Create and Manage Virtualized
  Development Environments}. [Online]. Available:
  \url{https://books.google.com/books?hl=en&lr=&id=7rJqqKCvdagC&oi=fnd&pg=PR2&dq=vagrant+up+and+running&ots=qDzDYo9MI5&sig=8NYhVHLCyNsW6ojyHL39KO4p5S8}
\BIBentrySTDinterwordspacing

\bibitem{virtualboxChapterVirtualNetworking}
\BIBentryALTinterwordspacing
VirtualBox. Chapter~6.~{{Virtual Networking}}. [Online]. Available:
  \url{https://www.virtualbox.org/manual/ch06.html}
\BIBentrySTDinterwordspacing

\end{thebibliography}



@article{adhikariCyberphysicalPowerSystem2014,
  title = {A Cyber-Physical Power System Test Bed for Intrusion Detection Systems},
  author = {Adhikari, U and Morris, TH and Meeting, S Pan - 2014 IEEE PES General and 2014, undefined},
  date = {2014},
  journaltitle = {ieeexplore.ieee.org},
  url = {https://ieeexplore.ieee.org/abstract/document/6939262/},
  urldate = {2022-03-01},
  abstract = {The rapid advancement of technology used in operation, monitoring, and control introduces several threats against power system. Cyber-physical power system vulnerabilities are increasing and the consequences of attack can be catastrophic. Understanding power system phenomena and attacks is vital to identifying and detecting such events. Researchers require a suitable power system test bed that can provide a platform for simulation of power system events and attacks. An essential part of such a test bed is the ability to provide software and hardware interaction to mimic real world scenarios. This paper presents a test bed for the development of an intrusion detection system (IDS) for power systems. The test bed consists of a power system modeled on a real time digital simulator (RTDS), a data collection and processing engine, and a MATLAB/RSCAD parameter calculation engine. This test bed provides a platform for hardware in the loop (HIL) simulation, power system attacks, and generates data sets required by cyber security researchers. Coordinated distance protection and overcurrent protection schemes are implemented on the IEEE 9 bus system and a 3-generator 4 bus system [11]. Fault, contingency and cyber-attack scenarios have been developed for both power systems. Selected relevant simulation results are presented.},
  isbn = {9781479964154},
  keywords = {contingencies,data,faults,IDS,Index Terms-attacks,power system,test bed},
  file = {C\:\\Users\\dassc\\Zotero\\storage\\6P8IKLDQ\\Adhikari et al_A cyber-physical power system test bed for intrusion detection systems.pdf}
}

@article{aristizabalModelingDistributedPower2016,
  title = {Modeling of {{Distributed Power Systems}} in 13 {{Nodes IEEE Electric Grids}}},
  author = {Aristizabal, Andres Julian and Pinilla, Henry Giovanni and Forero, Carlos Andres},
  date = {2016-08-31},
  journaltitle = {Periodicals of Engineering and Natural Sciences (PEN)},
  shortjournal = {PEN},
  volume = {4},
  number = {2},
  issn = {2303-4521},
  doi = {10.21533/pen.v4i2.55},
  url = {http://pen.ius.edu.ba/index.php/pen/article/view/55},
  urldate = {2022-06-20},
  abstract = {This work aims to develop a model capable of evaluating the behavior of distributed energy resources in 13-nodes IEEE systems as a result of the change in the disconnector’s opening protocol that creates a power generation island. The first scenario simulated a failure in the 632-671 line isolating the subsystem into two 375 kVA distributed generation units (DG) in the nodes 675 and 652. Likewise, a second scenario considered the aperture of the disconnector located between nodes 671 and 692 representing a 375 kVA DG feeding a 900 kVA load. The last scenario produced a threephase failure modeling two 500 kVA DG units in the nodes 634 and 646 supplying an 800 kVA load.},
  langid = {english},
  annotation = {1 citations (Crossref) [2022-06-20]},
  file = {C\:\\Users\\dassc\\Zotero\\storage\\GKNB5QZ5\\Aristizabal et al. - 2016 - Modeling of Distributed Power Systems in 13 Nodes .pdf}
}

@thesis{bandakPOWERSYSTEMSSTATE2013,
  title = {{{POWER SYSTEMS STATE ESTIMATION}}},
  author = {Bandak, Carlos Expedite},
  date = {2013},
  file = {C\:\\Users\\dassc\\Zotero\\storage\\V64W3G9Q\\Bandak_2013_POWER SYSTEMS STATE ESTIMATION.pdf}
}

@misc{biondiNetworkPacketManipulation2007,
  title = {Network Packet Manipulation with {{Scapy}}},
  author = {Biondi, Philippe},
  date = {2007},
  file = {C\:\\Users\\dassc\\Zotero\\storage\\BLMYNB8B\\m-api-75d70f1b-8066-fa3f-7c43-c084feed577e.pdf}
}

@inproceedings{chenStateEstimationDistribution2015,
  title = {State Estimation for Distribution Systems Using Micro-Synchrophasors},
  booktitle = {2015 {{IEEE PES Asia-Pacific Power}} and {{Energy Engineering Conference}} ({{APPEEC}})},
  author = {Chen, Xuebing and Tseng, King Jet and Amaratunga, Gehan},
  date = {2015-11},
  pages = {1--5},
  doi = {10.1109/APPEEC.2015.7381051},
  abstract = {Phasor Measurement Unit (PMU) has contributed greatly to power system state estimation because of its ability to directly measure voltage and current phase angle. PMUs have been used almost exclusively in transmission systems monitoring. Recent years, increasing penetration of renewable energy sources brings new characteristics into distribution system such as bi-directional power flows and voltage profile issues, which has necessitated continuous monitoring of distribution systems. Therefore, fast and accurate measurement device is needed for application in distribution networks. μPMU is such a device that can provide phasor measurements with high precision and low cost. This paper uses a linear three phase state estimator for applications in distribution systems. The proposed estimator can make use of synchrophasor measurements which can be realized by μPMU. This is tested on IEEE 13 bus feeder which has unbalanced three phase transmission lines and loads.},
  eventtitle = {2015 {{IEEE PES Asia-Pacific Power}} and {{Energy Engineering Conference}} ({{APPEEC}})},
  keywords = {✅,Current measurement,Phasor measurement units,Power measurement,State estimation,Transmission line matrix methods,Transmission line measurements,Voltage measurement},
  annotation = {13 citations (Crossref) [2022-05-30]},
  file = {C\:\\Users\\dassc\\Zotero\\storage\\VJMEU3IM\\Chen et al_2015_State estimation for distribution systems using micro-synchrophasors.pdf;C\:\\Users\\dassc\\Zotero\\storage\\K52W93CT\\7381051.html}
}

@article{cintugluSurveySmartGrid2017,
  title = {A Survey on Smart Grid Cyber-Physical System Testbeds},
  author = {Cintuglu, MH and Mohammed, OA and Tutorials, K Akkaya - … Surveys \& and 2016, undefined},
  date = {2017},
  journaltitle = {ieeexplore.ieee.org},
  url = {https://ieeexplore.ieee.org/abstract/document/7740849/},
  urldate = {2022-03-01},
  file = {C\:\\Users\\dassc\\Zotero\\storage\\55MGEJES\\full-text.pdf}
}

@article{cuiCyberPhysicalSystem2020,
  title = {Cyber‐physical System Testbed for Power System Monitoring and Wide‐area Control Verification},
  author = {Cui, H and Li, F and Integration, K Tomsovic - IET Energy Systems and 2020, undefined},
  date = {2020},
  journaltitle = {ieeexplore.ieee.org},
  doi = {10.1049/iet-esi.2019.0084},
  file = {C\:\\Users\\dassc\\Zotero\\storage\\JEZFIDB3\\Cui et al_Cyber‐physical system testbed for power system monitoring and wide‐area control.pdf}
}

@article{davisCyberphysicalModelingAssessment2015,
  title = {A Cyber-Physical Modeling and Assessment Framework for Power Grid Infrastructures},
  author = {Davis, KR and Davis, CM and on smart {grid}, SA Zonouz - … and 2015, undefined},
  date = {2015},
  journaltitle = {ieeexplore.ieee.org},
  url = {https://ieeexplore.ieee.org/abstract/document/7103368/},
  urldate = {2022-01-12},
  file = {C\:\\Users\\dassc\\Zotero\\storage\\G3PISGVI\\full-text.pdf}
}

@article{fanSynchrophasorDataCorrection2018,
  title = {Synchrophasor Data Correction under {{GPS}} Spoofing Attack: {{A}} State Estimation-Based Approach},
  author = {Fan, X and Du, L and Grid, D Duan - IEEE Transactions on Smart and 2017, undefined},
  date = {2018},
  journaltitle = {ieeexplore.ieee.org},
  url = {https://ieeexplore.ieee.org/abstract/document/7839276/},
  urldate = {2022-03-01},
  file = {C\:\\Users\\dassc\\Zotero\\storage\\3C6N46WG\\full-text.pdf}
}

@book{hashimotoVagrantRunningCreate2013,
  title = {Vagrant: Up and Running: Create and Manage Virtualized Development Environments},
  author = {Hashimoto, M},
  date = {2013},
  url = {https://books.google.com/books?hl=en&lr=&id=7rJqqKCvdagC&oi=fnd&pg=PR2&dq=vagrant+up+and+running&ots=qDzDYo9MI5&sig=8NYhVHLCyNsW6ojyHL39KO4p5S8},
  urldate = {2022-02-23}
}

@article{haughtonLinearStateEstimation2013,
  title = {A {{Linear State Estimation Formulation}} for {{Smart Distribution Systems}}},
  author = {Haughton, Daniel A. and Heydt, Gerald Thomas},
  date = {2013-05},
  journaltitle = {IEEE Transactions on Power Systems},
  volume = {28},
  number = {2},
  pages = {1187--1195},
  issn = {1558-0679},
  doi = {10.1109/TPWRS.2012.2212921},
  abstract = {This paper presents a linearized, three-phase, distribution class state estimation algorithm for applications in smart distribution systems. Unbalanced three-phase cases and single-phase cases are accommodated. The estimator follows a complex variable formulation and is intended to incorporate synchronized phasor measurements into distribution state estimation. Potential applications in smart distribution system control and management are discussed.},
  eventtitle = {{{IEEE Transactions}} on {{Power Systems}}},
  keywords = {��,Current measurement,Distribution management systems,distribution system monitoring and control,distribution system state estimation,Loading,power distribution engineering,Power measurement,Real time systems,State estimation,synchronized phasor measurements,three-phase unbalance,Vectors,Voltage measurement},
  file = {C\:\\Users\\dassc\\Zotero\\storage\\FJ9AECVJ\\Haughton_Heydt_2013_A Linear State Estimation Formulation for Smart Distribution Systems.pdf;C\:\\Users\\dassc\\Zotero\\storage\\HSB7IVTF\\Haughton and Heydt - 2013 - A Linear State Estimation Formulation for Smart Di.pdf;C\:\\Users\\dassc\\Zotero\\storage\\SLRUCQMQ\\6302216.html}
}

@article{ieeepowerandenergysocietyIEEEStdC372011,
  title = {{{IEEE Std C37}}.118.1™-2011},
  author = {IEEE Power {and} Energy Society},
  date = {2011-12},
  journaltitle = {IEEE Std C37.118.1-2011 (Revision of IEEE Std C37.118-2005)},
  pages = {1--61},
  doi = {10.1109/IEEESTD.2011.6111219},
  abstract = {Synchronized phasor (synchrophasor) measurements for power systems are presented. This standard defines synchrophasors, frequency, and rate of change of frequency (ROCOF) measurement under all operating conditions. It specifies methods for evaluating these measurements and requirements for compliance with the standard under both steady-state and dynamic conditions. Time tag and synchronization requirements are included. Performance requirements are confirmed with a reference model, provided in detail. This document defines a phasor measurement unit (PMU), which can be a stand-alone physical unit or a functional unit within another physical unit. This standard does not specify hardware, software, or a method for computing phasors, frequency, or ROCOF.},
  eventtitle = {{{IEEE Std C37}}.118.1-2011 ({{Revision}} of {{IEEE Std C37}}.118-2005)},
  keywords = {data concentrator,Data processing,Data storage,DC,Error statistics,FE,frequency error,IEEE C37.118.1,IEEE standards,IRIG-B,PDC,phasor,phasor measurement,phasor measurement unit,Phasor measurement units,PMU,RFE,ROCOF,ROCOF error,Synchronization,synchronized phasor,synchrophasor,total vector error,TVE,Vectors},
  annotation = {173 citations (Crossref) [2022-05-30]},
  file = {C\:\\Users\\dassc\\Zotero\\storage\\2W6YVCSU\\IEEE Power and Energy Society_2011_IEEE Std C37.pdf;C\:\\Users\\dassc\\Zotero\\storage\\8F99K5E2\\2011_IEEE Standard for Synchrophasor Measurements for Power Systems.pdf;C\:\\Users\\dassc\\Zotero\\storage\\2T9GDBEY\\6111219.html}
}

@article{ieeepowerandenergysocietyIEEEStdC372013,
  title = {{{IEEE Std C37}}.244™-2013},
  author = {{IEEE Power and Energy Society}},
  date = {2013},
  journaltitle = {IEEE Std C37.244-2013},
  pages = {1--65},
  doi = {10.1109/IEEESTD.2013.6514039},
  abstract = {The functional, performance, and testing guidelines for a phasor data concentrator are described in this guide. Supporting information is also provided.},
  isbn = {978-0-7381-8260-5},
  issue = {May},
  keywords = {Concentrators,GPS synchronization,IEEE C37.244TM,IEEE guide,IEEE standards,pdata concentrator (DC),phasor,phasor data concentrator (PDC),phasor data concentrator requirement,phasor measurement,phasor measurement unit (PMU),Phasor measurements,Phasors,power system control,power system monitoring,power system protection,Power system protection,Power system reliability,Synchronization,synchrophasor},
  file = {C\:\\Users\\dassc\\Zotero\\storage\\4K4YGBP5\\IEEE Power and Energy Society - 2013 - IEEE Guide for Phasor Data Concentrator Requiremen.pdf}
}

@article{ieeepowerenergysocietEEEStdC372011,
  title = {{{EEE Std C37}}.118.2™-2011},
  author = {IEEE Power \& Energy Societ},
  date = {2011-12},
  journaltitle = {IEEE Std C37.118.2-2011 (Revision of IEEE Std C37.118-2005)},
  pages = {1--53},
  doi = {10.1109/IEEESTD.2011.6111222},
  abstract = {A method for real-time exchange of synchronized phasor measurement data between power system equipment is defined. This standard specifies messaging that can be used with any suitable communication protocol for real-time communication between phasor measurement units (PMU), phasor data concentrators (PDC), and other applications. It defines message types, contents, and use. Data types and formats are specified. A typical measurement system is described. Communication options and requirements are described in annexes. Keywords: data concentrator, DC, IEEE C37.118.2, PDC, phasor, phasor data concentrator, phasor measurement unit, PMU, synchronized phasor, synchrophasor.},
  eventtitle = {{{IEEE Std C37}}.118.2-2011 ({{Revision}} of {{IEEE Std C37}}.118-2005)},
  keywords = {data concentrator,Data processing,Data storage,DC,IEEE C37.118.2,IEEE standards,PDC,phasor,phasor data concentrator,phasor measurement unit,Phasor measurement units,PMU,Power system dynamics,Synchronization,synchronized phasor,synchrophasor},
  file = {C\:\\Users\\dassc\\Zotero\\storage\\KBWZG3XA\\2011_IEEE Standard for Synchrophasor Data Transfer for Power Systems.pdf;C\:\\Users\\dassc\\Zotero\\storage\\HWIY58QF\\6111222.html}
}

@article{johnsonAssessingNetworkCybersecurity2020,
  title = {Assessing {{DER}} Network Cybersecurity Defences in a Power-Communication Co-Simulation Environment},
  author = {Johnson, J and Onunkwo, I and …, P Cordeiro - IET Cyber-Physical and 2020, undefined},
  date = {2020},
  journaltitle = {ieeexplore.ieee.org},
  file = {C\:\\Users\\dassc\\Zotero\\storage\\TXA33S25\\Johnson et al_Assessing DER network cybersecurity defences in a power-communication.pdf}
}

@inproceedings{kerstingRadialDistributionTest2001,
  title = {Radial Distribution Test Feeders},
  booktitle = {2001 {{IEEE Power Engineering Society Winter Meeting}}. {{Conference Proceedings}} ({{Cat}}. {{No}}.{{01CH37194}})},
  author = {Kersting, W.H.},
  date = {2001-01},
  volume = {2},
  pages = {908-912 vol.2},
  doi = {10.1109/PESW.2001.916993},
  abstract = {Many computer programs are available for the analysis of radial distribution feeders. In 1992 a paper was published that presented the complete data for three four-wire wye and one three-wire delta radial distribution test feeders. The purpose of publishing the data was to make available a common set of data that could be used by program developers and users to verify the correctness of their solutions. This paper presents an updated version of the same test feeders along with a simple system that can be used to test three-phase transformer models.},
  eventtitle = {2001 {{IEEE Power Engineering Society Winter Meeting}}. {{Conference Proceedings}} ({{Cat}}. {{No}}.{{01CH37194}})},
  keywords = {Aluminum,Capacitors,Conductors,Copper,Distributed computing,Impedance,Load modeling,Phase transformers,Shunt (electrical),System testing},
  annotation = {614 citations (Crossref) [2022-06-22]},
  file = {C\:\\Users\\dassc\\Zotero\\storage\\6UV7LH94\\Kersting_2001_Radial distribution test feeders.pdf;C\:\\Users\\dassc\\Zotero\\storage\\ZUK7L4LK\\Kersting_2001_Radial distribution test feeders.pdf;C\:\\Users\\dassc\\Zotero\\storage\\VKP2PR9Z\\916993.html}
}

@article{khanAnalysisIEEEC372016,
  title = {Analysis of {{IEEE C37}}. 118 and {{IEC}} 61850-90-5 Synchrophasor Communication Frameworks},
  author = {Khan, R and Mclaughlin, K and Laverty, D and Sezer, \& and Khan, Rafiullah and Mclaughlin, Kieran and Laverty, David and Sezer, Sakir},
  date = {2016},
  journaltitle = {ieeexplore.ieee.org},
  pages = {2016},
  publisher = {{PESGM}},
  doi = {10.1109/PESGM.2016.7741343},
  url = {https://ieeexplore.ieee.org/abstract/document/7741343/},
  urldate = {2022-03-01},
  abstract = {ICT in smart grid provides enormous opportunities for real-time and wide-area grid monitoring, protection and control. To this aim, synchrophasor technology was proposed for reliable and secure transmission of grid status information. IEEE C37.118 and IEC 61850-90-5 emerged as two well known communication frameworks for synchrophasor technology. However, literature lacks a comprehensive analysis of some key features and limitations. Further, knowledge of cyber vulnerabilities in both communication frameworks is still quite limited. This paper analyzes characteristics of both communication frameworks inferred from their complete implementation. In particular, it addresses their embedded features, required network character-istics/resources and their resilience against cyber attacks.},
  file = {C\:\\Users\\dassc\\Zotero\\storage\\J9U6TCZS\\full-text.pdf}
}

@article{khanDemonstratingCyberPhysicalAttacks2018,
  title = {Demonstrating {{Cyber-Physical Attacks}} and {{Defense}} for {{Synchrophasor Technology}} in {{Smart Grid}}},
  author = {Khan, Rafiullah and McLaughlin, Kieran and Laverty, John Hastings David and David, Hastings and Sezer, Sakir},
  date = {2018-10-29},
  journaltitle = {2018 16th Annual Conference on Privacy, Security and Trust, PST 2018},
  publisher = {{Institute of Electrical and Electronics Engineers Inc.}},
  doi = {10.1109/PST.2018.8514197},
  abstract = {Synchrophasor technology is used for real-time control and monitoring in smart grid. Previous works in literature identified critical vulnerabilities in IEEE C37.118.2 synchrophasor communication standard. To protect synchrophasor-based systems, stealthy cyber-attacks and effective defense mechanisms still need to be investigated.This paper investigates how an attacker can develop a custom tool to execute stealthy man-in-the-middle attacks against synchrophasor devices. In particular, four different types of attack capabilities have been demonstrated in a real synchrophasorbased synchronous islanding testbed in laboratory: (i) command injection attack, (ii) packet drop attack, (iii) replay attack and (iv) stealthy data manipulation attack. With deep technical understanding of the attack capabilities and potential physical impacts, this paper also develops and tests a distributed Intrusion Detection System (IDS) following NIST recommendations. The functionalities of the proposed IDS have been validated in the testbed for detecting aforementioned cyber-attacks. The paper identified that a distributed IDS with decentralized decision making capability and the ability to learn system behavior could effectively detect stealthy malicious activities and improve synchrophasor network security.},
  isbn = {9781538674932},
  file = {C\:\\Users\\dassc\\Zotero\\storage\\BN3ASKM6\\Khan et al_2018_Demonstrating Cyber-Physical Attacks and Defense for Synchrophasor Technology.pdf}
}

@article{khanIEEEC3711822016,
  title = {{{IEEE C37}}.118-2 {{Synchrophasor Communication Framework Overview}}, {{Cyber Vulnerabilities Analysis}} and {{Performance Evaluation}}},
  author = {Khan, R and McLaughlin, K and …, D Laverty - … Conference on Information and 2016, undefined},
  date = {2016},
  journaltitle = {scitepress.org},
  doi = {10.5220/0005745001670178},
  abstract = {Synchrophasors have become an important part of the modern power system and numerous applications have been developed covering wide-area monitoring, protection and control. Most applications demand continuous transmission of synchrophasor data across large geographical areas and require an efficient communication framework. IEEE C37.118-2 evolved as one of the most successful synchrophasor communication standards and is widely adopted. However, it lacks a predefined security mechanism and is highly vulnerable to cyber attacks. This paper analyzes different types of cyber attacks on IEEE C37.118-2 communication system and evaluates their possible impact on any developed synchrophasor application. Further, the paper also recommends an efficent security mechanism that can provide strong protection against cyber attacks. Although, IEEE C37.118-2 has been widely adopted, there is no clear understanding of the requirements and limitations. To this aim, the paper also presents detailed performance evaluation of IEEE C37.118-2 implementations which could help determine required resources and network characteristics before designing any synchropha-sor application.},
  isbn = {9789897581670},
  keywords = {��,Cyber Security,IEEE C37118,Smart Grid,Synchrophasor,Vulnerability},
  file = {C\:\\Users\\dassc\\Zotero\\storage\\EXRI74Q8\\m-api-2a480766-82be-e365-6013-1cb72f84a562.pdf}
}

@article{khanThreatAnalysisBlackEnergy2016,
  title = {Threat Analysis of {{BlackEnergy}} Malware for Synchrophasor Based Real-Time Control and Monitoring in Smart Grid},
  author = {Khan, Rafiullah and Maynard, Peter and McLaughlin, Kieran and Laverty, David and Sezer, Sakir},
  date = {2016-08-23},
  pages = {1--11},
  doi = {10.14236/ewic/ics2016.7},
  abstract = {The BlackEnergy malware targeting critical infrastructures has a long history. It evolved over time from a simple DDoS platform to a quite sophisticated plug-in based malware. The plug-in architecture has a persistent malware core with easily installable attack specific modules for DDoS, spamming, info-stealing, remote access, boot-sector formatting etc. BlackEnergy has been involved in several high profile cyber physical attacks including the recent Ukraine power grid attack in December 2015. This paper investigates the evolution of BlackEnergy and its cyber attack capabilities. It presents a basic cyber attack model used by BlackEnergy for targeting industrial control systems. In particular, the paper analyzes cyber threats of BlackEnergy for synchrophasor based systems which are used for real-time control and monitoring functionalities in smart grid. Several BlackEnergy based attack scenarios have been investigated by exploiting the vulnerabilities in two widely used synchrophasor communication standards: (i) IEEE C37.118 and (ii) IEC 61850-90-5. Further, the paper also investigates protection strategies for detection and prevention of BlackEnergy based cyber physical attacks.},
  keywords = {��,��},
  annotation = {MAG ID: 2514382028},
  file = {C\:\\Users\\dassc\\Zotero\\storage\\Y3PMFMAU\\Khan et al. - 2016 - Threat analysis of BlackEnergy malware for synchro.pdf}
}

@article{koltysSHaPeHoneypotElectric2015,
  title = {{{SHaPe}}: {{A Honeypot}} for {{Electric Power Substation}}},
  shorttitle = {{{SHaPe}}},
  author = {Kołtyś, K. and Gajewski, R.},
  date = {2015},
  journaltitle = {Journal of Telecommunications and Information Technology},
  pages = {37--43},
  issn = {1509-4553, 1899-8852},
  url = {https://www.infona.pl//resource/bwmeta1.element.baztech-dd7fe369-6682-4a1e-bd10-e89d3723cbb6},
  urldate = {2022-05-28},
  abstract = {Supervisory Control and Data Acquisition (SCADA) systems play a crucial role in national critical infrastructures, and any failure may result in severe damages. Initially SCADA networks were separated from other networks and used proprietary communications protocols that were well known only to the device manufacturers. At that time such isolation and obscurity ensured an acceptable security level. Nowadays, modern SCADA systems usually have direct or indirect Internet connection, use open protocols and commercial-off-the-shelf hardware and software. This trend is also noticeable in the power industry. Present substation automation systems (SASs) go beyond traditional SCADA and employ many solutions derived from Information and Communications Technology (ICT). As a result electric power substations have become more vulnerable for cybersecurity attacks and they need ICT security mechanisms adaptation. This paper shows the SCADA honeypot that allows detecting unauthorized or illicit trac in SAS which communication architecture is dened according to the IEC 61850 standard.},
  issue = {nr 4},
  langid = {english},
  keywords = {��},
  file = {C\:\\Users\\dassc\\Zotero\\storage\\Q24HJEK3\\Kołtyś_Gajewski_2015_SHaPe.pdf;C\:\\Users\\dassc\\Zotero\\storage\\FIXQFPMI\\bwmeta1.element.html}
}

@article{liang2015UkraineBlackout2017,
  title = {The 2015 {{Ukraine Blackout}}: {{Implications}} for {{False Data Injection Attacks}}},
  author = {Liang, Gaoqi and Weller, Steven R. and Zhao, Junhua and Luo, Fengji and Dong, Zhao Yang},
  date = {2017-07-01},
  journaltitle = {IEEE Transactions on Power Systems},
  volume = {32},
  number = {4},
  pages = {3317--3318},
  publisher = {{Institute of Electrical and Electronics Engineers Inc.}},
  issn = {08858950},
  doi = {10.1109/TPWRS.2016.2631891},
  abstract = {In a false data injection attack (FDIA), an adversary stealthily compromises measurements from electricity grid sensors in a coordinated fashion, with a view to evading detection by the power system bad data detection module. A successful FDIA can cause the system operator to perform control actions that compromise either the physical or economic operation of the power system. In this letter, we consider some implications for FDIAs arising from the late 2015 Ukraine Blackout event.},
  keywords = {Cyber-attacks,false data injection attacks,Ukraine blackout},
  file = {C\:\\Users\\dassc\\Zotero\\storage\\KELQGSH8\\full-text.pdf}
}

@article{liangReviewFalseData2017,
  title = {A {{Review}} of {{False Data Injection Attacks Against Modern Power Systems}}},
  author = {Liang, Gaoqi and Zhao, Junhua and Luo, Fengji and Weller, Steven R. and Dong, Zhao Yang},
  date = {2017-07-01},
  journaltitle = {IEEE Transactions on Smart Grid},
  volume = {8},
  number = {4},
  pages = {1630--1638},
  publisher = {{Institute of Electrical and Electronics Engineers Inc.}},
  issn = {19493053},
  doi = {10.1109/TSG.2015.2495133},
  abstract = {With rapid advances in sensor, computer, and communication networks, modern power systems have become complicated cyber-physical systems. Assessing and enhancing cyber-physical system security is, therefore, of utmost importance for the future electricity grid. In a successful false data injection attack (FDIA), an attacker compromises measurements from grid sensors in such a way that undetected errors are introduced into estimates of state variables such as bus voltage angles and magnitudes. In evading detection by commonly employed residue-based bad data detection tests, FDIAs are capable of severely threatening power system security. Since the first published research on FDIAs in 2009, research into FDIA-based cyber-attacks has been extensive. This paper gives a comprehensive review of state-of-the-art in FDIAs against modern power systems. This paper first summarizes the theoretical basis of FDIAs, and then discusses both the physical and the economic impacts of a successful FDIA. This paper presents the basic defense strategies against FDIAs and discusses some potential future research directions in this field.},
  keywords = {Cyber-physical security,false data injection attacks,power system,state estimation},
  file = {C\:\\Users\\dassc\\Zotero\\storage\\2793JKUD\\Liang et al_2017_A Review of False Data Injection Attacks Against Modern Power Systems.pdf}
}

@article{liLargeScaleTestbedVirtual2020,
  title = {A {{Large-Scale Testbed}} as a {{Virtual Power Grid}}: {{For Closed-Loop Controls}} in {{Research}} and {{Testing}}},
  shorttitle = {A {{Large-Scale Testbed}} as a {{Virtual Power Grid}}},
  author = {Li, Fangxing and Tomsovic, Kevin and Cui, Hantao},
  date = {2020-03},
  journaltitle = {IEEE Power and Energy Magazine},
  shortjournal = {IEEE Power and Energy Mag.},
  volume = {18},
  number = {2},
  pages = {60--68},
  issn = {1540-7977, 1558-4216},
  doi = {10.1109/MPE.2019.2959054},
  url = {https://ieeexplore.ieee.org/document/9007798/},
  urldate = {2022-06-12},
  langid = {english},
  annotation = {9 citations (Crossref) [2022-06-13]},
  file = {C\:\\Users\\dassc\\Zotero\\storage\\784DU5RE\\Li et al. - 2020 - A Large-Scale Testbed as a Virtual Power Grid For.pdf;C\:\\Users\\dassc\\Zotero\\storage\\P7I75GY8\\Li et al_2020_A Large-Scale Testbed as a Virtual Power Grid - For Closed-Loop Controls in.pdf;C\:\\Users\\dassc\\Zotero\\storage\\V9M73G2V\\Li et al_2020_A Large-Scale Testbed as a Virtual Power Grid.pdf}
}

@book{liuIntegratedSimulationAnalyze2015,
  title = {Integrated Simulation to Analyze the Impact of Cyber-Attacks on the Power Grid},
  author = {Liu, Ren and Srivastava, Anurag},
  date = {2015-04-01},
  pages = {6},
  doi = {10.1109/MSCPES.2015.7115395},
  abstract = {With the development of the smart grid technology, Information and Communication Technology (ICT) plays a sig- nificant role in the smart grid. ICT enables to realize the smart grid, but also brings cyber vulnerabilities. It is important to analyze the impact of possible cyber-attacks on the power grid. In this paper, a real-time, cyber-physical co-simulation testbed with hardware-in-the-loop capability is discussed. Real-time Digital Simulator (RTDS), Synchrophasor devices, DeterLab, and a wide- area monitoring application with closed-loop control are utilized in the developed testbed. Two different real life cyber-attacks, including TCP SYN flood attack, and man-in-the-middle attack, are simulated on an IEEE standard power system test case to analyze the the impact of these cyber-attacks on the power grid.},
  pagetotal = {1},
  file = {C\:\\Users\\dassc\\Zotero\\storage\\LEF6S2YG\\Liu et al_2015_Integrated simulation to analyze the impact of cyber-attacks on the power grid.pdf}
}

@article{monticelliElectricPowerSystem2000,
  title = {Electric Power System State Estimation},
  author = {Monticelli, A.},
  date = {2000-02},
  journaltitle = {Proceedings of the IEEE},
  volume = {88},
  number = {2},
  pages = {262--282},
  issn = {1558-2256},
  doi = {10.1109/5.824004},
  abstract = {This paper discusses the state of the art in electric power system state estimation. Within energy management systems, state estimation is a key function for building a network real-time model. A real-time model is a quasi-static mathematical representation of the current conditions in an interconnected power network. This model is extracted at intervals from snapshots of real-time measurements (both analog and status). The new modeling needs associated with the introduction of new control devices and the changes induced by emerging energy markets are making state estimation and its related functions more important than ever.},
  eventtitle = {Proceedings of the {{IEEE}}},
  keywords = {:book,��,Data analysis,Energy management,Load flow,Mathematical model,Network topology,Parameter estimation,Power system modeling,Power system reliability,State estimation,Voltage},
  file = {C\:\\Users\\dassc\\Zotero\\storage\\2B35LDPC\\Monticelli_2000_Electric power system state estimation.pdf;C\:\\Users\\dassc\\Zotero\\storage\\IYNFM9RC\\824004.html}
}

@article{morrisCybersecurityRiskTesting2011,
  title = {Cybersecurity Risk Testing of Substation Phasor Measurement Units and Phasor Data Concentrators},
  author = {Morris, Thomas and Pan, Shengyi and Lewis, Jeremy and Moorhead, Jonathan and Younan, Nicholas and King, Roger and Freund, Mark and Madani, Vahid},
  date = {2011},
  journaltitle = {ACM International Conference Proceeding Series},
  doi = {10.1145/2179298.2179324},
  abstract = {Future bulk electric transmission systems will include substation automation, synchrophasor measurement systems, and automated control algorithms which leverage wide area monitoring system to better control the grid. Prior to installation of new networked devices, utilities should perform cybersecurity testing and develop corrective actions for identified vulnerabilities. This paper outlines testing performed prior to the installation of a synchrophasor wide area monitoring system. Phasor measurement unit and phasor data concentrator devices from multiple vendors were subjected to laboratory testing including; device security feature identification, port scans, network congestion testing, denial of service testing, protocol mutation testing, and network traffic disclosure testing. This paper outlines the procedures used to perform the testing and discusses the types of results expected from testing. Copyright © 2011 ACM.},
  isbn = {9781450309455},
  keywords = {Cybersecurity,Experimentation,Security Keywords Smart Grid,Smart Grid},
  file = {C\:\\Users\\dassc\\Zotero\\storage\\WH3GHBK5\\Morris et al_2011_Cybersecurity risk testing of substation phasor measurement units and phasor.pdf}
}

@article{muslehSurveyDetectionAlgorithms2020,
  title = {A {{Survey}} on the {{Detection Algorithms}} for {{False Data Injection Attacks}} in {{Smart Grids}}},
  author = {Musleh, Ahmed S. and Chen, Guo and Dong, Zhao Yang},
  date = {2020-05-01},
  journaltitle = {IEEE Transactions on Smart Grid},
  volume = {11},
  number = {3},
  pages = {2218--2234},
  publisher = {{Institute of Electrical and Electronics Engineers Inc.}},
  issn = {19493061},
  doi = {10.1109/TSG.2019.2949998},
  abstract = {Cyber-physical attacks are the main substantial threats facing the utilization and development of the various smart grid technologies. Among these attacks, false data injection attack represents a main category with its widely varied types and impacts that have been extensively reported recently. In addressing this threat, several detection algorithms have been developed in the last few years. These were either model-based or data-driven algorithms. This paper provides an intensive summary of these algorithms by categorizing them and elaborating on the pros and cons of each category. The paper starts by introducing the various cyber-physical attacks along with the main reported incidents in history. The significance and the impacts of the false data injection attacks are then reported. The concluding remarks present the main criteria that should be considered in developing future detection algorithms for the false data injection attacks.},
  keywords = {��,Cyber-physical attacks,data-driven detection algorithms,false data injection,machine learning,model-based detection algorithms,smart grid,state estimation,stealth attacks},
  file = {C\:\\Users\\dassc\\Zotero\\storage\\JGFFSFUM\\full-text.pdf}
}

@inproceedings{mynamSynchrophasorsRedefiningSCADA2013,
  title = {Synchrophasors Redefining {{SCADA}} Systems},
  booktitle = {13th {{Annual Western Power Delivery Automation Conference}}, {{March}} 2011},
  author = {Mynam, Mangapathirao V. and Harikrishna, Ala and Singh, Vivek},
  date = {2013},
  volume = {26},
  pages = {22--28},
  publisher = {{International Association on Electricity Generation Transmission \& Distribution}},
  keywords = {},
  file = {C\:\\Users\\dassc\\Zotero\\storage\\XRKGJJY4\\Mynam et al_2013_Synchrophasors redefining SCADA systems.pdf;C\:\\Users\\dassc\\Zotero\\storage\\HG8JGVJ8\\ijor.html}
}


@article{nayak2010,
  title = {Different Flavours of {{Man-In-The-Middle}} Attack, Consequences and Feasible Solutions},
  author = {Nayak, Gopi Nath and Samaddar, Shefalika Ghosh},
  date = {2010},
  journaltitle = {Proceedings - 2010 3rd IEEE International Conference on Computer Science and Information Technology, ICCSIT 2010},
  volume = {5},
  pages = {491--495},
  doi = {10.1109/ICCSIT.2010.5563900},
  abstract = {Man-In-The-Middle (MITM) attack is one of the primary techniques employed in computer based hacking. MITM attack can successfully invoke attacks such as Denial of service (DoS), DNS spoofing and Port stealing. MITM attack is particularly suitable in a LAN environment, Where it is typically performed through ARP poisoning. MITM attack of every kind has lot of surprising consequences in store for users such as, stealing online account userid, password, stealing of local ftp id, ssh or telnet session etc. This paper emphasizes on different types of MITM attacks, their consequences and feasible solutions under different circumstances giving users options to choose one from various solutions. ARP spoofing and its effect in a LAN environment is studied in detail to achieve the stated objective. © 2010 IEEE.},
  isbn = {9781424455386},
  keywords = {Attacker,DNS,DoS,Feasible solution,Gateway,MITM,Packet,Victim,Vulnerability},
  file = {C\:\\Users\\dassc\\Zotero\\storage\\MEHU4QP7\\Nayak and Samaddar - 2010 - Different flavours of Man-In-The-Middle attack, co.pdf}
}



@article{palRealtimeDetectionPacket2014,
  title = {Real-Time Detection of Packet Drop Attacks on Synchrophasor Data},
  author = {Pal, S and Sikdar, B and Conference, J Chow - 2014 IEEE International and 2014, undefined},
  date = {2014},
  journaltitle = {ieeexplore.ieee.org},
  url = {https://ieeexplore.ieee.org/abstract/document/7007762/},
  urldate = {2022-03-01},
  abstract = {The importance of phasor measurement unit (PMU) or synchrophasor data towards the functioning of real-time monitoring and control of power generation and distribution systems makes them an attractive target for cyber-attacks. An attack with potential for significant damage is the packet drop attack, where the adversary arbitrarily drops packets with synchrophasor data. This paper develops a real-time mechanism for detecting packet drop attacks on synchrophasor data carried over the Internet. The proposed solution is receiver-based, and uses the one-way packet delays to extract features that are used to detect attacks. The proposed attack detection mechanism leads to lower detection delays and greater accuracy as compared to existing mechanisms.},
  file = {C\:\\Users\\dassc\\Zotero\\storage\\L47EFFTG\\full-text.pdf}
}

@article{paudelDataIntegrityAttacks2016,
  title = {Data Integrity Attacks in Smart Grid Wide Area Monitoring},
  author = {Paudel, S and Smith, P and …, T Zseby - for ICS \& SCADA Cyber Security and 2016, undefined},
  date = {2016},
  journaltitle = {scienceopen.com},
  publisher = {{BCS Learning \& Development}},
  doi = {10.14236/ewic/ICS2016.9},
  url = {https://www.scienceopen.com/hosted-document?doi=10.14236/ewic/ICS2016.9},
  urldate = {2022-03-01},
  abstract = {A smart grid requires the implementation of ICT technologies in order to incorporate new functions into electricity grid monitoring and control. Wide Area Monitoring Systems (WAMSs) are used to measure synchrophasor data at different locations and give operators a near-real-time picture of what is happening in the system. The measurement data is periodically collected via communication channels to monitor, predict and control the power consumption, and detect any problems in the power grid. Attacks on WAMSs can trigger wrong decisions and create dangerous failures in the smart grid system. In this paper, we investigate data integrity attacks at different attack entry points of a WAMS, their impacts on the smart grid system, and existing mitigation strategies. We conclude from our study that the existing techniques, methodologies and mechanisms are not effective enough to detect or mitigate some attacks.},
  keywords = {Cybersecurity,Data Integrity Attacks,Failure Scenarios,Wide Area Monitoring System},
  file = {C\:\\Users\\dassc\\Zotero\\storage\\CMG3W6V2\\full-text.pdf}
}

@article{pinillaModelingDistributedGenerators2016,
  title = {Modeling of Distributed Generators in 13 Nodes {{IEEE}} Test Feeder},
  author = {Pinilla, Henry Giovanni and Aristizábal, Andrés Julián},
  date = {2016},
  publisher = {{Universidad Jorge Tadeo Lozano}},
  file = {C\:\\Users\\dassc\\Zotero\\storage\\XGRFXXZV\\Pinilla et al_2016_Modeling of distributed generators in 13 nodes IEEE test feeder.pdf;C\:\\Users\\dassc\\Zotero\\storage\\K79BX4U9\\9365.html}
}

@article{prabhuStateoftheartReviewSynchrophasor2017,
  title = {A State-of-the-Art Review on Synchrophasor Applications to Power Network Protection},
  author = {Prabhu, M. S. and Nayak, Paresh Kumar},
  date = {2017},
  journaltitle = {Lecture Notes in Electrical Engineering},
  volume = {436},
  pages = {531--541},
  publisher = {{Springer Verlag}},
  issn = {18761119},
  doi = {10.1007/978-981-10-4394-9_52},
  abstract = {The demand for electricity supply has been increased many folds over the last few decades. However, the growth in the electric infrastructure has not been increased accordingly due to deregulation of the energy markets, economic and environmental reasons. In present days, power networks are most often operated closer to their stability limit to fulfill the growing electricity demand. As a result, the security and safety of the power system today is at risk. Investigation on large blackouts in the recent past show that maintaining system reliability and integrity becomes more and more difficult due to reduced transmission capacity margins and increased stress on the system. Under the stressed operating condition, the widely-used distance relaying based transmission line protection schemes are susceptible to maloperation. The use of series-compensated and multiterminal lines is another concern for the distance protection scheme. At the same time, the present advancements in the wide-area measurement systems (WAMS) using synchrophasors has shown potential for ensuring improved protection for different power networks operating even at critical conditions. In this paper, the authors first investigate the limitations of existing distance relays while protecting different power networks during stressed operating conditions. Then, an extensive review is made on the application of synchrophasor based WAMS technology for reliable power system protection. The objective of the present study is mainly to bring the attention of the researchers from academic institutions, industries and utility grid on the possible applications of synchrophasors based WAMS technology for ensuring improved protection to today’s power system.},
  keywords = {Distance relay,Multiterminal line,Series compensation,Synchrophasor,WAMS,Wide-area backup protection},
  file = {C\:\\Users\\dassc\\Zotero\\storage\\LX96ZWF4\\Prabhu et al_2017_A state-of-the-art review on synchrophasor applications to power network.pdf}
}

@online{RecentUpdatesGoodreads,
  title = {Recent Updates | {{Goodreads}}},
  url = {https://www.goodreads.com/},
  urldate = {2022-07-07},
  file = {C\:\\Users\\dassc\\Zotero\\storage\\HZWCCY7Z\\www.goodreads.com.html}
}

@report{rodofileRealTimeInteractiveAttacks2015,
  title = {Real-{{Time}} and {{Interactive Attacks}} on {{DNP3 Critical Infrastructure Using Scapy}}},
  author = {Rodofile, Nicholas R and Radke, Kenneth and Foo, Ernest},
  date = {2015},
  abstract = {The Distributed Network Protocol v3.0 (DNP3) is one of the most widely used protocols, to control national infrastructure. Widely used interactive packet manipulation tools, such as Scapy, have not yet been augmented to parse and create DNP3 frames (Biondi 2014). In this paper we extend Scapy to include DNP3, thus allowing us to perform attacks on DNP3 in real-time. Our contribution builds on East et al. (2009), who proposed a range of possible attacks on DNP3. We implement several of these attacks to validate our DNP3 extension to Scapy, then executed the attacks on real world equipment. We present our results , showing that many of these theoretical attacks would be unsuccessful in an Ethernet-based network.},
  keywords = {critical Infrastructure security,Distributed Network Proto-col 30,DNP3,Scapy,substations},
  file = {C\:\\Users\\dassc\\Zotero\\storage\\PNBQPY63\\DNP3 Implementation in Scapy.pdf}
}

@article{sahaModelingSimulationXendee2016,
  title = {Modeling \& Simulation in Xendee:                                     Ieee 13 Node Test Feeder},
  author = {Saha, Shammya and Johnson, Nathan},
  date = {2016},
  pages = {13},
  langid = {english},
  file = {C\:\\Users\\dassc\\Zotero\\storage\\B6PBUYZE\\2016 - modeling & simulation in xendee                  .pdf}
}

@article{sandiPypmuOpenSource2016,
  title = {Pypmu—Open Source Python Package for Synchrophasor Data Transfer},
  author = {Šandi, S and Krstajić, B and Telecommunications, T Popović - 2016 24th and 2016, undefined},
  date = {2016},
  journaltitle = {ieeexplore.ieee.org},
  url = {https://ieeexplore.ieee.org/abstract/document/7818916/},
  urldate = {2022-02-23},
  file = {C\:\\Users\\dassc\\Zotero\\storage\\J6QI9N2Y\\Šandi et al_2016_pypmu—open source python package for synchrophasor data transfer.pdf}
}

@article{sandiPYTHONIMPLEMENTATIONIEEE,
  title = {{{PYTHON IMPLEMENTATION OF IEEE C37}}.118 {{COMMUNICATION PROTOCOL}}},
  author = {Šandi, Stevan and Popovi, Tomo},
  volume = {21},
  number = {1},
  pages = {10},
  abstract = {The analysis of tools for supporting the measurement of synchrophasors revealed the need for open, customizable, and platform independent tools. One such software tool is the open implementation of the IEEE C37.118 communication protocol for data transfer. This paper describes an implementation process of the protocol in a form of a Python library module. The discussion illustrates its use and validation using third-party tools. Finally, the paper outlines future work on the improvements and possible practical applications.},
  langid = {english},
  file = {C\:\\Users\\dassc\\Zotero\\storage\\WMA6NCK3\\Šandi and Popovi - PYTHON IMPLEMENTATION OF IEEE C37.118 COMMUNICATIO.pdf}
}

@inproceedings{saraswatAnalyzingEffectsCyberattacks2021,
  title = {Analyzing the Effects of Cyberattacks on Distribution System State Estimation},
  booktitle = {2021 {{IEEE Power Energy Society Innovative Smart Grid Technologies Conference}} ({{ISGT}})},
  author = {Saraswat, Govind and Yang, Rui and Liu, Yajing and Zhang, Yingchen},
  date = {2021-02},
  pages = {01--05},
  issn = {2472-8152},
  doi = {10.1109/ISGT49243.2021.9372262},
  abstract = {Key components of power systems-such as energy management systems, automatic generation control, and state estimation-are under serious vulnerability from cyberattacks. Cyber threats in electric grids have increased significantly because of the increased interconnectivity of supervisory control and data acquisition systems and public network infrastructure. As the penetration level of distributed energy resources increases, it is imperative to employ system-monitoring techniques such as state estimation for the reliable operation of distribution systems. Recently, multiple methods have been developed that exploit the low rank property of distribution system state matrix and are robust to bad data, such as matrix completion. This paper analyzes the impact of various realistic cyberattack scenarios on matrix completion. Realistic cyberattack scenarios are converted into data corruption models that are used in an extensive simulation of a custom IEEE 123-bus system.},
  eventtitle = {2021 {{IEEE Power Energy Society Innovative Smart Grid Technologies Conference}} ({{ISGT}})},
  keywords = {��,Computer crime,Cyberattacks,Data models,distribution systems,electric grid,Matrix converters,Reliability,SCADA systems,security,Smart grids,state estimation,State estimation},
  annotation = {1 citations (Crossref) [2022-05-30]},
  file = {C\:\\Users\\dassc\\Zotero\\storage\\49AB5AWC\\Saraswat et al_2021_Analyzing the effects of cyberattacks on distribution system state estimation.pdf;C\:\\Users\\dassc\\Zotero\\storage\\9J5SV8DB\\Saraswat et al_2021_Analyzing the effects of cyberattacks on distribution system state estimation.pdf;C\:\\Users\\dassc\\Zotero\\storage\\PFGL4FZI\\9372262.html}
}

@unpublished{saygheSurveyMachineLearning2020,
  title = {A {{Survey}} of {{Machine Learning Methods}} for {{Detecting False Data Injection Attacks}} in {{Power Systems}}},
  author = {Sayghe, Ali and Hu, Yaodan and Zografopoulos, Ioannis and Liu, XiaoRui and Dutta, Raj Gautam and Jin, Yier and Konstantinou, Charalambos},
  date = {2020-08-16},
  number = {arXiv:2008.06926},
  eprint = {2008.06926},
  eprinttype = {arxiv},
  primaryclass = {cs, eess},
  publisher = {{arXiv}},
  doi = {10.48550/arXiv.2008.06926},
  url = {http://arxiv.org/abs/2008.06926},
  urldate = {2022-05-30},
  abstract = {Over the last decade, the number of cyberattacks targeting power systems and causing physical and economic damages has increased rapidly. Among them, False Data Injection Attacks (FDIAs) is a class of cyberattacks against power grid monitoring systems. Adversaries can successfully perform FDIAs in order to manipulate the power system State Estimation (SE) by compromising sensors or modifying system data. SE is an essential process performed by the Energy Management System (EMS) towards estimating unknown state variables based on system redundant measurements and network topology. SE routines include Bad Data Detection (BDD) algorithms to eliminate errors from the acquired measurements, e.g., in case of sensor failures. FDIAs can bypass BDD modules to inject malicious data vectors into a subset of measurements without being detected, and thus manipulate the results of the SE process. In order to overcome the limitations of traditional residual-based BDD approaches, data-driven solutions based on machine learning algorithms have been widely adopted for detecting malicious manipulation of sensor data due to their fast execution times and accurate results. This paper provides a comprehensive review of the most up-to-date machine learning methods for detecting FDIAs against power system SE algorithms.},
  archiveprefix = {arXiv},
  keywords = {Computer Science - Cryptography and Security,Electrical Engineering and Systems Science - Systems and Control},
  file = {C\:\\Users\\dassc\\Zotero\\storage\\HHBCDV2M\\iet-stg.2020.0015.pdf;C\:\\Users\\dassc\\Zotero\\storage\\JDPUYMD2\\Sayghe et al_2020_A Survey of Machine Learning Methods for Detecting False Data Injection Attacks.pdf;C\:\\Users\\dassc\\Zotero\\storage\\SUYSJ5IJ\\2008.html}
}

@article{singhStealthyCyberAttacks2016,
  title = {Stealthy Cyber Attacks and Impact Analysis on Wide-Area Protection of Smart Grid},
  author = {Singh, VK and …, A Ozen - 2016 North American and 2016, undefined},
  date = {2016},
  journaltitle = {ieeexplore.ieee.org},
  doi = {10.1109/NAPS.2016.7747927},
  url = {https://ieeexplore.ieee.org/abstract/document/7747927/},
  urldate = {2022-03-01},
  abstract = {Smart grid is vulnerable to many cyber attacks due to legacy nature of the infrastructure coupled with increase in control and monitoring functions through cyber advancements. Remedial Action Scheme (RAS), widely used for wide area protection, provides autonomous operations through the RAS controller. Due to its dependence on the timely cooperation of multiple devices and communication network, it is highly vulnerable to cyber attacks. In this paper, we present an impact analysis for the power system due to a class of malware-based coordinated cyber attacks targeting the RAS scheme. Specifically, we make the following two contributions. First, modeling a stealth attack vector based on malware and coordinated attack behavior. In particular, installing the malware (Trojan horse) in the RAS controller which turns the controller into an attacker's bot. Then, performing a coordinated attack which involves malicious tripping of one of the parallel lines connected to a generator followed by the continuous pulse attack on the generator. The pulse attack includes periodically changing the generations through RAS controller which remains undetected by the control center. Second, testbed-based implementation and evaluation to quantify system impacts. We have leveraged Iowa State's PowerCyber CPS security testbed for experimental evaluation. In our evaluation, we varied the duty cycle of the pulse attack to obtain different attack scenarios and consequent impacts are analyzed on modified IEEE 9-bus system in real-time simulation. Our studies show that the duty cycle of the pulse attack is a critical factor in determining the severity of the attack impacts on system stability.},
  isbn = {9781509032709},
  file = {C\:\\Users\\dassc\\Zotero\\storage\\HN8NYYJN\\full-text.pdf}
}

@thesis{taraliBadDataDetection2012,
  title = {Bad Data Detection in Two Stage Estimation Using Phasor Measurements},
  author = {Tarali, Aditya},
  date = {2012},
  institution = {{Northeastern University}},
  doi = {10.17760/d20002926},
  url = {http://hdl.handle.net/2047/d20002926},
  urldate = {2022-06-15},
  langid = {english},
  file = {C\:\\Users\\dassc\\Zotero\\storage\\KPKT33WM\\Tarali - 2012 - Bad data detection in two stage estimation using p.pdf}
}

@inproceedings{vasilomanolakisDidYouReally2015a,
  title = {Did You Really Hack a Nuclear Power Plant? {{An}} Industrial Control Mobile Honeypot},
  shorttitle = {Did You Really Hack a Nuclear Power Plant?},
  booktitle = {2015 {{IEEE Conference}} on {{Communications}} and {{Network Security}} ({{CNS}})},
  author = {Vasilomanolakis, Emmanouil and Srinivasa, Shreyas and Mühlhäuser, Max},
  date = {2015-09},
  pages = {729--730},
  doi = {10.1109/CNS.2015.7346907},
  abstract = {The emerge of sophisticated attackers and malware that target Industrial Control System (ICS) suggests that novel security mechanisms are required. Honeypots, can act as an additional line of defense, by providing early warnings for such attacks. We present a mobile ICS honeypot, that can be placed in various network positions to provide security administrators an on-the-go security status of their network. We discuss our system, its merits in comparison to other honeypots, and provide preliminary results towards a large-scale evaluation.},
  eventtitle = {2015 {{IEEE Conference}} on {{Communications}} and {{Network Security}} ({{CNS}})},
  keywords = {��,Industrial control,industrial control systems,Internet,Malware,Mobile communication,mobile honeypot,Protocols,Servers},
  file = {C\:\\Users\\dassc\\Zotero\\storage\\KLK7FCA5\\Vasilomanolakis et al_2015_Did you really hack a nuclear power plant.pdf;C\:\\Users\\dassc\\Zotero\\storage\\W633V8X6\\7346907.html}
}

@online{virtualboxChapterVirtualNetworking,
  title = {Chapter~6.~{{Virtual Networking}}},
  author = {VirtualBox},
  url = {https://www.virtualbox.org/manual/ch06.html},
  urldate = {2022-07-07},
  file = {C\:\\Users\\dassc\\Zotero\\storage\\CESW7EG3\\ch06.html}
}

@article{wangRevisedBranchCurrentbased2004,
  title = {A Revised Branch Current-Based Distribution System State Estimation Algorithm and Meter Placement Impact},
  author = {Wang, Haibin and Schulz, N.N.},
  date = {2004-02},
  journaltitle = {IEEE Transactions on Power Systems},
  volume = {19},
  number = {1},
  pages = {207--213},
  issn = {1558-0679},
  doi = {10.1109/TPWRS.2003.821426},
  abstract = {With the development of automation in distribution systems, distribution supervisory control and data acquisition (SCADA) and many automated meter reading (AMR) systems have been installed on distribution systems. Also distribution management system (DMS) have advanced and include more sophisticated analysis tools. The combination of these developments is providing a platform for development of distribution system state estimation (DSE). A branch-current-based three-phase state estimation algorithm for distribution systems has been developed and tested. This method chooses the magnitude and phase angle of the branch current as the state variables. Because of the limited number of real-time measurements in the distribution system, the state estimator can not acquire enough real-time measurements for convergence, so pseudo-measurements are necessary for distribution system state estimator. The load estimated at every node from the AMR systems is used as a pseudo-measurement for the state estimator. The algorithm has been tested on three IEEE radial test feeders. In addition to this new strategy for DSE, another issue is meter-placement. This topic includes the type of measurement as well as the location of the measurement. Our results show the impact of these two issues on accuracy. Several general meter rules based on this analysis are outlined.},
  eventtitle = {{{IEEE Transactions}} on {{Power Systems}}},
  keywords = {Automatic meter reading,Instruments,Jacobian matrices,Loss measurement,Meter reading,Real time systems,SCADA systems,State estimation,Testing,Voltage},
  annotation = {210 citations (Crossref) [2022-06-02]},
  file = {C\:\\Users\\dassc\\Zotero\\storage\\9MXQI3SB\\Wang_Schulz_2004_A revised branch current-based distribution system state estimation algorithm.pdf;C\:\\Users\\dassc\\Zotero\\storage\\BZMD9RD8\\Wang and Schulz - 2004 - A revised branch current-based distribution system.pdf;C\:\\Users\\dassc\\Zotero\\storage\\M5I325SU\\Wang et al_2004_A revised branch current-based distribution system state estimation algorithm.pdf;C\:\\Users\\dassc\\Zotero\\storage\\UP56TZZI\\1266570.html}
}

@unpublished{wlazloManinTheMiddleAttacksDefense2021,
  title = {Man-in-{{The-Middle Attacks}} and {{Defense}} in a {{Power System Cyber-Physical Testbed}}},
  author = {Wlazlo, Patrick and Sahu, Abhijeet and Mao, Zeyu and Huang, Hao and Goulart, Ana and Davis, Katherine and Zonouz, Saman},
  date = {2021-02-22},
  eprint = {2102.11455},
  eprinttype = {arxiv},
  doi = {10.1049/cps2.12014},
  url = {http://arxiv.org/abs/2102.11455},
  abstract = {Man-in-The-Middle (MiTM) attacks present numerous threats to a smart grid. In a MiTM attack, an intruder embeds itself within a conversation between two devices to either eavesdrop or impersonate one of the devices, making it appear to be a normal exchange of information. Thus, the intruder can perform false data injection (FDI) and false command injection (FCI) attacks that can compromise power system operations, such as state estimation, economic dispatch, and automatic generation control (AGC). Very few researchers have focused on MiTM methods that are difficult to detect within a smart grid. To address this, we are designing and implementing multi-stage MiTM intrusions in an emulation-based cyber-physical power system testbed against a large-scale synthetic grid model to demonstrate how such attacks can cause physical contingencies such as misguided operation and false measurements. MiTM intrusions create FCI, FDI, and replay attacks in this synthetic power grid. This work enables stakeholders to defend against these stealthy attacks, and we present detection mechanisms that are developed using multiple alerts from intrusion detection systems and network monitoring tools. Our contribution will enable other smart grid security researchers and industry to develop further detection mechanisms for inconspicuous MiTM attacks.},
  archiveprefix = {arXiv},
  file = {C\:\\Users\\dassc\\Zotero\\storage\\KN8YK326\\reference_paper_1.pdf}
}

@article{wuPowerSystemState1990,
  title = {Power System State Estimation: A Survey},
  shorttitle = {Power System State Estimation},
  author = {Wu, Felix F.},
  date = {1990-04-01},
  journaltitle = {International Journal of Electrical Power \& Energy Systems},
  shortjournal = {International Journal of Electrical Power \& Energy Systems},
  volume = {12},
  number = {2},
  pages = {80--87},
  issn = {0142-0615},
  doi = {10.1016/0142-0615(90)90003-T},
  url = {https://www.sciencedirect.com/science/article/pii/014206159090003T},
  urldate = {2022-05-28},
  abstract = {Recent developments in the solution methods for state estimation are reviewed. Concepts of decoupling, ill-conditioning and robustness in state estimation are discussed. Derivations of decoupled estimators, stable estimators and robust estimators are reviwed. Future directions for research are suggested.},
  langid = {english},
  keywords = {security monitoring,state estimation,weighted least square estimation},
  file = {C\:\\Users\\dassc\\Zotero\\storage\\8KVNQPQZ\\Wu_1990_Power system state estimation.pdf;C\:\\Users\\dassc\\Zotero\\storage\\X84IS4JQ\\014206159090003T.html}
}

@article{yangCybersecurityTestbedIEC2015,
  title = {Cybersecurity Test-Bed for {{IEC}} 61850 Based Smart Substations},
  author = {Yang, Y and Jiang, HT and McLaughlin, K and …, L Gao - 2015 IEEE Power \& and 2015, undefined},
  date = {2015},
  journaltitle = {ieeexplore.ieee.org},
  pages = {1--5},
  publisher = {{IEEE}},
  doi = {10.1109/PESGM.2015.7286357},
  url = {https://ieeexplore.ieee.org/abstract/document/7286357/},
  urldate = {2022-03-01},
  abstract = {With the development and deployment of IEC 61850 based smart substations, cybersecurity vulnerabilities of supervisory control and data acquisition (SCADA) systems are increasingly emerging. In response to the emergence of cybersecurity vulnerabilities in smart substations, a test-bed is indispensable to enable cybersecurity experimentation. In this paper, a comprehensive and realistic cyber-physical test-bed has been built to investigate potential cybersecurity vulnerabilities and the impact of cyber-attacks on IEC 61850 based smart substations. This test-bed is close to a real production type environment, and has the ability to carry out end-to-end testing of cyber-attacks and physical consequences. A fuzz testing approach is proposed for detecting IEC 61850 based intelligent electronic devices (IEDs) and validated in the proposed test-bed.},
  keywords = {Cybersecurity,Fuzz testing,IEC 61850,Index Terms-Smart Substation,Test-bed},
  file = {C\:\\Users\\dassc\\Zotero\\storage\\FBDX2ZMW\\Yang et al_2015_Cybersecurity test-bed for IEC 61850 based smart substations.pdf}
}

@article{yangManinthemiddleAttackTestbed2012,
  title = {Man-in-the-Middle Attack Test-Bed Investigating Cyber-Security Vulnerabilities in Smart Grid {{SCADA}} Systems},
  author = {Yang, Y. and McLaughlin, K. and Littler, T. and Sezer, S. and Im, Eul Gyu and Yao, Z. Q. and Pranggono, B. and Wang, H. F.},
  date = {2012},
  journaltitle = {IET Conference Publications},
  volume = {2012},
  doi = {10.1049/CP.2012.1831},
  abstract = {The increased complexity and interconnectivity of Supervisory Control and Data Acquisition (SCADA) systems in the Smart Grid has exposed them to a wide range of cyber-security issues, and there are a multitude of potential access points for cyber attackers. This paper presents a SCADA-specific cyber-security test-bed which contains SCADA software and communication infrastructure. This test-bed is used to investigate an Address Resolution Protocol (ARP) spoofing based man-in-the-middle attack. Finally, the paper proposes a future work plan which focuses on applying intrusion detection and prevention technology to address cyber-security issues in SCADA systems.},
  isbn = {9781849196734},
  issue = {611 CP},
  keywords = {Cyber-security,Man-in-the-middle attack,SCADA,Smart Grid}
}

@article{zhangDesignTestingImplementation2017a,
  title = {Design, {{Testing}}, and {{Implementation}} of a {{Linear State Estimator}} in a {{Real Power System}}},
  author = {Zhang, Lin and Bose, Anjan and Jampala, Anil and Madani, Vahid and Giri, Jay},
  date = {2017-07},
  journaltitle = {IEEE Transactions on Smart Grid},
  shortjournal = {IEEE Trans. Smart Grid},
  volume = {8},
  number = {4},
  pages = {1782--1789},
  issn = {1949-3053, 1949-3061},
  doi = {10.1109/TSG.2015.2508283},
  url = {http://ieeexplore.ieee.org/document/7373664/},
  urldate = {2022-06-14},
  abstract = {As more Phasor Measurement Units (PMUs) are installed, portions of the power grid become observable with just phasor measurements, making feasible the estimation of the state of these observable portions at much faster rates than the traditional state estimator. Although such a Linear State Estimator (LSE) was proposed over a decade ago, the first field implementation was completed at Dominion Virginia Power in 2013. Although this LSE is a stand-alone function not integrated to their Energy Management System (EMS), it demonstrated the feasibility of the LSE. In this paper, we present the design, development and implementation of an LSE that is fully integrated with the existing EMS and can estimate the state of the Extra High Voltage (EHV) portion of a power system at 30 times per second. Integration of the LSE to the existing EMS environment and some of the issues in the design and testing are presented in this paper. The work paved way for LSE to supply cleansed PMU data to other synchrophasor applications that are sensitive to data quality.},
  langid = {english},
  annotation = {41 citations (Crossref) [2022-06-14]},
  file = {C\:\\Users\\dassc\\Zotero\\storage\\9X9YVX8X\\Zhang et al. - 2017 - Design, Testing, and Implementation of a Linear St.pdf;C\:\\Users\\dassc\\Zotero\\storage\\LK95XZLG\\Zhang et al_2017_Design, Testing, and Implementation of a Linear State Estimator in a Real Power.pdf}
}



@article{ieeepowerenergysocietEEEStdC372011,
	title = {{EEE} {Std} {C37}.118.2™-2011},
	doi = {10.1109/IEEESTD.2011.6111222},
	abstract = {A method for real-time exchange of synchronized phasor measurement data between power system equipment is defined. This standard specifies messaging that can be used with any suitable communication protocol for real-time communication between phasor measurement units (PMU), phasor data concentrators (PDC), and other applications. It defines message types, contents, and use. Data types and formats are specified. A typical measurement system is described. Communication options and requirements are described in annexes. Keywords: data concentrator, DC, IEEE C37.118.2, PDC, phasor, phasor data concentrator, phasor measurement unit, PMU, synchronized phasor, synchrophasor.},
	journal = {IEEE Std C37.118.2-2011 (Revision of IEEE Std C37.118-2005)},
	author = {IEEE Power \& Energy Societ},
	month = dec,
	year = {2011},
	note = {Conference Name: IEEE Std C37.118.2-2011 (Revision of IEEE Std C37.118-2005)},
	keywords = {Power system dynamics, data concentrator, Data processing, Data storage, DC, IEEE C37.118.2, IEEE standards, PDC, phasor, phasor data concentrator, phasor measurement unit, Phasor measurement units, PMU, Synchronization, synchronized phasor, synchrophasor},
	pages = {1--53},
	file = {2011_IEEE Standard for Synchrophasor Data Transfer for Power Systems.pdf:C\:\\Users\\dassc\\Zotero\\storage\\KBWZG3XA\\2011_IEEE Standard for Synchrophasor Data Transfer for Power Systems.pdf:application/pdf;IEEE Xplore Abstract Record:C\:\\Users\\dassc\\Zotero\\storage\\HWIY58QF\\6111222.html:text/html},
}

@book{liuIntegratedSimulationAnalyze2015,
	title = {Integrated simulation to analyze the impact of cyber-attacks on the power grid},
	abstract = {With the development of the smart grid technology, Information and Communication Technology (ICT) plays a sig- nificant role in the smart grid. ICT enables to realize the smart grid, but also brings cyber vulnerabilities. It is important to analyze the impact of possible cyber-attacks on the power grid. In this paper, a real-time, cyber-physical co-simulation testbed with hardware-in-the-loop capability is discussed. Real-time Digital Simulator (RTDS), Synchrophasor devices, DeterLab, and a wide- area monitoring application with closed-loop control are utilized in the developed testbed. Two different real life cyber-attacks, including TCP SYN flood attack, and man-in-the-middle attack, are simulated on an IEEE standard power system test case to analyze the the impact of these cyber-attacks on the power grid.},
	author = {Liu, Ren and Srivastava, Anurag},
	month = apr,
	year = {2015},
	doi = {10.1109/MSCPES.2015.7115395},
	note = {Pages: 6},
	annote = {Annotations(5/17/2022, 2:22:10 PM)
“In this paper, a real-time, cyber-physical co-simulation testbed with hardware-in-the-loop capability is discussed.” (Liu and Srivastava, 2015, p. 1)“TCP SYN flood attack, and man-in-the-middle attack, are simulated” (Liu and Srivastava, 2015, p. 1)“According to IEEE definition of smart grid, the main difference between smart grid and conventional electrical power grid is the increased use of communication and information technology” (Liu and Srivastava, 2015, p. 1) \#smart-grid“There are number of efforts by other researchers to develop cyberphysical testbed to analyze the interdependencies of different domains within the smart grid. Each of these testbeds has their own unique advantages and limitations” (Liu and Srivastava, 2015, p. 1) \#PMU\_TEST\_BED\_PAPER“Comparing with these testbed, the cyber-physical co-simulation testbed presented in this paper has hardware-in-the-loop capability, ability of realtime simulation, ease of cyber-attack modeling, and end-toend system modeling. With the real industry power system hardwares involved into the cyber-physical testbed, there is a opportunity to test whether the attacker can manipulate those devices and which function can be controlled by the attacker.” (Liu and Srivastava, 2015, p. 1)“(i) Power system simulation and sensors (ii) Communication network (iii) Smart grid application.” (Liu and Srivastava, 2015, p. 1)},
	file = {Liu et al_2015_Integrated simulation to analyze the impact of cyber-attacks on the power grid.pdf:C\:\\Users\\dassc\\Zotero\\storage\\LEF6S2YG\\Liu et al_2015_Integrated simulation to analyze the impact of cyber-attacks on the power grid.pdf:application/pdf},
}

@article{wlazloManinTheMiddleAttacksDefense2021,
	title = {Man-in-{The}-{Middle} {Attacks} and {Defense} in a {Power} {System} {Cyber}-{Physical} {Testbed}},
	url = {http://arxiv.org/abs/2102.11455},
	doi = {https://doi.org/10.1049/cps2.12014},
	abstract = {Man-in-The-Middle (MiTM) attacks present numerous threats to a smart grid. In a MiTM attack, an intruder embeds itself within a conversation between two devices to either eavesdrop or impersonate one of the devices, making it appear to be a normal exchange of information. Thus, the intruder can perform false data injection (FDI) and false command injection (FCI) attacks that can compromise power system operations, such as state estimation, economic dispatch, and automatic generation control (AGC). Very few researchers have focused on MiTM methods that are difficult to detect within a smart grid. To address this, we are designing and implementing multi-stage MiTM intrusions in an emulation-based cyber-physical power system testbed against a large-scale synthetic grid model to demonstrate how such attacks can cause physical contingencies such as misguided operation and false measurements. MiTM intrusions create FCI, FDI, and replay attacks in this synthetic power grid. This work enables stakeholders to defend against these stealthy attacks, and we present detection mechanisms that are developed using multiple alerts from intrusion detection systems and network monitoring tools. Our contribution will enable other smart grid security researchers and industry to develop further detection mechanisms for inconspicuous MiTM attacks.},
	author = {Wlazlo, Patrick and Sahu, Abhijeet and Mao, Zeyu and Huang, Hao and Goulart, Ana and Davis, Katherine and Zonouz, Saman},
	month = feb,
	year = {2021},
	note = {arXiv: 2102.11455},
	file = {Wlazlo et al_2021_Man-in-The-Middle Attacks and Defense in a Power System Cyber-Physical Testbed.pdf:C\:\\Users\\dassc\\Zotero\\storage\\KN8YK326\\reference_paper_1.pdf:application/pdf},
}

@article{davisCyberphysicalModelingAssessment2015,
	title = {A cyber-physical modeling and assessment framework for power grid infrastructures},
	url = {https://ieeexplore.ieee.org/abstract/document/7103368/},
	urldate = {2022-01-12},
	journal = {ieeexplore.ieee.org},
	author = {Davis, KR and Davis, CM and grid, SA Zonouz - … on smart and 2015, undefined},
	year = {2015},
	file = {Davis et al_A cyber-physical modeling and assessment framework for power grid.pdf:C\:\\Users\\dassc\\Zotero\\storage\\G3PISGVI\\full-text.pdf:application/pdf},
}

@article{nayakDifferentFlavoursManInTheMiddle2010,
	title = {Different flavours of {Man}-{In}-{The}-{Middle} attack, consequences and feasible solutions},
	volume = {5},
	doi = {10.1109/ICCSIT.2010.5563900},
	abstract = {Man-In-The-Middle (MITM) attack is one of the primary techniques employed in computer based hacking. MITM attack can successfully invoke attacks such as Denial of service (DoS), DNS spoofing and Port stealing. MITM attack is particularly suitable in a LAN environment, Where it is typically performed through ARP poisoning. MITM attack of every kind has lot of surprising consequences in store for users such as, stealing online account userid, password, stealing of local ftp id, ssh or telnet session etc. This paper emphasizes on different types of MITM attacks, their consequences and feasible solutions under different circumstances giving users options to choose one from various solutions. ARP spoofing and its effect in a LAN environment is studied in detail to achieve the stated objective. © 2010 IEEE.},
	urldate = {2022-01-12},
	journal = {Proceedings - 2010 3rd IEEE International Conference on Computer Science and Information Technology, ICCSIT 2010},
	author = {Nayak, Gopi Nath and Samaddar, Shefalika Ghosh},
	year = {2010},
	note = {ISBN: 9781424455386},
	keywords = {Attacker, DNS, DoS, Feasible solution, Gateway, MITM, Packet, Victim, Vulnerability},
	pages = {491--495},
	file = {Nayak et al_2010_Different flavours of Man-In-The-Middle attack, consequences and feasible.pdf:C\:\\Users\\dassc\\Zotero\\storage\\MEHU4QP7\\Nayak and Samaddar - 2010 - Different flavours of Man-In-The-Middle attack, co.pdf:application/pdf},
}

@article{yangManinthemiddleAttackTestbed2012,
	title = {Man-in-the-middle attack test-bed investigating cyber-security vulnerabilities in smart grid {SCADA} systems},
	volume = {2012},
	doi = {10.1049/CP.2012.1831},
	abstract = {The increased complexity and interconnectivity of Supervisory Control and Data Acquisition (SCADA) systems in the Smart Grid has exposed them to a wide range of cyber-security issues, and there are a multitude of potential access points for cyber attackers. This paper presents a SCADA-specific cyber-security test-bed which contains SCADA software and communication infrastructure. This test-bed is used to investigate an Address Resolution Protocol (ARP) spoofing based man-in-the-middle attack. Finally, the paper proposes a future work plan which focuses on applying intrusion detection and prevention technology to address cyber-security issues in SCADA systems.},
	number = {611 CP},
	urldate = {2022-01-12},
	journal = {IET Conference Publications},
	author = {Yang, Y. and McLaughlin, K. and Littler, T. and Sezer, S. and Im, Eul Gyu and Yao, Z. Q. and Pranggono, B. and Wang, H. F.},
	year = {2012},
	note = {ISBN: 9781849196734},
	keywords = {Cyber-security, Man-in-the-middle attack, SCADA, Smart Grid},
}

@article{liang2015UkraineBlackout2017,
	title = {The 2015 {Ukraine} {Blackout}: {Implications} for {False} {Data} {Injection} {Attacks}},
	volume = {32},
	issn = {08858950},
	doi = {10.1109/TPWRS.2016.2631891},
	abstract = {In a false data injection attack (FDIA), an adversary stealthily compromises measurements from electricity grid sensors in a coordinated fashion, with a view to evading detection by the power system bad data detection module. A successful FDIA can cause the system operator to perform control actions that compromise either the physical or economic operation of the power system. In this letter, we consider some implications for FDIAs arising from the late 2015 Ukraine Blackout event.},
	number = {4},
	urldate = {2022-01-12},
	journal = {IEEE Transactions on Power Systems},
	author = {Liang, Gaoqi and Weller, Steven R. and Zhao, Junhua and Luo, Fengji and Dong, Zhao Yang},
	month = jul,
	year = {2017},
	note = {Publisher: Institute of Electrical and Electronics Engineers Inc.},
	keywords = {Cyber-attacks, false data injection attacks, Ukraine blackout},
	pages = {3317--3318},
	file = {Liang et al_2017_The 2015 Ukraine Blackout - Implications for False Data Injection Attacks.pdf:C\:\\Users\\dassc\\Zotero\\storage\\KELQGSH8\\full-text.pdf:application/pdf},
}

@article{liangReviewFalseData2017,
	title = {A {Review} of {False} {Data} {Injection} {Attacks} {Against} {Modern} {Power} {Systems}},
	volume = {8},
	issn = {19493053},
	doi = {10.1109/TSG.2015.2495133},
	abstract = {With rapid advances in sensor, computer, and communication networks, modern power systems have become complicated cyber-physical systems. Assessing and enhancing cyber-physical system security is, therefore, of utmost importance for the future electricity grid. In a successful false data injection attack (FDIA), an attacker compromises measurements from grid sensors in such a way that undetected errors are introduced into estimates of state variables such as bus voltage angles and magnitudes. In evading detection by commonly employed residue-based bad data detection tests, FDIAs are capable of severely threatening power system security. Since the first published research on FDIAs in 2009, research into FDIA-based cyber-attacks has been extensive. This paper gives a comprehensive review of state-of-the-art in FDIAs against modern power systems. This paper first summarizes the theoretical basis of FDIAs, and then discusses both the physical and the economic impacts of a successful FDIA. This paper presents the basic defense strategies against FDIAs and discusses some potential future research directions in this field.},
	number = {4},
	urldate = {2022-01-12},
	journal = {IEEE Transactions on Smart Grid},
	author = {Liang, Gaoqi and Zhao, Junhua and Luo, Fengji and Weller, Steven R. and Dong, Zhao Yang},
	month = jul,
	year = {2017},
	note = {Publisher: Institute of Electrical and Electronics Engineers Inc.},
	keywords = {false data injection attacks, Cyber-physical security, power system, state estimation},
	pages = {1630--1638},
	file = {Liang et al_2017_A Review of False Data Injection Attacks Against Modern Power Systems.pdf:C\:\\Users\\dassc\\Zotero\\storage\\2793JKUD\\Liang et al_2017_A Review of False Data Injection Attacks Against Modern Power Systems.pdf:application/pdf},
}

@article{muslehSurveyDetectionAlgorithms2020,
	title = {A {Survey} on the {Detection} {Algorithms} for {False} {Data} {Injection} {Attacks} in {Smart} {Grids}},
	volume = {11},
	issn = {19493061},
	doi = {10.1109/TSG.2019.2949998},
	abstract = {Cyber-physical attacks are the main substantial threats facing the utilization and development of the various smart grid technologies. Among these attacks, false data injection attack represents a main category with its widely varied types and impacts that have been extensively reported recently. In addressing this threat, several detection algorithms have been developed in the last few years. These were either model-based or data-driven algorithms. This paper provides an intensive summary of these algorithms by categorizing them and elaborating on the pros and cons of each category. The paper starts by introducing the various cyber-physical attacks along with the main reported incidents in history. The significance and the impacts of the false data injection attacks are then reported. The concluding remarks present the main criteria that should be considered in developing future detection algorithms for the false data injection attacks.},
	number = {3},
	urldate = {2022-01-12},
	journal = {IEEE Transactions on Smart Grid},
	author = {Musleh, Ahmed S. and Chen, Guo and Dong, Zhao Yang},
	month = may,
	year = {2020},
	note = {Publisher: Institute of Electrical and Electronics Engineers Inc.},
	keywords = {state estimation, Cyber-physical attacks, data-driven detection algorithms, false data injection, machine learning, model-based detection algorithms, smart grid, stealth attacks, ��},
	pages = {2218--2234},
	file = {Musleh et al_2020_A Survey on the Detection Algorithms for False Data Injection Attacks in Smart.pdf:C\:\\Users\\dassc\\Zotero\\storage\\JGFFSFUM\\full-text.pdf:application/pdf},
}

@article{khanDemonstratingCyberPhysicalAttacks2018,
	title = {Demonstrating {Cyber}-{Physical} {Attacks} and {Defense} for {Synchrophasor} {Technology} in {Smart} {Grid}},
	doi = {10.1109/PST.2018.8514197},
	abstract = {Synchrophasor technology is used for real-time control and monitoring in smart grid. Previous works in literature identified critical vulnerabilities in IEEE C37.118.2 synchrophasor communication standard. To protect synchrophasor-based systems, stealthy cyber-attacks and effective defense mechanisms still need to be investigated.This paper investigates how an attacker can develop a custom tool to execute stealthy man-in-the-middle attacks against synchrophasor devices. In particular, four different types of attack capabilities have been demonstrated in a real synchrophasorbased synchronous islanding testbed in laboratory: (i) command injection attack, (ii) packet drop attack, (iii) replay attack and (iv) stealthy data manipulation attack. With deep technical understanding of the attack capabilities and potential physical impacts, this paper also develops and tests a distributed Intrusion Detection System (IDS) following NIST recommendations. The functionalities of the proposed IDS have been validated in the testbed for detecting aforementioned cyber-attacks. The paper identified that a distributed IDS with decentralized decision making capability and the ability to learn system behavior could effectively detect stealthy malicious activities and improve synchrophasor network security.},
	urldate = {2022-01-25},
	journal = {2018 16th Annual Conference on Privacy, Security and Trust, PST 2018},
	author = {Khan, Rafiullah and McLaughlin, Kieran and Laverty, John Hastings David and David, Hastings and Sezer, Sakir},
	month = oct,
	year = {2018},
	note = {Publisher: Institute of Electrical and Electronics Engineers Inc.
ISBN: 9781538674932},
	file = {Khan et al_2018_Demonstrating Cyber-Physical Attacks and Defense for Synchrophasor Technology.pdf:C\:\\Users\\dassc\\Zotero\\storage\\BN3ASKM6\\Khan et al_2018_Demonstrating Cyber-Physical Attacks and Defense for Synchrophasor Technology.pdf:application/pdf},
}

@techreport{rodofileRealTimeInteractiveAttacks2015,
	title = {Real-{Time} and {Interactive} {Attacks} on {DNP3} {Critical} {Infrastructure} {Using} {Scapy}},
	abstract = {The Distributed Network Protocol v3.0 (DNP3) is one of the most widely used protocols, to control national infrastructure. Widely used interactive packet manipulation tools, such as Scapy, have not yet been augmented to parse and create DNP3 frames (Biondi 2014). In this paper we extend Scapy to include DNP3, thus allowing us to perform attacks on DNP3 in real-time. Our contribution builds on East et al. (2009), who proposed a range of possible attacks on DNP3. We implement several of these attacks to validate our DNP3 extension to Scapy, then executed the attacks on real world equipment. We present our results , showing that many of these theoretical attacks would be unsuccessful in an Ethernet-based network.},
	author = {Rodofile, Nicholas R and Radke, Kenneth and Foo, Ernest},
	year = {2015},
	keywords = {critical Infrastructure security, Distributed Network Proto-col 30, DNP3, Scapy, substations},
	file = {Rodofile et al_Real-Time and Interactive Attacks on DNP3 Critical Infrastructure Using Scapy.pdf:C\:\\Users\\dassc\\Zotero\\storage\\PNBQPY63\\DNP3 Implementation in Scapy.pdf:application/pdf},
}

@article{ieeepowerandenergysocietyIEEEStdC372013,
	title = {{IEEE} {Std} {C37}.244™-2013},
	doi = {https://doi.org/10.1109/IEEESTD.2013.6514039},
	abstract = {The functional, performance, and testing guidelines for a phasor data concentrator are described in this guide. Supporting information is also provided.},
	number = {May},
	urldate = {2022-02-16},
	journal = {IEEE Std C37.244-2013},
	author = {{IEEE Power and Energy Society}},
	year = {2013},
	note = {ISBN: 978-0-7381-8260-5},
	keywords = {IEEE standards, phasor, Synchronization, synchrophasor, Concentrators, GPS synchronization, IEEE C37.244TM, IEEE guide, pdata concentrator (DC), phasor data concentrator (PDC), phasor data concentrator requirement, phasor measurement, phasor measurement unit (PMU), Phasor measurements, Phasors, power system control, power system monitoring, power system protection, Power system protection, Power system reliability},
	pages = {1 -- 65},
	file = {IEEE Power and Energy Society_2013_IEEE Guide for Phasor Data Concentrator Requirements for Power System.pdf:C\:\\Users\\dassc\\Zotero\\storage\\4K4YGBP5\\IEEE Power and Energy Society - 2013 - IEEE Guide for Phasor Data Concentrator Requiremen.pdf:application/pdf},
}

@misc{biondiNetworkPacketManipulation2007,
	title = {Network packet manipulation with {Scapy}},
	urldate = {2022-02-17},
	author = {Biondi, Philippe},
	year = {2007},
	file = {Biondi_2007_Network packet manipulation with Scapy.pdf:C\:\\Users\\dassc\\Zotero\\storage\\BLMYNB8B\\m-api-75d70f1b-8066-fa3f-7c43-c084feed577e.pdf:application/pdf},
}

@book{hashimotoVagrantRunningCreate2013,
	title = {Vagrant: up and running: create and manage virtualized development environments},
	url = {https://books.google.com/books?hl=en&lr=&id=7rJqqKCvdagC&oi=fnd&pg=PR2&dq=vagrant+up+and+running&ots=qDzDYo9MI5&sig=8NYhVHLCyNsW6ojyHL39KO4p5S8},
	urldate = {2022-02-23},
	author = {Hashimoto, M},
	year = {2013},
}

@article{sandiPypmuOpenSource2016,
	title = {pypmu—open source python package for synchrophasor data transfer},
	url = {https://ieeexplore.ieee.org/abstract/document/7818916/},
	urldate = {2022-02-23},
	journal = {ieeexplore.ieee.org},
	author = {Šandi, S and Krstajić, B and Telecommunications, T Popović - 2016 24th and 2016, undefined},
	year = {2016},
	file = {Šandi et al_2016_pypmu—open source python package for synchrophasor data transfer.pdf:C\:\\Users\\dassc\\Zotero\\storage\\J6QI9N2Y\\Šandi et al_2016_pypmu—open source python package for synchrophasor data transfer.pdf:application/pdf},
}

@article{prabhuStateoftheartReviewSynchrophasor2017,
	title = {A state-of-the-art review on synchrophasor applications to power network protection},
	volume = {436},
	issn = {18761119},
	doi = {10.1007/978-981-10-4394-9_52},
	abstract = {The demand for electricity supply has been increased many folds over the last few decades. However, the growth in the electric infrastructure has not been increased accordingly due to deregulation of the energy markets, economic and environmental reasons. In present days, power networks are most often operated closer to their stability limit to fulfill the growing electricity demand. As a result, the security and safety of the power system today is at risk. Investigation on large blackouts in the recent past show that maintaining system reliability and integrity becomes more and more difficult due to reduced transmission capacity margins and increased stress on the system. Under the stressed operating condition, the widely-used distance relaying based transmission line protection schemes are susceptible to maloperation. The use of series-compensated and multiterminal lines is another concern for the distance protection scheme. At the same time, the present advancements in the wide-area measurement systems (WAMS) using synchrophasors has shown potential for ensuring improved protection for different power networks operating even at critical conditions. In this paper, the authors first investigate the limitations of existing distance relays while protecting different power networks during stressed operating conditions. Then, an extensive review is made on the application of synchrophasor based WAMS technology for reliable power system protection. The objective of the present study is mainly to bring the attention of the researchers from academic institutions, industries and utility grid on the possible applications of synchrophasors based WAMS technology for ensuring improved protection to today’s power system.},
	urldate = {2022-03-01},
	journal = {Lecture Notes in Electrical Engineering},
	author = {Prabhu, M. S. and Nayak, Paresh Kumar},
	year = {2017},
	note = {Publisher: Springer Verlag},
	keywords = {Distance relay, Multiterminal line, Series compensation, Synchrophasor, WAMS, Wide-area backup protection},
	pages = {531--541},
	file = {Prabhu et al_2017_A state-of-the-art review on synchrophasor applications to power network.pdf:C\:\\Users\\dassc\\Zotero\\storage\\LX96ZWF4\\Prabhu et al_2017_A state-of-the-art review on synchrophasor applications to power network.pdf:application/pdf},
}

@article{khanAnalysisIEEEC372016,
	title = {Analysis of {IEEE} {C37}. 118 and {IEC} 61850-90-5 synchrophasor communication frameworks},
	url = {https://ieeexplore.ieee.org/abstract/document/7741343/},
	doi = {10.1109/PESGM.2016.7741343},
	abstract = {ICT in smart grid provides enormous opportunities for real-time and wide-area grid monitoring, protection and control. To this aim, synchrophasor technology was proposed for reliable and secure transmission of grid status information. IEEE C37.118 and IEC 61850-90-5 emerged as two well known communication frameworks for synchrophasor technology. However, literature lacks a comprehensive analysis of some key features and limitations. Further, knowledge of cyber vulnerabilities in both communication frameworks is still quite limited. This paper analyzes characteristics of both communication frameworks inferred from their complete implementation. In particular, it addresses their embedded features, required network character-istics/resources and their resilience against cyber attacks.},
	urldate = {2022-03-01},
	journal = {ieeexplore.ieee.org},
	author = {Khan, R and Mclaughlin, K and Laverty, D and Sezer, \& and Khan, Rafiullah and Mclaughlin, Kieran and Laverty, David and Sezer, Sakir},
	year = {2016},
	note = {Publisher: PESGM},
	pages = {2016},
	file = {Khan et al_2016_Analysis of IEEE C37. 118 and IEC 61850-90-5 synchrophasor communication.pdf:C\:\\Users\\dassc\\Zotero\\storage\\J9U6TCZS\\full-text.pdf:application/pdf},
}

@article{khanIEEEC3711822016,
	title = {{IEEE} {C37}.118-2 {Synchrophasor} {Communication} {Framework} {Overview}, {Cyber} {Vulnerabilities} {Analysis} and {Performance} {Evaluation}},
	doi = {10.5220/0005745001670178},
	abstract = {Synchrophasors have become an important part of the modern power system and numerous applications have been developed covering wide-area monitoring, protection and control. Most applications demand continuous transmission of synchrophasor data across large geographical areas and require an efficient communication framework. IEEE C37.118-2 evolved as one of the most successful synchrophasor communication standards and is widely adopted. However, it lacks a predefined security mechanism and is highly vulnerable to cyber attacks. This paper analyzes different types of cyber attacks on IEEE C37.118-2 communication system and evaluates their possible impact on any developed synchrophasor application. Further, the paper also recommends an efficent security mechanism that can provide strong protection against cyber attacks. Although, IEEE C37.118-2 has been widely adopted, there is no clear understanding of the requirements and limitations. To this aim, the paper also presents detailed performance evaluation of IEEE C37.118-2 implementations which could help determine required resources and network characteristics before designing any synchropha-sor application.},
	urldate = {2022-03-01},
	journal = {scitepress.org},
	author = {Khan, R and McLaughlin, K and …, D Laverty - … Conference on Information and 2016, undefined},
	year = {2016},
	note = {ISBN: 9789897581670},
	keywords = {Vulnerability, Smart Grid, Synchrophasor, Cyber Security, IEEE C37118, ��},
	file = {Khan et al_Ieee c37. 118-2 synchrophasor communication framework-overview, cyber.pdf:C\:\\Users\\dassc\\Zotero\\storage\\EXRI74Q8\\m-api-2a480766-82be-e365-6013-1cb72f84a562.pdf:application/pdf},
}

@article{morrisCybersecurityRiskTesting2011,
	title = {Cybersecurity risk testing of substation phasor measurement units and phasor data concentrators},
	doi = {10.1145/2179298.2179324},
	abstract = {Future bulk electric transmission systems will include substation automation, synchrophasor measurement systems, and automated control algorithms which leverage wide area monitoring system to better control the grid. Prior to installation of new networked devices, utilities should perform cybersecurity testing and develop corrective actions for identified vulnerabilities. This paper outlines testing performed prior to the installation of a synchrophasor wide area monitoring system. Phasor measurement unit and phasor data concentrator devices from multiple vendors were subjected to laboratory testing including; device security feature identification, port scans, network congestion testing, denial of service testing, protocol mutation testing, and network traffic disclosure testing. This paper outlines the procedures used to perform the testing and discusses the types of results expected from testing. Copyright © 2011 ACM.},
	urldate = {2022-03-01},
	journal = {ACM International Conference Proceeding Series},
	author = {Morris, Thomas and Pan, Shengyi and Lewis, Jeremy and Moorhead, Jonathan and Younan, Nicholas and King, Roger and Freund, Mark and Madani, Vahid},
	year = {2011},
	note = {ISBN: 9781450309455},
	keywords = {Smart Grid, Cybersecurity, Experimentation, Security Keywords Smart Grid},
	file = {Morris et al_2011_Cybersecurity risk testing of substation phasor measurement units and phasor.pdf:C\:\\Users\\dassc\\Zotero\\storage\\WH3GHBK5\\Morris et al_2011_Cybersecurity risk testing of substation phasor measurement units and phasor.pdf:application/pdf},
}

@article{palRealtimeDetectionPacket2014,
	title = {Real-time detection of packet drop attacks on synchrophasor data},
	url = {https://ieeexplore.ieee.org/abstract/document/7007762/},
	abstract = {The importance of phasor measurement unit (PMU) or synchrophasor data towards the functioning of real-time monitoring and control of power generation and distribution systems makes them an attractive target for cyber-attacks. An attack with potential for significant damage is the packet drop attack, where the adversary arbitrarily drops packets with synchrophasor data. This paper develops a real-time mechanism for detecting packet drop attacks on synchrophasor data carried over the Internet. The proposed solution is receiver-based, and uses the one-way packet delays to extract features that are used to detect attacks. The proposed attack detection mechanism leads to lower detection delays and greater accuracy as compared to existing mechanisms.},
	urldate = {2022-03-01},
	journal = {ieeexplore.ieee.org},
	author = {Pal, S and Sikdar, B and Conference, J Chow - 2014 IEEE International and 2014, undefined},
	year = {2014},
	file = {Pal et al_Real-time detection of packet drop attacks on synchrophasor data.pdf:C\:\\Users\\dassc\\Zotero\\storage\\L47EFFTG\\full-text.pdf:application/pdf},
}

@article{fanSynchrophasorDataCorrection2018,
	title = {Synchrophasor data correction under {GPS} spoofing attack: {A} state estimation-based approach},
	url = {https://ieeexplore.ieee.org/abstract/document/7839276/},
	urldate = {2022-03-01},
	journal = {ieeexplore.ieee.org},
	author = {Fan, X and Du, L and Grid, D Duan - IEEE Transactions on Smart and 2017, undefined},
	year = {2018},
	file = {Fan et al_Synchrophasor data correction under GPS spoofing attack - A state.pdf:C\:\\Users\\dassc\\Zotero\\storage\\3C6N46WG\\full-text.pdf:application/pdf},
}

@article{paudelDataIntegrityAttacks2016,
	title = {Data integrity attacks in smart grid wide area monitoring},
	url = {https://www.scienceopen.com/hosted-document?doi=10.14236/ewic/ICS2016.9},
	doi = {10.14236/ewic/ICS2016.9},
	abstract = {A smart grid requires the implementation of ICT technologies in order to incorporate new functions into electricity grid monitoring and control. Wide Area Monitoring Systems (WAMSs) are used to measure synchrophasor data at different locations and give operators a near-real-time picture of what is happening in the system. The measurement data is periodically collected via communication channels to monitor, predict and control the power consumption, and detect any problems in the power grid. Attacks on WAMSs can trigger wrong decisions and create dangerous failures in the smart grid system. In this paper, we investigate data integrity attacks at different attack entry points of a WAMS, their impacts on the smart grid system, and existing mitigation strategies. We conclude from our study that the existing techniques, methodologies and mechanisms are not effective enough to detect or mitigate some attacks.},
	urldate = {2022-03-01},
	journal = {scienceopen.com},
	author = {Paudel, S and Smith, P and …, T Zseby - for ICS \& SCADA Cyber Security and 2016, undefined},
	year = {2016},
	note = {Publisher: BCS Learning \& Development},
	keywords = {Cybersecurity, Data Integrity Attacks, Failure Scenarios, Wide Area Monitoring System},
	file = {Paudel et al_2016_Data integrity attacks in smart grid wide area monitoring.pdf:C\:\\Users\\dassc\\Zotero\\storage\\CMG3W6V2\\full-text.pdf:application/pdf},
}

@article{singhStealthyCyberAttacks2016,
	title = {Stealthy cyber attacks and impact analysis on wide-area protection of smart grid},
	url = {https://ieeexplore.ieee.org/abstract/document/7747927/},
	doi = {10.1109/NAPS.2016.7747927},
	abstract = {Smart grid is vulnerable to many cyber attacks due to legacy nature of the infrastructure coupled with increase in control and monitoring functions through cyber advancements. Remedial Action Scheme (RAS), widely used for wide area protection, provides autonomous operations through the RAS controller. Due to its dependence on the timely cooperation of multiple devices and communication network, it is highly vulnerable to cyber attacks. In this paper, we present an impact analysis for the power system due to a class of malware-based coordinated cyber attacks targeting the RAS scheme. Specifically, we make the following two contributions. First, modeling a stealth attack vector based on malware and coordinated attack behavior. In particular, installing the malware (Trojan horse) in the RAS controller which turns the controller into an attacker's bot. Then, performing a coordinated attack which involves malicious tripping of one of the parallel lines connected to a generator followed by the continuous pulse attack on the generator. The pulse attack includes periodically changing the generations through RAS controller which remains undetected by the control center. Second, testbed-based implementation and evaluation to quantify system impacts. We have leveraged Iowa State's PowerCyber CPS security testbed for experimental evaluation. In our evaluation, we varied the duty cycle of the pulse attack to obtain different attack scenarios and consequent impacts are analyzed on modified IEEE 9-bus system in real-time simulation. Our studies show that the duty cycle of the pulse attack is a critical factor in determining the severity of the attack impacts on system stability.},
	urldate = {2022-03-01},
	journal = {ieeexplore.ieee.org},
	author = {Singh, VK and …, A Ozen - 2016 North American and 2016, undefined},
	year = {2016},
	note = {ISBN: 9781509032709},
	file = {Singh et al_2016_Stealthy cyber attacks and impact analysis on wide-area protection of smart grid.pdf:C\:\\Users\\dassc\\Zotero\\storage\\HN8NYYJN\\full-text.pdf:application/pdf},
}

@article{johnsonAssessingNetworkCybersecurity2020,
	title = {Assessing {DER} network cybersecurity defences in a power-communication co-simulation environment},
	urldate = {2022-03-01},
	journal = {ieeexplore.ieee.org},
	author = {Johnson, J and Onunkwo, I and …, P Cordeiro - IET Cyber-Physical and 2020, undefined},
	year = {2020},
	file = {Johnson et al_Assessing DER network cybersecurity defences in a power-communication.pdf:C\:\\Users\\dassc\\Zotero\\storage\\TXA33S25\\Johnson et al_Assessing DER network cybersecurity defences in a power-communication.pdf:application/pdf},
}

@article{cuiCyberPhysicalSystem2020,
	title = {Cyber‐physical system testbed for power system monitoring and wide‐area control verification},
	doi = {https://doi.org/10.1049/iet-esi.2019.0084},
	urldate = {2022-03-01},
	journal = {ieeexplore.ieee.org},
	author = {Cui, H and Li, F and Integration, K Tomsovic - IET Energy Systems and 2020, undefined},
	year = {2020},
	file = {Cui et al_Cyber‐physical system testbed for power system monitoring and wide‐area control.pdf:C\:\\Users\\dassc\\Zotero\\storage\\JEZFIDB3\\Cui et al_Cyber‐physical system testbed for power system monitoring and wide‐area control.pdf:application/pdf},
}

@article{adhikariCyberphysicalPowerSystem2014,
	title = {A cyber-physical power system test bed for intrusion detection systems},
	url = {https://ieeexplore.ieee.org/abstract/document/6939262/},
	abstract = {The rapid advancement of technology used in operation, monitoring, and control introduces several threats against power system. Cyber-physical power system vulnerabilities are increasing and the consequences of attack can be catastrophic. Understanding power system phenomena and attacks is vital to identifying and detecting such events. Researchers require a suitable power system test bed that can provide a platform for simulation of power system events and attacks. An essential part of such a test bed is the ability to provide software and hardware interaction to mimic real world scenarios. This paper presents a test bed for the development of an intrusion detection system (IDS) for power systems. The test bed consists of a power system modeled on a real time digital simulator (RTDS), a data collection and processing engine, and a MATLAB/RSCAD parameter calculation engine. This test bed provides a platform for hardware in the loop (HIL) simulation, power system attacks, and generates data sets required by cyber security researchers. Coordinated distance protection and overcurrent protection schemes are implemented on the IEEE 9 bus system and a 3-generator 4 bus system [11]. Fault, contingency and cyber-attack scenarios have been developed for both power systems. Selected relevant simulation results are presented.},
	urldate = {2022-03-01},
	journal = {ieeexplore.ieee.org},
	author = {Adhikari, U and Morris, TH and Meeting, S Pan - 2014 IEEE PES General and 2014, undefined},
	year = {2014},
	note = {ISBN: 9781479964154},
	keywords = {power system, contingencies, data, faults, IDS, Index Terms-attacks, test bed},
	file = {Adhikari et al_A cyber-physical power system test bed for intrusion detection systems.pdf:C\:\\Users\\dassc\\Zotero\\storage\\6P8IKLDQ\\Adhikari et al_A cyber-physical power system test bed for intrusion detection systems.pdf:application/pdf},
}

@article{cintugluSurveySmartGrid2017,
	title = {A survey on smart grid cyber-physical system testbeds},
	url = {https://ieeexplore.ieee.org/abstract/document/7740849/},
	urldate = {2022-03-01},
	journal = {ieeexplore.ieee.org},
	author = {Cintuglu, MH and Mohammed, OA and Tutorials, K Akkaya - … Surveys \& and 2016, undefined},
	year = {2017},
	file = {Cintuglu et al_A survey on smart grid cyber-physical system testbeds.pdf:C\:\\Users\\dassc\\Zotero\\storage\\55MGEJES\\full-text.pdf:application/pdf},
}

@article{yangCybersecurityTestbedIEC2015,
	title = {Cybersecurity test-bed for {IEC} 61850 based smart substations},
	url = {https://ieeexplore.ieee.org/abstract/document/7286357/},
	doi = {10.1109/PESGM.2015.7286357},
	abstract = {With the development and deployment of IEC 61850 based smart substations, cybersecurity vulnerabilities of supervisory control and data acquisition (SCADA) systems are increasingly emerging. In response to the emergence of cybersecurity vulnerabilities in smart substations, a test-bed is indispensable to enable cybersecurity experimentation. In this paper, a comprehensive and realistic cyber-physical test-bed has been built to investigate potential cybersecurity vulnerabilities and the impact of cyber-attacks on IEC 61850 based smart substations. This test-bed is close to a real production type environment, and has the ability to carry out end-to-end testing of cyber-attacks and physical consequences. A fuzz testing approach is proposed for detecting IEC 61850 based intelligent electronic devices (IEDs) and validated in the proposed test-bed.},
	urldate = {2022-03-01},
	journal = {ieeexplore.ieee.org},
	author = {Yang, Y and Jiang, HT and McLaughlin, K and …, L Gao - 2015 IEEE Power \& and 2015, undefined},
	year = {2015},
	note = {Publisher: IEEE},
	keywords = {Cybersecurity, Fuzz testing, IEC 61850, Index Terms-Smart Substation, Test-bed},
	pages = {1--5},
	file = {Yang et al_2015_Cybersecurity test-bed for IEC 61850 based smart substations.pdf:C\:\\Users\\dassc\\Zotero\\storage\\FBDX2ZMW\\Yang et al_2015_Cybersecurity test-bed for IEC 61850 based smart substations.pdf:application/pdf},
}

@article{khanThreatAnalysisBlackEnergy2016,
	title = {Threat analysis of {BlackEnergy} malware for synchrophasor based real-time control and monitoring in smart grid},
	doi = {10.14236/ewic/ics2016.7},
	abstract = {The BlackEnergy malware targeting critical infrastructures has a long history. It evolved over time from a simple DDoS platform to a quite sophisticated plug-in based malware. The plug-in architecture has a persistent malware core with easily installable attack specific modules for DDoS, spamming, info-stealing, remote access, boot-sector formatting etc. BlackEnergy has been involved in several high profile cyber physical attacks including the recent Ukraine power grid attack in December 2015. This paper investigates the evolution of BlackEnergy and its cyber attack capabilities. It presents a basic cyber attack model used by BlackEnergy for targeting industrial control systems. In particular, the paper analyzes cyber threats of BlackEnergy for synchrophasor based systems which are used for real-time control and monitoring functionalities in smart grid. Several BlackEnergy based attack scenarios have been investigated by exploiting the vulnerabilities in two widely used synchrophasor communication standards: (i) IEEE C37.118 and (ii) IEC 61850-90-5. Further, the paper also investigates protection strategies for detection and prevention of BlackEnergy based cyber physical attacks.},
	author = {Khan, Rafiullah and Maynard, Peter and McLaughlin, Kieran and Laverty, David and Sezer, Sakir},
	month = aug,
	year = {2016},
	doi = {10.14236/ewic/ics2016.7},
	note = {MAG ID: 2514382028},
	keywords = {��, ��},
	pages = {1--11},
	file = {Khan et al_2016_Threat analysis of BlackEnergy malware for synchrophasor based real-time.pdf:C\:\\Users\\dassc\\Zotero\\storage\\Y3PMFMAU\\Khan et al. - 2016 - Threat analysis of BlackEnergy malware for synchro.pdf:application/pdf},
}

@article{monticelliElectricPowerSystem2000,
	title = {Electric power system state estimation},
	volume = {88},
	issn = {1558-2256},
	doi = {10.1109/5.824004},
	abstract = {This paper discusses the state of the art in electric power system state estimation. Within energy management systems, state estimation is a key function for building a network real-time model. A real-time model is a quasi-static mathematical representation of the current conditions in an interconnected power network. This model is extracted at intervals from snapshots of real-time measurements (both analog and status). The new modeling needs associated with the introduction of new control devices and the changes induced by emerging energy markets are making state estimation and its related functions more important than ever.},
	number = {2},
	journal = {Proceedings of the IEEE},
	author = {Monticelli, A.},
	month = feb,
	year = {2000},
	note = {Conference Name: Proceedings of the IEEE},
	keywords = {Mathematical model, Load flow, Power system reliability, ��, Data analysis, Energy management, Network topology, Parameter estimation, Power system modeling, State estimation, Voltage, :book},
	pages = {262--282},
	file = {IEEE Xplore Abstract Record:C\:\\Users\\dassc\\Zotero\\storage\\IYNFM9RC\\824004.html:text/html;Monticelli_2000_Electric power system state estimation.pdf:C\:\\Users\\dassc\\Zotero\\storage\\2B35LDPC\\Monticelli_2000_Electric power system state estimation.pdf:application/pdf},
}

@article{koltysSHaPeHoneypotElectric2015,
	title = {{SHaPe}: {A} {Honeypot} for {Electric} {Power} {Substation}},
	issn = {1509-4553, 1899-8852},
	shorttitle = {{SHaPe}},
	url = {https://www.infona.pl//resource/bwmeta1.element.baztech-dd7fe369-6682-4a1e-bd10-e89d3723cbb6},
	abstract = {Supervisory Control and Data Acquisition (SCADA) systems play a crucial role in national critical infrastructures, and any failure may result in severe damages. Initially SCADA networks were separated from other networks and used proprietary communications protocols that were well known only to the device manufacturers. At that time such isolation and obscurity ensured an acceptable security level. Nowadays, modern SCADA systems usually have direct or indirect Internet connection, use open protocols and commercial-off-the-shelf hardware and software. This trend is also noticeable in the power industry. Present substation automation systems (SASs) go beyond traditional SCADA and employ many solutions derived from Information and Communications Technology (ICT). As a result electric power substations have become more vulnerable for cybersecurity attacks and they need ICT security mechanisms adaptation. This paper shows the SCADA honeypot that allows detecting unauthorized or illicit trac in SAS which communication architecture is dened according to the IEC 61850 standard.},
	language = {English},
	number = {nr 4},
	urldate = {2022-05-28},
	journal = {Journal of Telecommunications and Information Technology},
	author = {Kołtyś, K. and Gajewski, R.},
	year = {2015},
	keywords = {��},
	pages = {37--43},
	file = {Kołtyś_Gajewski_2015_SHaPe.pdf:C\:\\Users\\dassc\\Zotero\\storage\\Q24HJEK3\\Kołtyś_Gajewski_2015_SHaPe.pdf:application/pdf;Snapshot:C\:\\Users\\dassc\\Zotero\\storage\\FIXQFPMI\\bwmeta1.element.html:text/html},
}

@inproceedings{vasilomanolakisDidYouReally2015a,
	title = {Did you really hack a nuclear power plant? {An} industrial control mobile honeypot},
	shorttitle = {Did you really hack a nuclear power plant?},
	doi = {10.1109/CNS.2015.7346907},
	abstract = {The emerge of sophisticated attackers and malware that target Industrial Control System (ICS) suggests that novel security mechanisms are required. Honeypots, can act as an additional line of defense, by providing early warnings for such attacks. We present a mobile ICS honeypot, that can be placed in various network positions to provide security administrators an on-the-go security status of their network. We discuss our system, its merits in comparison to other honeypots, and provide preliminary results towards a large-scale evaluation.},
	booktitle = {2015 {IEEE} {Conference} on {Communications} and {Network} {Security} ({CNS})},
	author = {Vasilomanolakis, Emmanouil and Srinivasa, Shreyas and Mühlhäuser, Max},
	month = sep,
	year = {2015},
	keywords = {Protocols, ��, Industrial control, industrial control systems, Internet, Malware, Mobile communication, mobile honeypot, Servers},
	pages = {729--730},
	file = {IEEE Xplore Abstract Record:C\:\\Users\\dassc\\Zotero\\storage\\W633V8X6\\7346907.html:text/html;Vasilomanolakis et al_2015_Did you really hack a nuclear power plant.pdf:C\:\\Users\\dassc\\Zotero\\storage\\KLK7FCA5\\Vasilomanolakis et al_2015_Did you really hack a nuclear power plant.pdf:application/pdf},
}

@article{haughtonLinearStateEstimation2013,
	title = {A {Linear} {State} {Estimation} {Formulation} for {Smart} {Distribution} {Systems}},
	volume = {28},
	issn = {1558-0679},
	doi = {10.1109/TPWRS.2012.2212921},
	abstract = {This paper presents a linearized, three-phase, distribution class state estimation algorithm for applications in smart distribution systems. Unbalanced three-phase cases and single-phase cases are accommodated. The estimator follows a complex variable formulation and is intended to incorporate synchronized phasor measurements into distribution state estimation. Potential applications in smart distribution system control and management are discussed.},
	number = {2},
	journal = {IEEE Transactions on Power Systems},
	author = {Haughton, Daniel A. and Heydt, Gerald Thomas},
	month = may,
	year = {2013},
	note = {Conference Name: IEEE Transactions on Power Systems},
	keywords = {��, State estimation, Current measurement, Distribution management systems, distribution system monitoring and control, distribution system state estimation, Loading, power distribution engineering, Power measurement, Real time systems, synchronized phasor measurements, three-phase unbalance, Vectors, Voltage measurement},
	pages = {1187--1195},
	file = {Haughton_Heydt_2013_A Linear State Estimation Formulation for Smart Distribution Systems.pdf:C\:\\Users\\dassc\\Zotero\\storage\\FJ9AECVJ\\Haughton_Heydt_2013_A Linear State Estimation Formulation for Smart Distribution Systems.pdf:application/pdf;IEEE Xplore Abstract Record:C\:\\Users\\dassc\\Zotero\\storage\\SLRUCQMQ\\6302216.html:text/html;IEEE Xplore Full Text PDF:C\:\\Users\\dassc\\Zotero\\storage\\HSB7IVTF\\Haughton and Heydt - 2013 - A Linear State Estimation Formulation for Smart Di.pdf:application/pdf},
}

@article{wuPowerSystemState1990,
	title = {Power system state estimation: a survey},
	volume = {12},
	issn = {0142-0615},
	shorttitle = {Power system state estimation},
	url = {https://www.sciencedirect.com/science/article/pii/014206159090003T},
	doi = {10.1016/0142-0615(90)90003-T},
	abstract = {Recent developments in the solution methods for state estimation are reviewed. Concepts of decoupling, ill-conditioning and robustness in state estimation are discussed. Derivations of decoupled estimators, stable estimators and robust estimators are reviwed. Future directions for research are suggested.},
	language = {en},
	number = {2},
	urldate = {2022-05-28},
	journal = {International Journal of Electrical Power \& Energy Systems},
	author = {Wu, Felix F.},
	month = apr,
	year = {1990},
	keywords = {state estimation, security monitoring, weighted least square estimation},
	pages = {80--87},
	file = {ScienceDirect Snapshot:C\:\\Users\\dassc\\Zotero\\storage\\X84IS4JQ\\014206159090003T.html:text/html;Wu_1990_Power system state estimation.pdf:C\:\\Users\\dassc\\Zotero\\storage\\8KVNQPQZ\\Wu_1990_Power system state estimation.pdf:application/pdf},
}

@inproceedings{chenStateEstimationDistribution2015,
	title = {State estimation for distribution systems using micro-synchrophasors},
	doi = {10.1109/APPEEC.2015.7381051},
	abstract = {Phasor Measurement Unit (PMU) has contributed greatly to power system state estimation because of its ability to directly measure voltage and current phase angle. PMUs have been used almost exclusively in transmission systems monitoring. Recent years, increasing penetration of renewable energy sources brings new characteristics into distribution system such as bi-directional power flows and voltage profile issues, which has necessitated continuous monitoring of distribution systems. Therefore, fast and accurate measurement device is needed for application in distribution networks. μPMU is such a device that can provide phasor measurements with high precision and low cost. This paper uses a linear three phase state estimator for applications in distribution systems. The proposed estimator can make use of synchrophasor measurements which can be realized by μPMU. This is tested on IEEE 13 bus feeder which has unbalanced three phase transmission lines and loads.},
	booktitle = {2015 {IEEE} {PES} {Asia}-{Pacific} {Power} and {Energy} {Engineering} {Conference} ({APPEEC})},
	author = {Chen, Xuebing and Tseng, King Jet and Amaratunga, Gehan},
	month = nov,
	year = {2015},
	note = {13 citations (Crossref) [2022-05-30]},
	keywords = {Phasor measurement units, ✅, State estimation, Current measurement, Power measurement, Voltage measurement, Transmission line matrix methods, Transmission line measurements},
	pages = {1--5},
	annote = {Annotations(5/29/2022, 10:54:04 PM)
 Abstract “Phasor Measurement Unit (PMU) has contributed greatly to power system state estimation because of its ability to directly measure voltage and current phase angle” “This paper uses a linear three phase state estimator for applications in distribution systems.” “This is tested on IEEE 13 bus feeder which has unbalanced three phase transmission lines and loads” I. INTRODUCTION “However, PMU that has been used in transmission networks is not ideal choice for distribution system application due to economic and technical constraints. Micro-synchrophasors(��PMUs) are high-precision measurement units that can work well in distribution networks [8].” “There are lots of potential applications of ��PMU to distribution networks such as phase identification, reverse power flow detection and state estimation [8], [9]. State estimation is a possible application of ��PMU.”[[Micro-PMU]] “single phase state estimator for transmission systems no longer applicable in distribution system”\#questionNeed to verity this statement. That means transmission PMUs only measure single phase. “Several distribution level state estimation methods have already been proposed to account for imbalances in system operation. Switching to full three-phase representation of the network is the most straightforward method. Decoupled state estimation using sequence network could be more efficient by decreasing problem dimension [11].” II. PHASOR MEASUREMENT UNITS AND MICRO SYNCHROPHASOR “Synchronized phasor measurement units (PMUs) were developed in the mid-1980s” “Due to the short transmission lines and low power flow, voltage angle between locations on a distribution network will be up to two orders of magnitude smaller than those on the transmission network (tenths of a degree, rather than tens of degrees) [8].” III. STATE ESTIMATION “The state of a power system can be described by complex voltage (magnitudes and phase angles) at every bus.” “In traditional state estimation, measurement data are usually voltage magnitude, power flows and power injections. These measurements have nonlinear relationship with power grid states (complex voltages), resulting non-linear and time consuming state estimator. Non-linear state estimator can be solved by iterative method, among which weighted least squares is the most common method [14].”[[Linear State Estimation]] “Mean Absolute Percentage Error (MAPE) and Mean Absolute Error (MAE) can be used to evaluate the accuracy of voltage magnitude and phase angle respectively.”[[MAPE]], [[MAE]](Chen et al., 2015, p. 2) “Transmission networks transfer large amounts of power through mesh topologies made up with high reactance-resistance ratios transmission line.” “In three-phase state estimation, three-phase bus voltage are designated as the system states. �� = ���� + �� (7) Measurement vector �� contains three phase bus voltage phasor and branch current phasors. �� is the measurement Jacobian matrix. �� is the system state vector and �� is measurement error vector.” “From this equation, it can be seen off-diagonal elements are zeros, so this matrix equation can be written as three separate equations and solved one by one. Hence the computational time can be reduced greatly with help of parallel processors [11].”So distribution system can be decomposed as three separate state estimation problem. IV. SIMULATION RESULTS “IEEE 13 bus feeder was selected to be the test feeder.” “In order to develop linear three phase state estimator, the network must be fully observable by ��PMUs. For this test feeder, five ��PMUs were necessary to make all bus voltage” “These five ��PMUs are placed at bus 1, 3, 5, 9, 10.” “these five ��PMUs returned voltage phasor at five buses and current phasor of nine transmission lines.” “13 voltage phasor measurements and 20 current phasor measurements.” “Bias were added to all measurements. According to specification [20], guaranteed amplitude accuracy is ±0.05\%, and guaranteed angle accuracy is ±0.01∘. Bias that added to measurements were random number within the given accuracy.”(Chen et al., 2015, p. 4) “In order to analysis the effect of redundancy on estimation results, another ��PMU was assumed to be installed at bus 2.” “The state estimator gave more accurate results with this six ��PMUs.”(Chen et al., 2015, p. 5) V. C ONCLUSION “Future work could include comparison of decoupled state estimator and phase state estimator. Theoretically decoupled state estimator can decrease the computation time if parallel processor is available. However, the decoupled state estimator is not as accurate as phase estimator because the admittance of transmission line is strictly symmetrical.”},
	file = {Chen et al_2015_State estimation for distribution systems using micro-synchrophasors.pdf:C\:\\Users\\dassc\\Zotero\\storage\\VJMEU3IM\\Chen et al_2015_State estimation for distribution systems using micro-synchrophasors.pdf:application/pdf;IEEE Xplore Abstract Record:C\:\\Users\\dassc\\Zotero\\storage\\K52W93CT\\7381051.html:text/html},
}

@techreport{saygheSurveyMachineLearning2020,
	title = {A {Survey} of {Machine} {Learning} {Methods} for {Detecting} {False} {Data} {Injection} {Attacks} in {Power} {Systems}},
	url = {http://arxiv.org/abs/2008.06926},
	abstract = {Over the last decade, the number of cyberattacks targeting power systems and causing physical and economic damages has increased rapidly. Among them, False Data Injection Attacks (FDIAs) is a class of cyberattacks against power grid monitoring systems. Adversaries can successfully perform FDIAs in order to manipulate the power system State Estimation (SE) by compromising sensors or modifying system data. SE is an essential process performed by the Energy Management System (EMS) towards estimating unknown state variables based on system redundant measurements and network topology. SE routines include Bad Data Detection (BDD) algorithms to eliminate errors from the acquired measurements, e.g., in case of sensor failures. FDIAs can bypass BDD modules to inject malicious data vectors into a subset of measurements without being detected, and thus manipulate the results of the SE process. In order to overcome the limitations of traditional residual-based BDD approaches, data-driven solutions based on machine learning algorithms have been widely adopted for detecting malicious manipulation of sensor data due to their fast execution times and accurate results. This paper provides a comprehensive review of the most up-to-date machine learning methods for detecting FDIAs against power system SE algorithms.},
	number = {arXiv:2008.06926},
	urldate = {2022-05-30},
	institution = {arXiv},
	author = {Sayghe, Ali and Hu, Yaodan and Zografopoulos, Ioannis and Liu, XiaoRui and Dutta, Raj Gautam and Jin, Yier and Konstantinou, Charalambos},
	month = aug,
	year = {2020},
	doi = {10.48550/arXiv.2008.06926},
	note = {arXiv:2008.06926 [cs, eess]
type: article},
	keywords = {Electrical Engineering and Systems Science - Systems and Control, Computer Science - Cryptography and Security},
	file = {arXiv.org Snapshot:C\:\\Users\\dassc\\Zotero\\storage\\SUYSJ5IJ\\2008.html:text/html;iet-stg.2020.0015.pdf:C\:\\Users\\dassc\\Zotero\\storage\\HHBCDV2M\\iet-stg.2020.0015.pdf:application/pdf;Sayghe et al_2020_A Survey of Machine Learning Methods for Detecting False Data Injection Attacks.pdf:C\:\\Users\\dassc\\Zotero\\storage\\JDPUYMD2\\Sayghe et al_2020_A Survey of Machine Learning Methods for Detecting False Data Injection Attacks.pdf:application/pdf},
}

@inproceedings{saraswatAnalyzingEffectsCyberattacks2021,
	title = {Analyzing the effects of cyberattacks on distribution system state estimation},
	doi = {10.1109/ISGT49243.2021.9372262},
	abstract = {Key components of power systems-such as energy management systems, automatic generation control, and state estimation-are under serious vulnerability from cyberattacks. Cyber threats in electric grids have increased significantly because of the increased interconnectivity of supervisory control and data acquisition systems and public network infrastructure. As the penetration level of distributed energy resources increases, it is imperative to employ system-monitoring techniques such as state estimation for the reliable operation of distribution systems. Recently, multiple methods have been developed that exploit the low rank property of distribution system state matrix and are robust to bad data, such as matrix completion. This paper analyzes the impact of various realistic cyberattack scenarios on matrix completion. Realistic cyberattack scenarios are converted into data corruption models that are used in an extensive simulation of a custom IEEE 123-bus system.},
	booktitle = {2021 {IEEE} {Power} {Energy} {Society} {Innovative} {Smart} {Grid} {Technologies} {Conference} ({ISGT})},
	author = {Saraswat, Govind and Yang, Rui and Liu, Yajing and Zhang, Yingchen},
	month = feb,
	year = {2021},
	note = {1 citations (Crossref) [2022-05-30]
ISSN: 2472-8152},
	keywords = {state estimation, ��, State estimation, Computer crime, Cyberattacks, Data models, distribution systems, electric grid, Matrix converters, Reliability, SCADA systems, security, Smart grids},
	pages = {01--05},
	file = {IEEE Xplore Abstract Record:C\:\\Users\\dassc\\Zotero\\storage\\PFGL4FZI\\9372262.html:text/html;Saraswat et al_2021_Analyzing the effects of cyberattacks on distribution system state estimation.pdf:C\:\\Users\\dassc\\Zotero\\storage\\9J5SV8DB\\Saraswat et al_2021_Analyzing the effects of cyberattacks on distribution system state estimation.pdf:application/pdf;Saraswat et al_2021_Analyzing the effects of cyberattacks on distribution system state estimation.pdf:C\:\\Users\\dassc\\Zotero\\storage\\49AB5AWC\\Saraswat et al_2021_Analyzing the effects of cyberattacks on distribution system state estimation.pdf:application/pdf},
}

@article{ieeepowerandenergysocietyIEEEStdC372011,
	title = {{IEEE} {Std} {C37}.118.1™-2011},
	doi = {10.1109/IEEESTD.2011.6111219},
	abstract = {Synchronized phasor (synchrophasor) measurements for power systems are presented. This standard defines synchrophasors, frequency, and rate of change of frequency (ROCOF) measurement under all operating conditions. It specifies methods for evaluating these measurements and requirements for compliance with the standard under both steady-state and dynamic conditions. Time tag and synchronization requirements are included. Performance requirements are confirmed with a reference model, provided in detail. This document defines a phasor measurement unit (PMU), which can be a stand-alone physical unit or a functional unit within another physical unit. This standard does not specify hardware, software, or a method for computing phasors, frequency, or ROCOF.},
	journal = {IEEE Std C37.118.1-2011 (Revision of IEEE Std C37.118-2005)},
	author = {IEEE Power {and} Energy Society},
	month = dec,
	year = {2011},
	note = {173 citations (Crossref) [2022-05-30]
Conference Name: IEEE Std C37.118.1-2011 (Revision of IEEE Std C37.118-2005)},
	keywords = {data concentrator, Data processing, Data storage, DC, IEEE standards, PDC, phasor, phasor measurement unit, Phasor measurement units, PMU, Synchronization, synchronized phasor, synchrophasor, phasor measurement, Vectors, Error statistics, FE, frequency error, IEEE C37.118.1, IRIG-B, RFE, ROCOF, ROCOF error, total vector error, TVE},
	pages = {1--61},
	file = {2011_IEEE Standard for Synchrophasor Measurements for Power Systems.pdf:C\:\\Users\\dassc\\Zotero\\storage\\8F99K5E2\\2011_IEEE Standard for Synchrophasor Measurements for Power Systems.pdf:application/pdf;IEEE Power and Energy Society_2011_IEEE Std C37.pdf:C\:\\Users\\dassc\\Zotero\\storage\\2W6YVCSU\\IEEE Power and Energy Society_2011_IEEE Std C37.pdf:application/pdf;IEEE Xplore Abstract Record:C\:\\Users\\dassc\\Zotero\\storage\\2T9GDBEY\\6111219.html:text/html},
}

@inproceedings{mynamSynchrophasorsRedefiningSCADA2013,
	title = {Synchrophasors redefining {SCADA} systems},
	volume = {26},
	booktitle = {13th {Annual} {Western} {Power} {Delivery} {Automation} {Conference}, {March} 2011},
	author = {Mynam, Mangapathirao V. and Harikrishna, Ala and Singh, Vivek},
	year = {2013},
	note = {Publisher: International Association on Electricity Generation Transmission \& Distribution},
	keywords = {��},
	pages = {22--28},
	file = {Mynam et al_2013_Synchrophasors redefining SCADA systems.pdf:C\:\\Users\\dassc\\Zotero\\storage\\XRKGJJY4\\Mynam et al_2013_Synchrophasors redefining SCADA systems.pdf:application/pdf;Snapshot:C\:\\Users\\dassc\\Zotero\\storage\\HG8JGVJ8\\ijor.html:text/html},
}

@article{wangRevisedBranchCurrentbased2004,
	title = {A revised branch current-based distribution system state estimation algorithm and meter placement impact},
	volume = {19},
	issn = {1558-0679},
	doi = {10.1109/TPWRS.2003.821426},
	abstract = {With the development of automation in distribution systems, distribution supervisory control and data acquisition (SCADA) and many automated meter reading (AMR) systems have been installed on distribution systems. Also distribution management system (DMS) have advanced and include more sophisticated analysis tools. The combination of these developments is providing a platform for development of distribution system state estimation (DSE). A branch-current-based three-phase state estimation algorithm for distribution systems has been developed and tested. This method chooses the magnitude and phase angle of the branch current as the state variables. Because of the limited number of real-time measurements in the distribution system, the state estimator can not acquire enough real-time measurements for convergence, so pseudo-measurements are necessary for distribution system state estimator. The load estimated at every node from the AMR systems is used as a pseudo-measurement for the state estimator. The algorithm has been tested on three IEEE radial test feeders. In addition to this new strategy for DSE, another issue is meter-placement. This topic includes the type of measurement as well as the location of the measurement. Our results show the impact of these two issues on accuracy. Several general meter rules based on this analysis are outlined.},
	number = {1},
	journal = {IEEE Transactions on Power Systems},
	author = {Wang, Haibin and Schulz, N.N.},
	month = feb,
	year = {2004},
	note = {210 citations (Crossref) [2022-06-02]
Conference Name: IEEE Transactions on Power Systems},
	keywords = {State estimation, Voltage, Real time systems, SCADA systems, Automatic meter reading, Instruments, Jacobian matrices, Loss measurement, Meter reading, Testing},
	pages = {207--213},
	file = {IEEE Xplore Abstract Record:C\:\\Users\\dassc\\Zotero\\storage\\UP56TZZI\\1266570.html:text/html;IEEE Xplore Full Text PDF:C\:\\Users\\dassc\\Zotero\\storage\\BZMD9RD8\\Wang and Schulz - 2004 - A revised branch current-based distribution system.pdf:application/pdf;Wang et al_2004_A revised branch current-based distribution system state estimation algorithm.pdf:C\:\\Users\\dassc\\Zotero\\storage\\M5I325SU\\Wang et al_2004_A revised branch current-based distribution system state estimation algorithm.pdf:application/pdf;Wang_Schulz_2004_A revised branch current-based distribution system state estimation algorithm.pdf:C\:\\Users\\dassc\\Zotero\\storage\\9MXQI3SB\\Wang_Schulz_2004_A revised branch current-based distribution system state estimation algorithm.pdf:application/pdf},
}

@article{liLargeScaleTestbedVirtual2020,
	title = {A {Large}-{Scale} {Testbed} as a {Virtual} {Power} {Grid}: {For} {Closed}-{Loop} {Controls} in {Research} and {Testing}},
	volume = {18},
	issn = {1540-7977, 1558-4216},
	shorttitle = {A {Large}-{Scale} {Testbed} as a {Virtual} {Power} {Grid}},
	url = {https://ieeexplore.ieee.org/document/9007798/},
	doi = {10.1109/MPE.2019.2959054},
	language = {en},
	number = {2},
	urldate = {2022-06-12},
	journal = {IEEE Power and Energy Magazine},
	author = {Li, Fangxing and Tomsovic, Kevin and Cui, Hantao},
	month = mar,
	year = {2020},
	note = {9 citations (Crossref) [2022-06-13]},
	pages = {60--68},
	file = {Li et al_2020_A Large-Scale Testbed as a Virtual Power Grid - For Closed-Loop Controls in.pdf:C\:\\Users\\dassc\\Zotero\\storage\\P7I75GY8\\Li et al_2020_A Large-Scale Testbed as a Virtual Power Grid - For Closed-Loop Controls in.pdf:application/pdf;Li et al_2020_A Large-Scale Testbed as a Virtual Power Grid.pdf:C\:\\Users\\dassc\\Zotero\\storage\\V9M73G2V\\Li et al_2020_A Large-Scale Testbed as a Virtual Power Grid.pdf:application/pdf;Li et al. - 2020 - A Large-Scale Testbed as a Virtual Power Grid For.pdf:C\:\\Users\\dassc\\Zotero\\storage\\784DU5RE\\Li et al. - 2020 - A Large-Scale Testbed as a Virtual Power Grid For.pdf:application/pdf},
}

@article{sandiPYTHONIMPLEMENTATIONIEEE,
	title = {{PYTHON} {IMPLEMENTATION} {OF} {IEEE} {C37}.118 {COMMUNICATION} {PROTOCOL}},
	volume = {21},
	abstract = {The analysis of tools for supporting the measurement of synchrophasors revealed the need for open, customizable, and platform independent tools. One such software tool is the open implementation of the IEEE C37.118 communication protocol for data transfer. This paper describes an implementation process of the protocol in a form of a Python library module. The discussion illustrates its use and validation using third-party tools. Finally, the paper outlines future work on the improvements and possible practical applications.},
	language = {en},
	number = {1},
	author = {Šandi, Stevan and Popovi, Tomo},
	pages = {10},
	file = {Šandi and Popovi - PYTHON IMPLEMENTATION OF IEEE C37.118 COMMUNICATIO.pdf:C\:\\Users\\dassc\\Zotero\\storage\\WMA6NCK3\\Šandi and Popovi - PYTHON IMPLEMENTATION OF IEEE C37.118 COMMUNICATIO.pdf:application/pdf},
}

@article{zhangDesignTestingImplementation2017a,
	title = {Design, {Testing}, and {Implementation} of a {Linear} {State} {Estimator} in a {Real} {Power} {System}},
	volume = {8},
	issn = {1949-3053, 1949-3061},
	url = {http://ieeexplore.ieee.org/document/7373664/},
	doi = {10.1109/TSG.2015.2508283},
	abstract = {As more Phasor Measurement Units (PMUs) are installed, portions of the power grid become observable with just phasor measurements, making feasible the estimation of the state of these observable portions at much faster rates than the traditional state estimator. Although such a Linear State Estimator (LSE) was proposed over a decade ago, the ﬁrst ﬁeld implementation was completed at Dominion Virginia Power in 2013. Although this LSE is a stand-alone function not integrated to their Energy Management System (EMS), it demonstrated the feasibility of the LSE. In this paper, we present the design, development and implementation of an LSE that is fully integrated with the existing EMS and can estimate the state of the Extra High Voltage (EHV) portion of a power system at 30 times per second. Integration of the LSE to the existing EMS environment and some of the issues in the design and testing are presented in this paper. The work paved way for LSE to supply cleansed PMU data to other synchrophasor applications that are sensitive to data quality.},
	language = {en},
	number = {4},
	urldate = {2022-06-14},
	journal = {IEEE Transactions on Smart Grid},
	author = {Zhang, Lin and Bose, Anjan and Jampala, Anil and Madani, Vahid and Giri, Jay},
	month = jul,
	year = {2017},
	note = {41 citations (Crossref) [2022-06-14]},
	pages = {1782--1789},
	file = {Zhang et al_2017_Design, Testing, and Implementation of a Linear State Estimator in a Real Power.pdf:C\:\\Users\\dassc\\Zotero\\storage\\LK95XZLG\\Zhang et al_2017_Design, Testing, and Implementation of a Linear State Estimator in a Real Power.pdf:application/pdf;Zhang et al. - 2017 - Design, Testing, and Implementation of a Linear St.pdf:C\:\\Users\\dassc\\Zotero\\storage\\9X9YVX8X\\Zhang et al. - 2017 - Design, Testing, and Implementation of a Linear St.pdf:application/pdf},
}

@phdthesis{bandakPOWERSYSTEMSSTATE2013,
	title = {{POWER} {SYSTEMS} {STATE} {ESTIMATION}},
	author = {Bandak, Carlos Expedite},
	year = {2013},
	file = {Bandak_2013_POWER SYSTEMS STATE ESTIMATION.pdf:C\:\\Users\\dassc\\Zotero\\storage\\V64W3G9Q\\Bandak_2013_POWER SYSTEMS STATE ESTIMATION.pdf:application/pdf},
}

@phdthesis{taraliBadDataDetection2012,
	title = {Bad data detection in two stage estimation using phasor measurements},
	url = {http://hdl.handle.net/2047/d20002926},
	language = {en},
	urldate = {2022-06-15},
	school = {Northeastern University},
	author = {Tarali, Aditya},
	year = {2012},
	doi = {10.17760/d20002926},
	file = {Tarali - 2012 - Bad data detection in two stage estimation using p.pdf:C\:\\Users\\dassc\\Zotero\\storage\\KPKT33WM\\Tarali - 2012 - Bad data detection in two stage estimation using p.pdf:application/pdf},
}


% Generated by IEEEtran.bst, version: 1.14 (2015/08/26)
	
	\iffalse
	
	\section*{Acknowledgments}
	This should be a simple paragraph before the References to thank those individuals and institutions who have supported your work on this article.

	\section{Appendix}
	
	\begin{figure*}[t]
		\includegraphics[width=\linewidth]{pySynphasor_dissection.pdf}
		\caption{Scapy dissection of configuration packet}
		\label{fig:scapy dissection}
	\end{figure*}
	\fi
	
\end{document}